\documentclass[twocolumn,showpacs,preprintnumbers,amsmath,amssymb,floatfix]{revtex4}
\usepackage{graphicx}
\usepackage{dcolumn}
\usepackage{bm}
\begin{document}

\title{Systematical study of optical potential strengths in reactions involving strongly, weakly bound and exotic nuclei on $^{120}$Sn}
      
\author{M. A. G. Alvarez, J. P. Fern\'{a}ndez-Garc\'{i}a, J. L. Le\'{o}n-Garc\'{i}a}
\affiliation{Departamento FAMN, Universidad de Sevilla, Apartado 1065, 41080 Sevilla, Spain}
\author{M. Rodr\'{i}guez-Gallardo}
\affiliation{Departamento FAMN, Universidad de Sevilla, Apartado 1065, 41080 Sevilla, Spain and \\ Instituto Carlos I de F\'{i}sica Te\'{o}rica y Computacional, Universidad de Sevilla, Spain}
\author{L. R. Gasques, L. C. Chamon, V. A. B. Zagatto, A. L\'{e}pine-Szily, J. R. B. Oliveira}   
\affiliation{Universidade de Sao Paulo, Instituto de Fisica, Rua do Matao, 1371, 05508-090, Sao Paulo, SP, Brazil} 
\author{V. Scarduelli}   
\affiliation{Instituto de F\'{i}sica, Universidade Federal Fluminense, 24210-340 Niter\'{o}i, Rio de Janeiro, Brazil and \\ Instituto de F\'{i}sica da Universidade de S\~ao Paulo, 05508-090, S\~ao Paulo, SP, Brazil}
\author{B. V. Carlson}  
\affiliation{Departamento de F\'{i}sica, Instituto Tecnol\'{o}gico de Aeron\'{a}utica, S\~{a}o Jos\'{e} dos Campos, SP, Brazil}
\author{J. Casal}  
\affiliation {Dipartimento di Fisica e Astronomia "G. Galilei" and INFN-Sezione di Padova, Via Marzolo, 8, I-35131, Padova, Italy}
\author{A. Arazi}  
\affiliation{Laboratorio TANDAR, Comisi\'{o}n Nacional de Energ\'{i}a At\'{o}mica, Av. Gral. Paz 1499, BKNA1650 San Mart\'{i}n, Argentina}
\author{D. A. Torres, F. Ramirez}
\affiliation{Departamento de F\'{\i}sica, Universidad  Nacional de Colombia, Bogot\'a, Colombia}

\date{\today}     
 
\begin{abstract} 
We present new experimental angular distributions for the elastic scattering 
of $^{6}$Li + $^{120}$Sn at three bombarding energies. We include these data 
in a wide systematic involving the elastic scattering of $^{4,6}$He, $^{7}$Li, 
$^{9}$Be, $^{10}$B and $^{16,18}$O projectiles on the same target at energies 
around the respective Coulomb barriers. Considering this data set, we report on 
optical model analyses based on the double-folding S\~ao Paulo Potential. Within 
this approach, we study the sensitivity of the data fit to different models for 
the nuclear matter densities and to variations in the optical potential 
strengths.
\end{abstract}      
      
      
\maketitle      
      
\section{Introduction}      
\label{introduct}      

Nuclei present cluster structures \cite{Hori13}. Light, strongly or weakly bound, stable or exotic, nuclei such as $^{6}$He, $^{6,7}$Li, $^{7,8,9}$Be, $^{12,13,14}$C, $^{16,18}$O, among others (isotopes and nuclei), can be considered as results of n, $^{1,2,3}$H and $^{3,4}$He combinations. It has been evidenced by experimental observations on break-up or transfer reactions (e.g. \cite{Luo13,Esc07,DiPietro2004,jp11li}). 

The $^{4}$He possesses a significantly higher binding energy per nucleon than its light neighbors (see Table I), and a first excited state with very high excitation energy (20.6 MeV) that makes it a rather robust and inert nucleus. 

Unlike $^{4}$He, $^{6}$He is an exotic nucleus that decays, by beta minus emission, in $^{6}$Li, with a half-life of 806.7(15) ms \cite{Till02}. It is a Borromean nucleus, i.e., the two sub-systems, $^{4}$He-$n$ and $n$-$n$, are not bound. Reactions induced by $^{6}$He on different targets, at energies around the Coulomb barrier, exhibit a remarkable large cross section for $\alpha$ particles production \cite{Esc07,DiPietro2004}. It confirms a break-up picture, which is associated to the weak binding of the halo neutrons ($S_{2n}$ = 0.98 MeV - Table I) \cite{Till02}, that favours the dissociation of the $^{6}$He projectile.

\begin{table}[h]
\caption{\label{tab:qvalueBU} Binding energy per nucleon, proton and neutron
separation energies, possible mode of break-up and corresponding $Q$ value for some 
nuclei. All energies are provided in MeV.}
\begin{ruledtabular}
\begin{tabular}{ c c c c c c }
nucleus & $BE/A$ & $S_{1p}$ & $S_{1n}$ & cluster & $Q$ \\
\hline \\
 $^4$He  & 7.07 & 19.81 & 20.58 & & \\
 $^6$He  & 4.88 & 22.59 & 1.71 & $\alpha + n + n$ & $-0.98$ \\
 $^6$Li  & 5.33 & 4.43 & 5.66 & $\alpha + d$ & $-1.47$ \\
 $^7$Li  & 5.61 & 9.97 & 7.25 & $\alpha + t$ & $-2.47$ \\  
 $^9$Be  & 6.46 & 16.89 & 1.66 & $\alpha + \alpha+n$ & $-1.57$ \\
 $^{10}$B & 6.47 & 6.59 & 8.44 & $^6$Li + $\alpha$ & $-4.46$ \\
 $^{11}$B & 6.93 & 11.23 & 11.54 & $^7$Li + $\alpha$ & $-8.66$ \\
\end{tabular}
\end{ruledtabular}
\end{table} 

$^{7}$Li is one of the heaviest nuclides formed with very small yields during the 
primordial Big-Bang nucleosynthesis. Stable nuclei heavier than $^{7}$Li were 
formed much later through light nuclei reacting during stellar evolution or explosions. 
Despite small amounts of $^{6}$Li and $^{7}$Li being produced in stars, they are 
expected to be burned very fast. Additional small amounts of both, $^{6}$Li and 
$^{7}$Li, may be generated from cosmic ray spallation on heavier atoms in the 
interstellar medium, from solar wind and from early solar system $^{7}$Be and 
$^{10}$Be radioactive decays \cite{Chau06}.
 
Both $^{6}$Li and $^{7}$Li have an anomalous low nuclear binding energy per nucleon 
compared to their stable neighbors (see Table I). In fact, these lithium isotopes 
have lower binding energy per nucleon than any other stable nuclide with $Z>3$. As a 
consequence, even being light, $^{6,7}$Li are less common in the solar system than 25 
of the first 32 chemical elements \cite{Lod03}. The $^{6}$Li and $^{7}$Li nuclei are stable weakly bound isotopes for which strong break-up effects are expected in collisions with other nuclei. These isotopes can be considered as $\alpha + d$ and $\alpha + t$ clusters, with small $Q$ values (see Table I). 

Luong {\it et al.} \cite{Luo13} showed that break-up of $^{6}$Li into its $\alpha + d$ 
constituents dominates in reactions with heavy targets. However, break-up triggered 
by nucleon transfer is highly probable. As an example, in the case of a $^{6}$Li beam
focusing on a $^{120}$Sn target these processes could be: \\
$^{6}$Li + $^{120}$Sn $\rightarrow$ $^{121}$Sn + $^{4}$He + $p$; \\
$^{6}$Li + $^{120}$Sn $\rightarrow$ $^{121}$Sb + $^{4}$He + $n$. \\
These strong break-up mechanisms triggered by nucleon transfer help in explaining the 
large number of $\alpha$ particles observed in different $^{6}$Li reactions 
\cite{Luo13,Ost72}. In Table II, we present $Q$ values of possible break-up processes 
triggered by transfer for systems involving some weakly bound projectiles on a 
$^{120}$Sn target.

\begin{table}[h]
\caption{\label{tab:qvalue}  $Q$ values of some break-up processes triggered (or not) 
by transfer for weakly bound projectiles colliding with a $^{120}$Sn target.}
\begin{ruledtabular}
\begin{tabular}{ c c c }
 projectile & reaction products & $Q$(MeV) \\
\hline \\
 $^6$Li  & $^{121}$Sn + $\alpha+p$ & 2.472 \\
 $^6$Li  & $^{121}$Sb + $\alpha+n$ & 2.092 \\
 \hline \\
 $^7$Li  & $^{122}$Sn + $\alpha+p$ & 4.036 \\
 $^7$Li  & $^{122}$Sb + $\alpha+n$ & 1.247 \\
 \hline \\
 $^9$Be  & $^{121}$Sn + $\alpha+\alpha$ & 4.597 \\
 $^9$Be  & $^{120}$Sn + $^8$Be + $n$ &  4.505 \\
 \hline \\
 $^{10}$B  & $^{121}$Sn + 2$\alpha+p$ & -1.989 \\
 $^{10}$B  & $^{121}$Sb + 2$\alpha+n$ & -2.368 \\
\end{tabular}
\end{ruledtabular}
\end{table} 

Unlike $^{6}$Li, $^{7}$Li presents a first excited state with relatively low excitation energy ($E^* = 0.48$ MeV). The $^{7}$Li nucleus also has a small 
binding energy for the $\alpha + t$ break-up, which is, however, about 1 MeV higher 
than that for $^{6}$Li (see Table I). Even so, in reactions of $^{7}$Li, the 
break-up 
channel of the $\alpha + t$ cluster is relevant \cite{Luo13}. Notwithstanding, $^{8}$Be 
formation (with subsequent $\alpha + \alpha$ decay) through a proton pick-up 
transfer process ($Q=6.658$ MeV) is more probable.
          
The $^{9}$Be nucleus presents a Borromean structure composed of two $\alpha$ 
particles and one weakly bound neutron \cite{Cas14}. It has a binding energy for the 
$\alpha + \alpha + n$ break-up comparable to that for $^{6}$Li (see Table I). 
The 1n-separation energy of $^{9}$Be is quite small in comparison 
with those for the other nuclei of Table I. Thus, when colliding with a target 
nucleus, $^{9}$Be tends (with high probability) to transfer its weakly bound 
neutron, with $\alpha + \alpha$ or $^{8}$Be formation (the later followed by 
$\alpha + \alpha$ decay). In \cite{Ara18}, Arazi {\it et al.}
demonstrated the importance of couplings to unbound states to obtain theoretical 
agreement with the $^{9}$Be + $^{120}$Sn data set, at energies around the Coulomb barrier, corroborating break-up as an important process.

Similar to $^{7}$Li, $^{10}$B also presents a first excited state with low 
excitation energy ($E^*=0.72$ MeV). However, compared to $^{6,7}$Li and 
$^{9}$Be 
(Table I), its most favorable break-up channel, $^{10}$B $\rightarrow$ $^{6}$Li + 
$^{4}$He, is energetically higher and, therefore, less probable. In addition, 
considering the different values of the 1n-separation energy (Table I), 
break-up triggered by nucleon transfer is not as favored for $^{10}$B as it is 
for $^{9}$Be. In \cite{Alv18}, we demonstrated that couplings to the continuum 
states are not important to obtain a good agreement between theoretical 
calculations and experimental data for $^{10}$B + $^{120}$Sn, at energies around 
the Coulomb barrier, indicating that 
break-up is not an important process in this case. The above mentioned features 
indicate a very different reaction dynamics for $^{9}$Be and $^{10}$B weakly
bound projectiles reacting with $^{120}$Sn.

Studying reactions involving weakly bound stable nuclei is a crucial step towards a 
better understanding of their abundances. The structural models of these nuclei are 
fundamental to determine how they interact and, therefore, to shed light on such abundances. Weakly bound nuclei, in general, have fundamental structural 
characteristics, such as the above mentioned low break-up thresholds and cluster 
structures. Break-up can lead to a complex problem of three or more bodies, and can 
occur by direct excitation of the weakly bound projectile into continuum states or by 
populating continuum states of the target 
\cite{Esc07,San08,Aco09,Aco11,Raf10,Luo11,Kal16}. 

Weakly bound stable nuclei can easily be produced and accelerated, with high 
intensities, in conventional particle accelerators. Within this context, complementary
experimental campaigns are being developed in two laboratories: the 8 MV tandem 
accelerator of the Open Laboratory of Nuclear Physics (LAFN, acronym in Portuguese) 
in the Institute of Physics of the University of S\~ao Paulo (Brazil), and the 20 MV 
tandem accelerator TANDAR (Buenos Aires, Argentina). The aim of the joint collaboration 
is to study the scattering involving stable, strongly and weakly bound, nuclei on the 
same target ($^{120}$Sn), at energies around the respective Coulomb barriers. These 
measurements allow systematic studies that involve the comparison of behavior 
for the different projectiles. 

Many data, obtained in our experiments, with $^{120}$Sn as target, have already been 
published \cite{Zag17,Ara18,Gas18,Alv18}. In the present paper, we present new 
experimental angular distributions for the elastic scattering of the $^{6}$Li + 
$^{120}$Sn system, at three bombarding energies. We include these data in a wide 
systematic involving the elastic scattering of $^{4,6}$He, $^{7}$Li, $^{9}$Be, 
$^{10}$B 
and $^{16,18}$O projectiles, on the same target, at energies around the respective 
Coulomb barriers. We analyze the complete data set within the approach of the 
optical model (OM), assuming the double-folding S\~ao Paulo Potential (SPP) \cite{Cha02} for 
the real part of the optical potential (OP) and two different models for the 
imaginary part.  With this, we study the behavior of the OP as a function of 
the energy for the different projectiles. 

In the next section, we present a summarized review of the experiments. It will be 
followed by the explanation of the theoretical approach and corresponding 
application to the experimental data. Then, we discuss and compare the behaviors 
of the OPs that fit the data for different projectiles. Finally, we present 
our main conclusions.  
          
\section{The experiments}      
\label{Data}      

The measurements for the $^{6,7}$Li, $^{10,11}$B + $^{120}$Sn systems are part 
of the E-125 experimental campaign, developed at the LAFN, and correspond to the 
following energies: 1) $^{6}$Li at $E_{\rm LAB}=$ 19, 24 and 27 MeV, reported 
for the the first time in this paper; 2) $^{7}$Li at $E_{\rm LAB}=$ 20, 22, 24 
and 26 MeV \cite{Zag17}; 3) $^{10}$B at $E_{\rm LAB}=$ 31.35, 33.35, 34.85 and 
37.35 MeV \cite{Gas18,Alv18}. 
The experimental setup is based on SATURN (Silicon Array 
based on Telescopes of USP for Reactions and Nuclear applications). SATURN 
is installed in the 30B experimental beam line of the laboratory, which contains 
a scattering chamber connected to the accelerator. The SATURN detection system has 
been mounted with 9 surface barrier detectors in angular intervals of 
5$^{\rm o}$. With this, in 3 runs we cover an angular range of 120$^{\rm o}$, 
from 40$^{\rm o}$ to 160$^{\rm o}$. The targets contained $^{120}$Sn and $^{197}$Au, the latter 
used for the purpose of normalization. Further details are found in 
\cite{Alv18}, \cite{Zag17} and \cite{Gas18}.

The experimental data for $^{9}$Be+$^{120}$Sn were obtained at the TANDAR laboratory, at 
$E_{\rm LAB}=$ 26, 27, 28, 29.5, 31, 42 and 50 MeV. An array of eight surface barrier 
detectors, with an angular separation of 5$^\circ$ between adjacent detectors, was used to 
distinguish scattering products. All details about data acquisition and analysis are 
presented in \cite{Ara18}.

In addition to our data, other experimental elastic scattering cross sections, for systems involving $^{120}$Sn as target, were obtained from \cite{Mohr10, Kum68, Cel01, Boh75, Far10, Appa19, Zer12, Kun17}.
         
\section{The theoretical approach}
\label{Data}      

Data of heavy-ion nuclear reactions have been successfully described in many 
works assuming double-folding theoretical models for the nuclear potential 
\cite{Sat79,Sat83,Sat91,Sat87,Khoa88,Bran88,Len89,Sat94,Bran97,Bran97rep}. 
Among these models, the SPP \cite{Cha02} associates the nuclear interaction to 
a dependence on the local velocity. The model includes a systematic of nuclear 
densities obtained for stable strongly bound nuclei and, in this context, it 
does not contain any free parameter. The SPP is related to the double-folding 
potential through:
\begin{equation} 
V_{\text{SPP}}(R)=V_{\text{Fold}}(R) \, e^{-4v^{2}/c^{2}}, \label{eq:1}
\end{equation}
where $c$  is the speed of light and $v(R)$ is the local relative velocity 
between projectile and target. At energies around the Coulomb barrier (as in the
present analysis) the velocity is much smaller than the speed of light and we
have: $V_{\text{SPP}}(R) \approx V_{\text{Fold}}(R)$. The folding potential is 
represented as: 
\begin{equation}
V_{\text{Fold}}(R)=\int \int \rho_{1}(\vec{r}_{1})\rho_{2}(\vec{r}_{2})V_{0}
\, \delta(\vec{R}-\vec{r}_1+\vec{r}_2) \, d\vec{r}_1 \, d\vec{r}_2 . 
\label{eq:2}
\end{equation}
Here, $\rho_{1}$ and $\rho_{2}$ are the projectile and target matter 
distributions, and  $V_{0} \, \delta(\vec{r})$ is the zero-range effective 
interaction (with $V_{0}=-456$ MeV fm$^{3}$). This $V_{0}$ value was obtained 
in \cite{Cha02}, through a very wide systematic involving phenomenological 
potentials extracted from elastic scattering data analyses for many systems. 
For a particular nucleus, the respective nucleon distribution is folded with the 
matter density of one nucleon to obtain the corresponding matter density of the 
nucleus (see \cite{Cha02}). 

An important point that stands out against obtaining a systematical description 
of the elastic scattering process with an OP (within the OM) is the 
difficulty in describing the imaginary part of the interaction from fundamental
grounds. A fully microscopic description based on the Feshbach theory is 
specially difficult at energies where collective as well as single particle 
excitations are important in the scattering process \cite{Pol81, Pol83, Sak87}. 
To face this problem within a simple model, an extension of the SPP to the OP 
imaginary part was proposed in \cite{Alv03}, considering the following OP:
\begin{equation} 
U_{OP}(R)= V_{\text{SPP}}(R) + i \, N_I \, V_{\text{SPP}}(R). 
\end{equation}
Elastic scattering data for many systems, at high energies, have been  
described using $N_I \approx 0.78$ \cite{Alv03}. At energies around the Coulomb 
barrier, the SPP has also been valuable in coupled channel calculations for
systems involving strongly (see e.g. \cite{Per06}) and weakly bound  
(e.g. \cite{Zag17,Gas18,Alv18}) projectiles. Furthermore, the SPP has accounted for 
data of  
systems with exotic nuclei (e.g. \cite{Fer10, Fer15}). Besides being 
successful in elastic scattering data analyses, the SPP has also provided 
good descriptions of data for the fusion process of many systems (e.g. 
\cite{Gas04,Can09,Nob76,Nob78,Nob07}).

In the present work, we propose the SPP theoretical approach in the context of 
the OM to systematically study the elastic scattering data for the 
$^{4,6}$He, $^{6,7}$Li, $^{9}$Be, $^{10}$B, $^{16,18}$O + $^{120}$Sn systems, at energies around the Coulomb barrier. We assume equation (4) to 
describe the OP:  
\begin{equation} 
U_{OP}(R) = N_R \, V_{\text{SPP}}(R) + i \, N_I \, V_{\text{SPP}}(R), 
\label{eq:3}
\end{equation}
where $N_R$ and $N_I$ represent multiplicative factors that determine the 
strengths of the OP (real and imaginary parts) and simulate, in a simple form, 
the effects of the polarization potential. The polarization arises from 
nonelastic couplings. According to Feshbach's theory \cite{Fesh92, Bran97}, 
it is energy dependent and complex. The imaginary part comes from transitions 
to open non-elastic channels that absorb flux from the elastic channel. The 
real part arises from virtual transitions to intermediate states (inelastic 
excitations, nucleon transfer, among others). As already commented, standard 
average values obtained in \cite{Alv03} are $N_R=1$ and $N_I=0.78$.

For the purpose of comparison and with the aim of accounting only for the 
internal absorption (fusion) from barrier penetration, without taking into 
account the effect of the couplings, we also perform OM calculations based on 
equation (5):
\begin{equation} 
U_{OP}(R) = V_{\text{SPP}}(R) + i \, W(R), 
\label{eq:4}
\end{equation} 
where $W(R)$ has a Woods-Saxon (WS) shape, 
\begin{equation} W(R) = W_{0}/\left[1+ \exp{(R-R_{0})/a}\right], 
\end{equation}
with $W_{0} = -100$ MeV, $R_{0} = r_0 \left( A_1^{1/3} + A_2^{1/3} \right)$, 
$r_0= 1.06$ fm and $a=0.25$ fm. Due to the small diffuseness value, such an internal imaginary potential just simulates the fusion process (without 
couplings) and does not take into account the absorption by the peripheral 
channels.  

Before proceeding with the OM analyses, we first examine the effects of the
densities on the nuclear interaction. As already commented, the SPP involves a 
systematics of densities that makes the interaction a parameter-free model.
However, one can question if the use of this systematics for weakly bound nuclei
is appropriate. Thus, we have calculated nuclear densities through theoretical 
Hartree-Bogoliubov (HB) calculations \cite{Carl00}, assuming two different 
interactions: the NL3 and DDME1 models \cite{Lala02,Nik02}. Figure 1 shows a comparison of different approaches for the matter densities of light weakly bound nuclei: the 
two-parameter Fermi systematic of the SPP and the theoretical HB. In the 
cases of $^{6}$Li and $^{10}$B (where $N = Z$), we also present in Fig. 1 the 
experimental charge density (obtained from electron scattering) multiplied by 2. 
Except for $^6$He, all these densities are very similar, and therefore the use 
of the systematics for densities of the SPP is justified. We have also verified 
that very similar values of cross sections are obtained from OM calculations 
using these different models for the densities. In the $^6$He case, the
theoretical HB density is rather different from that of the systematics at the
surface region. Thus, we have taken an ``experimental'' density for this nucleus,
obtained from data analyses of proton scattering at high energies \cite{Alk97}.
The dashed-dotted orange line in Fig. 1(d) represents this ``experimental'' matter density 
(obtained from folding the nucleon distribution with the matter density of the 
nucleon, according to \cite{Cha02}). The ``experimental'' density is quite similar 
to that from the systematics of the SPP (blue line). Thus, we consider that, 
even in the $^6$He case, the use of the SPP systematics for densities is 
justified.

\begin{figure}[h]
\begin{minipage}{20pc}
\includegraphics[width=23pc]{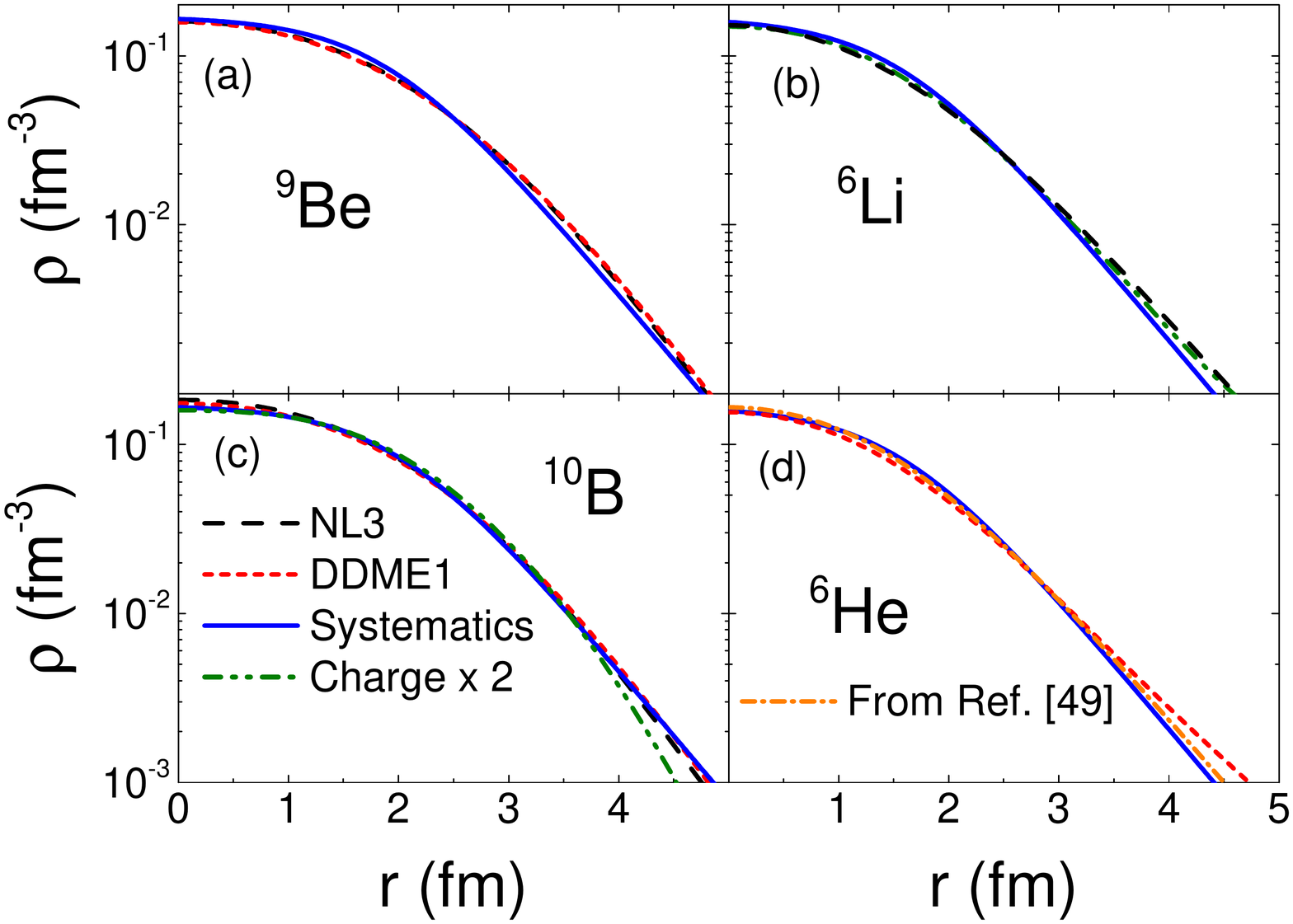}
\caption{\label{dens} (Color online) Matter densities for $^{10}$B, $^{9}$Be,
$^{6}$Li and $^{6}$He, calculated through different models (see text for
details).}
\end{minipage}\hspace{2pc}%
\end{figure}

\section{Standard Optical Model Calculations}      
\label{theo}      

Before providing the results of the elastic scattering data fits, we present a 
comparison of the experimental angular distributions with OM cross sections
obtained assuming the standard models for the OP. By standard models we mean 
Equation (4) with $N_R=1$ and $N_I=0.78$, and Equation (5) (internal 
imaginary potential). From now on, we refer to these standard models as Strong 
Surface Absorption (SSA) and Only Internal Absorption (OIA), respectively. 
In order to illustrate the region of energy of the data, for each angular 
distribution we provide the value of the reduced energy, defined as:
\begin{equation}
E_{Red} = E_{c.m.} - V_B ,
\end{equation}
where $E_{c.m.}$ represents the center of mass energy and $V_B$ is the s-wave
barrier height, obtained for the respective system with the SPP. In Table III 
we present the barrier heights, radii and curvatures ($\hbar w$) \cite{Gas04}, for the systems studied in the present work.

\begin{table}[h]
\caption{Values of the s-wave barrier parameters obtained with the SPP for
systems composed by projectiles focusing on $^{120}$Sn.}
\begin{ruledtabular}
\begin{tabular}{ c c c c }
 projectile & $V_B$(MeV) & $R_B$(fm) & $\hbar w$(MeV) \\
\hline \\
 $^4$He & 14.22 & 9.48 & 4.92 \\
 $^6$He & 12.78 & 10.52 & 3.35 \\
 $^6$Li & 19.76 & 10.16 & 4.20 \\
 $^7$Li & 19.45 & 10.34 & 3.86 \\
 $^9$Be & 25.78 & 10.40 & 3.93 \\
 $^{10}$B & 32.38 & 10.34 & 4.17 \\
 $^{16}$O & 50.79 & 10.56 & 4.14 \\
 $^{18}$O & 50.05 & 10.74 & 3.86 \\
\end{tabular}
\end{ruledtabular}
\end{table}

Figure 2 presents four experimental angular distributions for the strongly bound $^{4}$He projectile \cite{Mohr10}. The energies of the angular distributions vary from 5.1 to 19.1 MeV above the barrier ($5.1 \le E_{Red} \le 19.1$ MeV).
To avoid overlapping results, the cross sections for two angular 
distributions have been displaced by a constant factor of 0.5. The solid blue and dashed green lines represent the theoretical results obtained with SSA and OIA, respectively. Both standard models provide rather similar results, but the SSA accounts for the data with slightly better accuracy.

\begin{figure}[h]
\begin{minipage}{20pc}
\includegraphics[width=20pc]{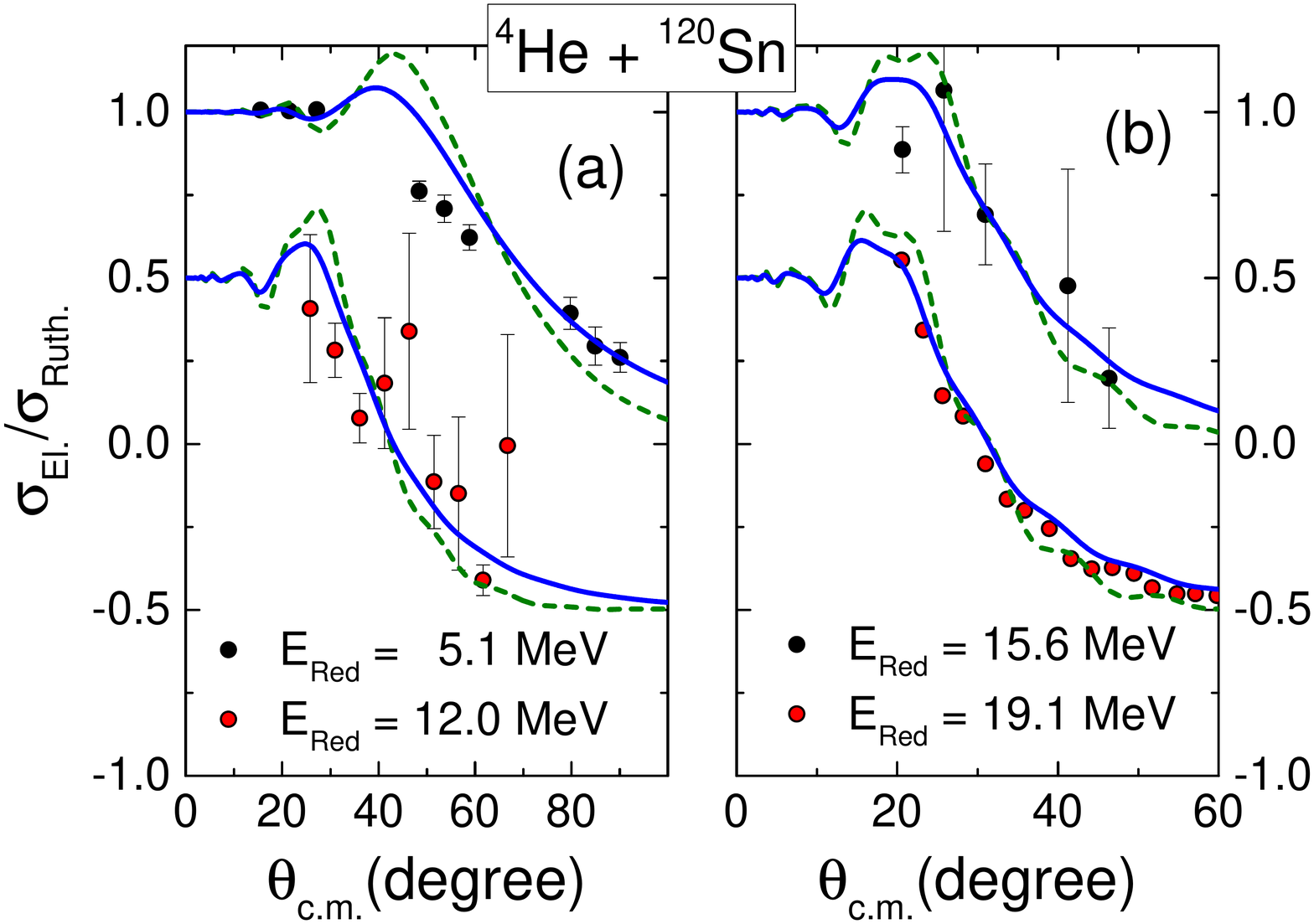}
\caption{\label{dens} (Color online) Experimental angular distributions for the elastic scattering of the $^{4}$He + $^{120}$Sn system \cite{Mohr10, Kum68}. To avoid overlapping results, the cross sections for $E_{Red} = 12.0$ and 19.1 MeV have been displaced by a constant factor of 0.5. The solid and dashed lines represent theoretical OM cross sections obtained with the SSA and OIA models, respectively.}
\end{minipage}\hspace{2pc}%
\end{figure}

Figure 3 presents experimental and theoretical (SSA and OIA) angular 
distributions for the strongly bound $^{16}$O and $^{18}$O projectiles. In the $^{16}$O case, all energies are below the corresponding barrier height. At the  lowest energy ($E_{Red} = -4.0$ MeV) the data are compatible with internal absorption (OIA), while for higher energies they approach to the results of strong surface absorption (SSA). In the case of $^{18}$O, the energy is slightly above the barrier and the SSA reproduces well the data set, except at the rainbow region ($\theta_{c.m.} \approx 90^{\rm o}$).

\begin{figure}[h]
\begin{minipage}{20pc}
\includegraphics[width=20pc]{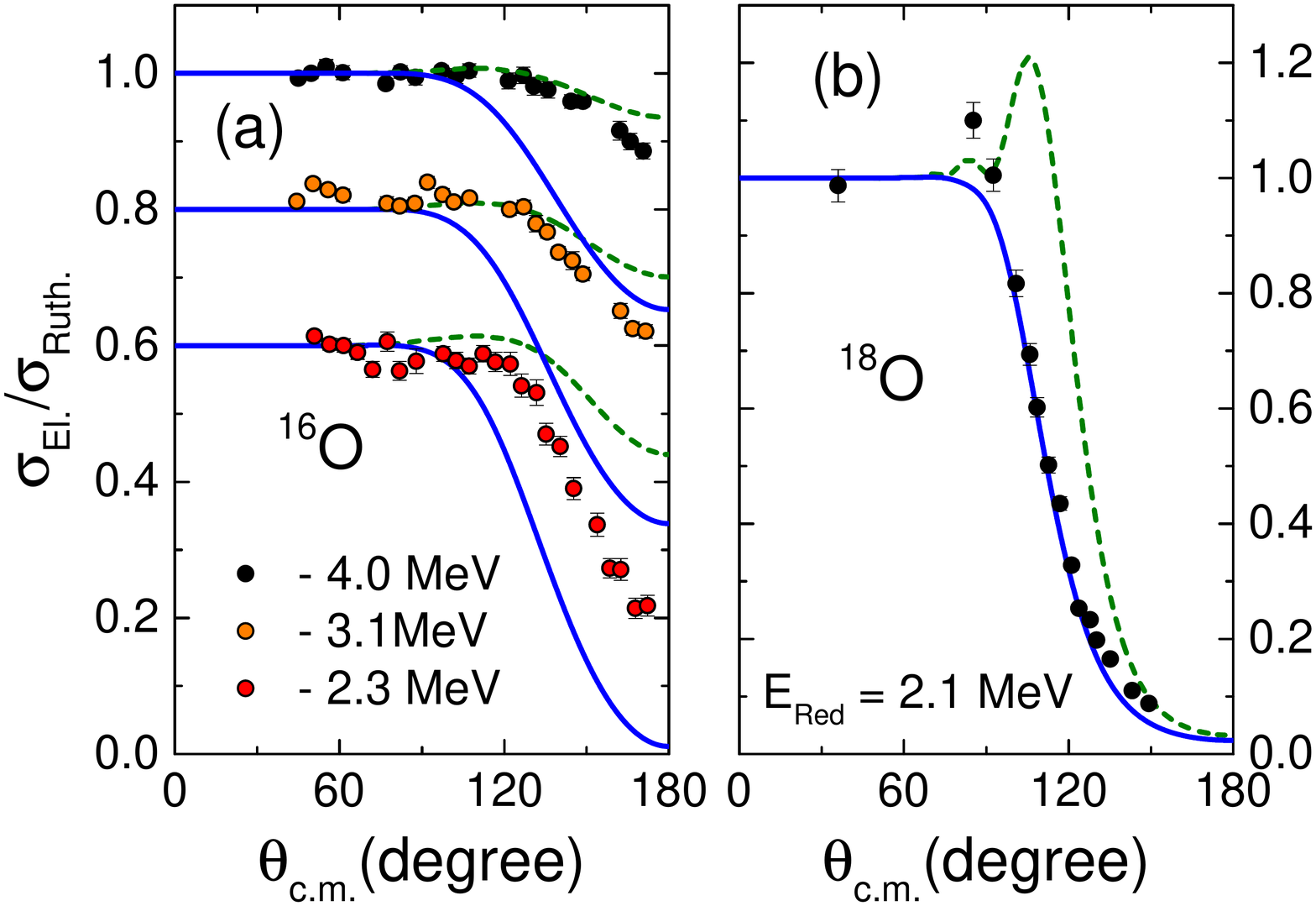}
\caption{\label{dens} (Color online) Experimental angular distributions for the $^{16,18}$O + $^{120}$Sn systems \cite{Cel01, Boh75}. To avoid overlapping results, the cross sections for some distributions have been displaced by a constant factor. The solid and dashed lines represent theoretical OM cross sections obtained with the SSA (solid) and OIA(dashed) models, respectively}
\end{minipage}\hspace{2pc}%
\end{figure}

Figure 4 presents data and theoretical predictions (SSA and OIA) for the elastic
scattering of the exotic $^{6}$He on $^{120}$Sn \cite{Far10, Appa19}. Again the SSA provides a good
description of the data, with some deviation for the lowest $E_{Red} = 3.8$ and 4.4 MeV, due to transfer/break-up channels \cite{Appa19}.

\begin{figure}[h]
\begin{minipage}{20pc}
\includegraphics[width=20pc]{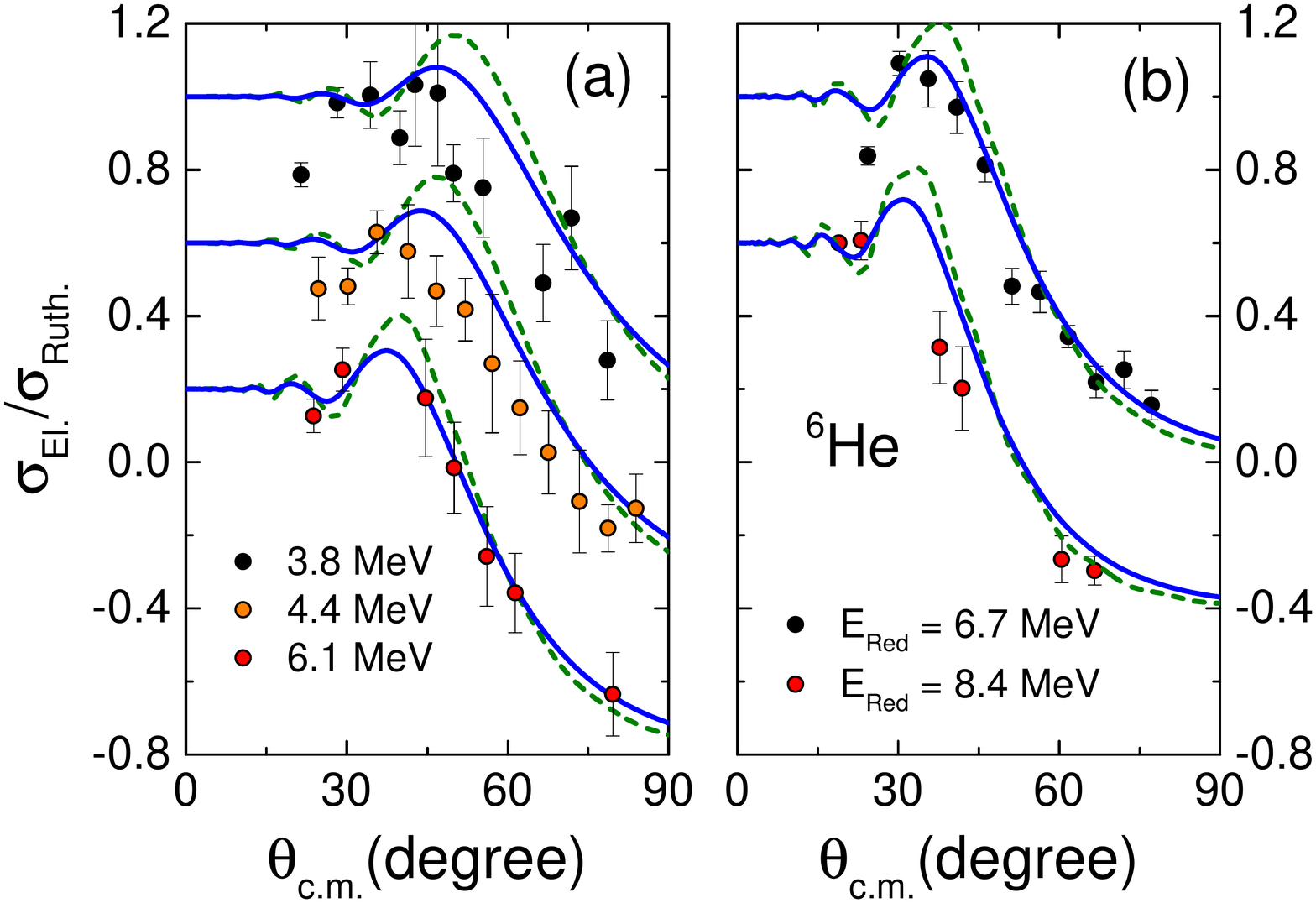}
\caption{\label{dens} (Color online) The same as Fig. 3, for the
$^6$He + $^{120}$Sn system. Data were extracted from \cite{Far10, Appa19}.}
\end{minipage}\hspace{2pc}%
\end{figure}

In Fig. 5, with the present new data, we show theoretical predictions for $^{6}$Li + $^{120}$Sn, at energies around the barrier, in linear (a) and logarithmic (b) scales. The SSA cross sections (solid blue lines) are in good agreement with the data, including at $E_{Red} = -1.66$ MeV, which indicates strong surface absorption even in the sub-barrier energy region. For comparison, in Fig. 6 we present an excitation function for the
elastic scattering of $^{6}$Li + $^{120}$Sn from earlier measurements 
\cite{Zer12}. The data correspond to an angular range of $160^{\rm o} \le \theta_{Lab} \le 170^{\rm o}$. The  solid line represents the SSA 
cross sections at the average angle $\theta_{c.m.}= 165.7^{\rm o}$. There is a reasonable agreement between experimental and theoretical results, but the slope of the data is somewhat different from that of the OM calculations.

\begin{figure}[h]
\begin{minipage}{20pc}
\includegraphics[width=20pc]{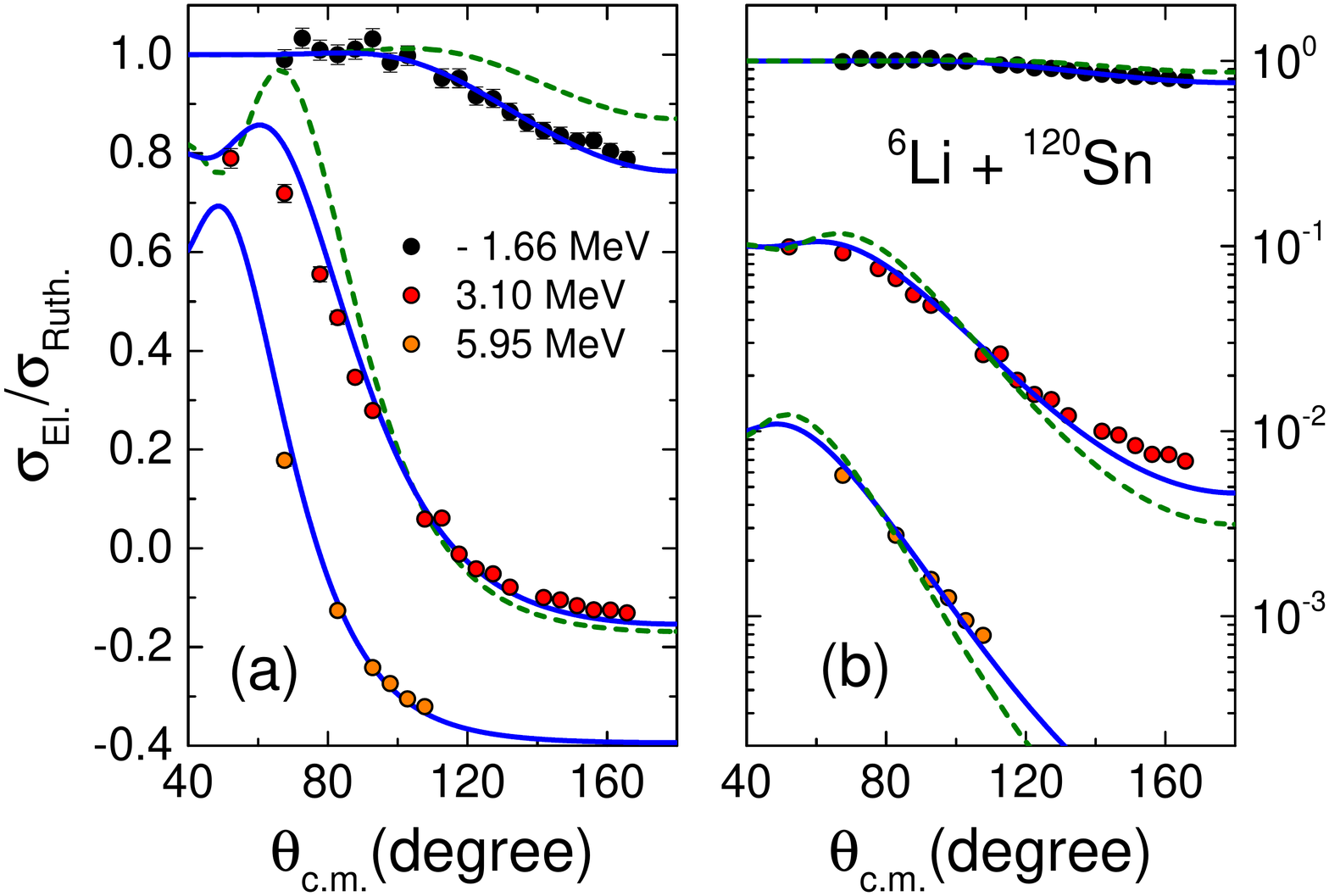}
\caption{\label{dens} (Color online) Experimental and theoretical SSA (solid
lines) and OIA (dashed lines)
elastic scattering angular distributions for $^6$Li + $^{120}$Sn. Note the 
change from linear (a) to logarithmic (b) scale. To avoid superposition, 
the cross sections for two distributions are displaced (a) or divided (b)
by constant factors.}
\end{minipage}\hspace{2pc}%
\end{figure}

\begin{figure}[h]
\begin{minipage}{20pc}
\includegraphics[width=20pc]{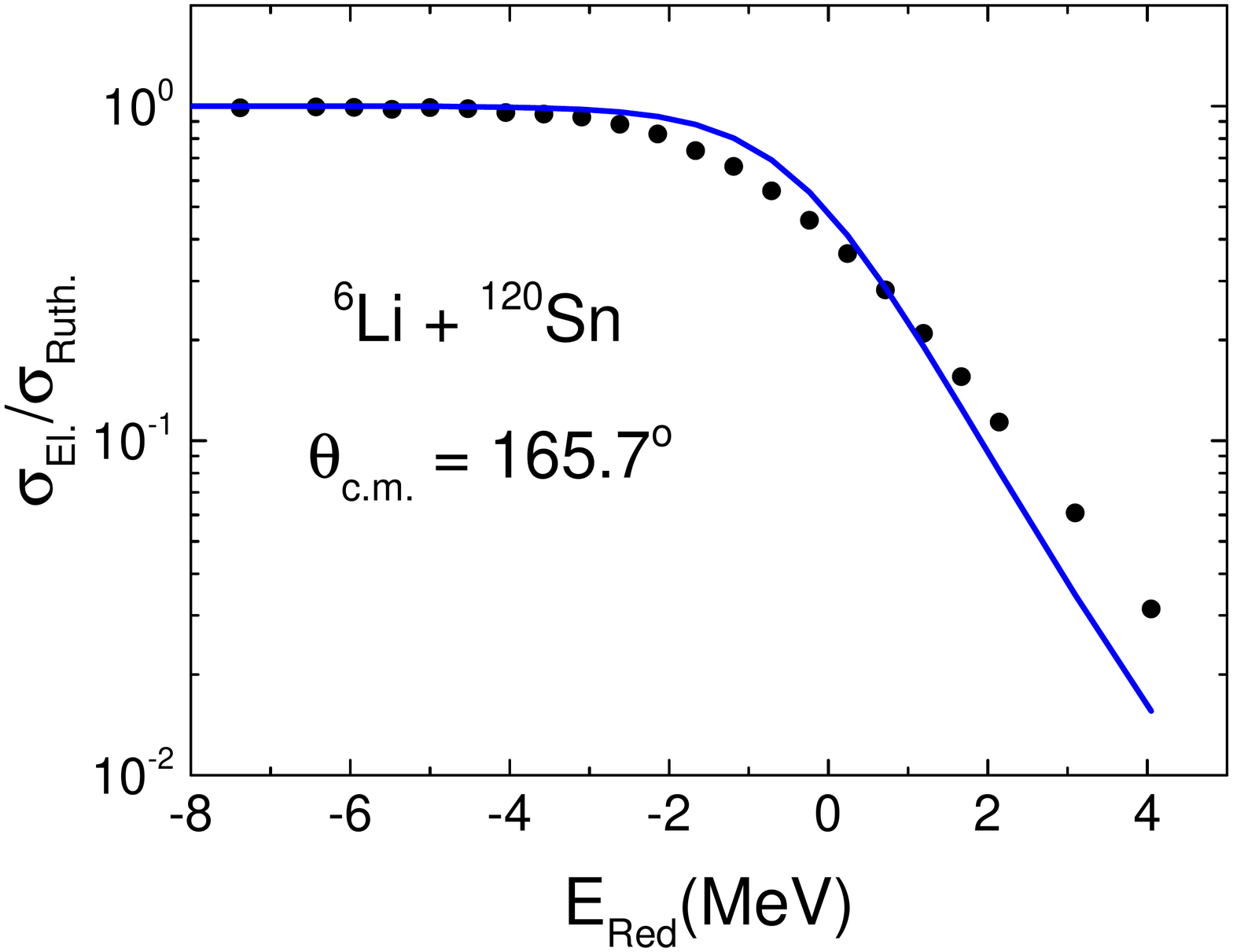}
\caption{\label{dens} (Color online) Experimental excitation function for $^6$Li + $^{120}$Sn \cite{Zer12} (see text for details). The solid line represents the SSA cross sections at $\theta_{c.m.}= 165.7^o$.}
\end{minipage}\hspace{2pc}%
\end{figure}

Figures 7 and 8 present results for $^{7}$Li + $^{120}$Sn \cite{Zag17}. In this case, the SSA provides even better agreement between data and theory than for $^{6}$Li.

\begin{figure}[h]
\begin{minipage}{20pc}
\includegraphics[width=20pc]{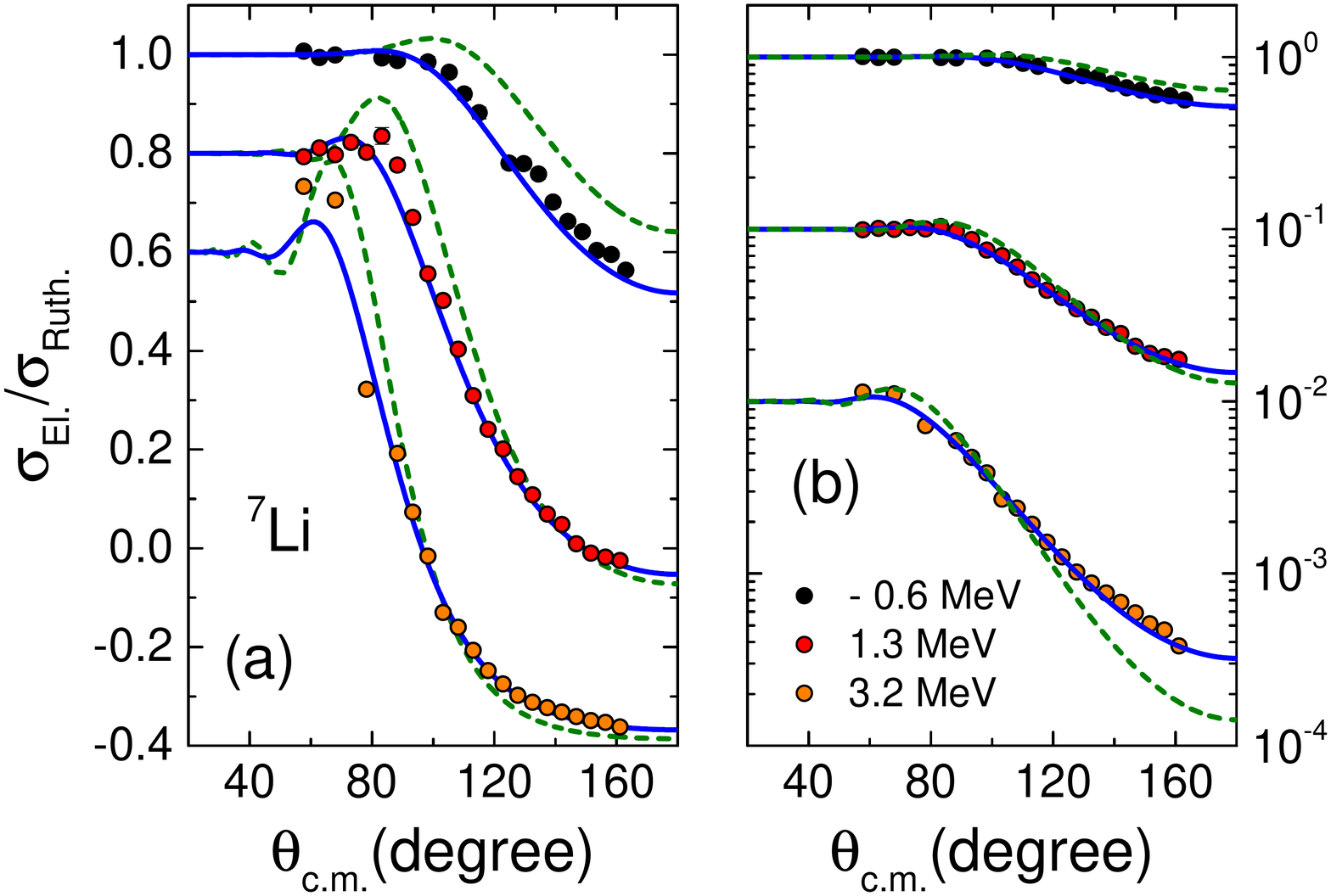}
\caption{\label{dens} (Color online) The same as Fig.5, for 
$^7$Li + $^{120}$Sn.  Data were extracted from \cite{Zag17}.}
\end{minipage}\hspace{2pc}%
\end{figure}

\begin{figure}[h]
\begin{minipage}{20pc}
\includegraphics[width=20pc]{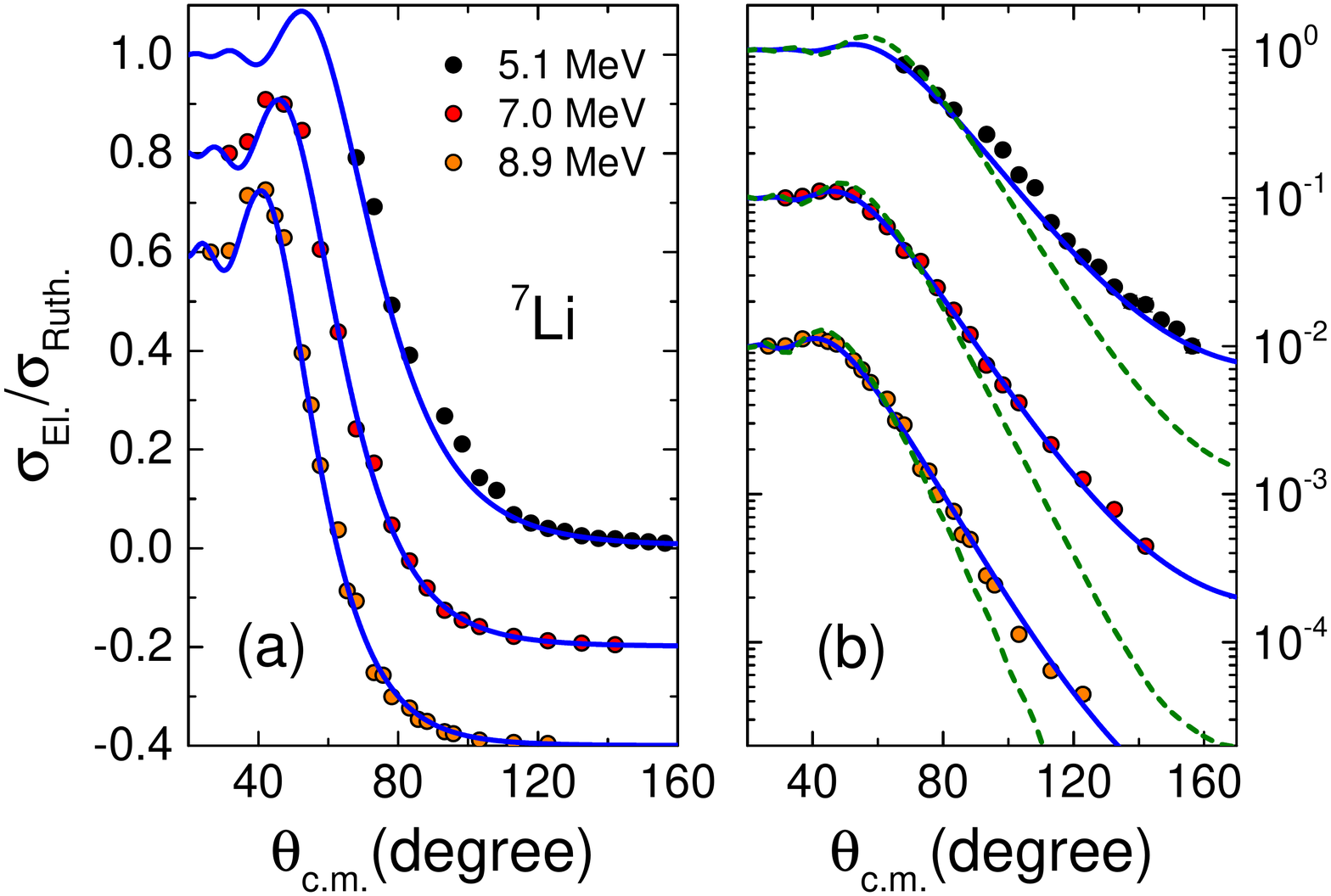}
\caption{\label{dens} (Color online) The same as Fig.5, for 
$^7$Li + $^{120}$Sn.  Data were extracted from \cite{Zag17,Kun17}.}
\end{minipage}\hspace{2pc}%
\end{figure}

Figures 9 and 10 present results for $^{9}$Be + $^{120}$Sn. Again, the 
SSA provides cross sections in reasonable agreement with the data.

\begin{figure}[h]
\begin{minipage}{20pc}
\includegraphics[width=20pc]{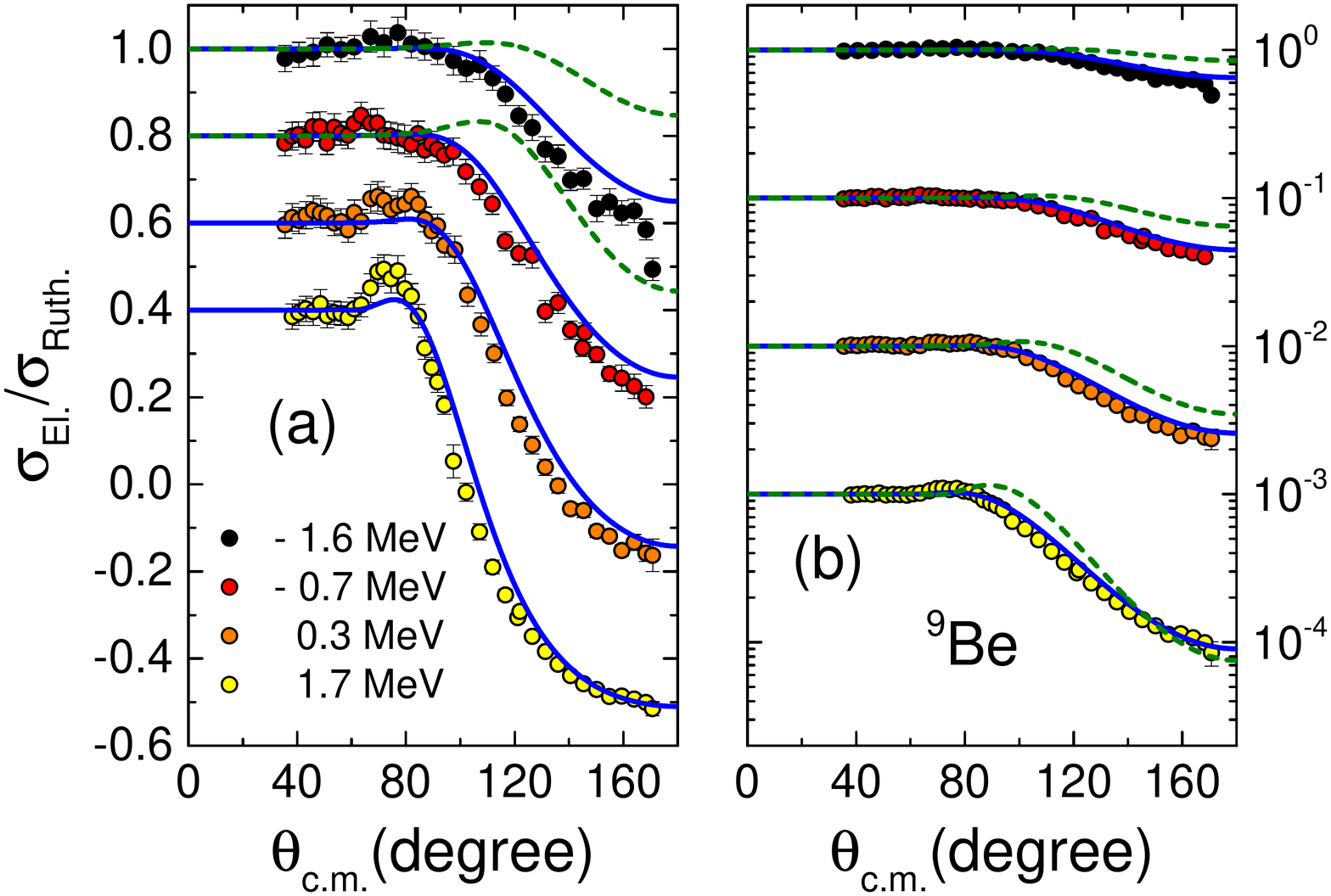}
\caption{\label{dens} (Color online) The same as Fig.5, for 
$^9$Be + $^{120}$Sn. Data were extracted from \cite{Ara18}.} 
\end{minipage}\hspace{2pc}%
\end{figure}

\begin{figure}[h]
\begin{minipage}{20pc}
\includegraphics[width=20pc]{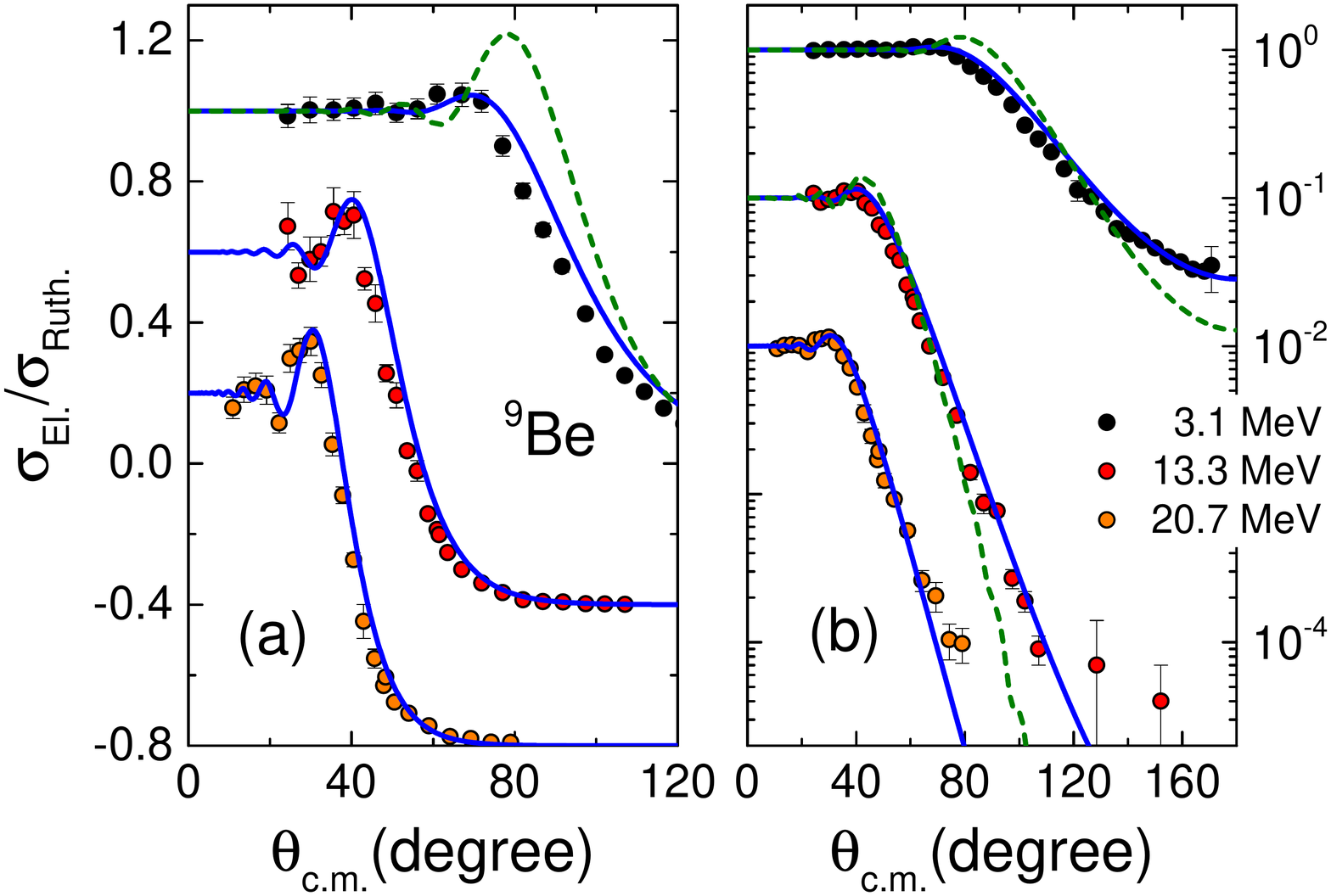}
\caption{\label{dens} (Color online) The same as Fig.5, for 
$^9$Be + $^{120}$Sn. Data were extracted from \cite{Ara18}.}
\end{minipage}\hspace{2pc}%
\end{figure}

Finally, Fig. 11 presents results for $^{10}$B + $^{120}$Sn. The SSA 
does not work as well as in other cases of weakly bound nuclei. However, the 
reduced energy region in the case of $^{10}$B is low and the results for this
nucleus are similar to those shown for $^{16}$O in Fig. 3(a).

\begin{figure}[h]
\begin{minipage}{20pc}
\includegraphics[width=20pc]{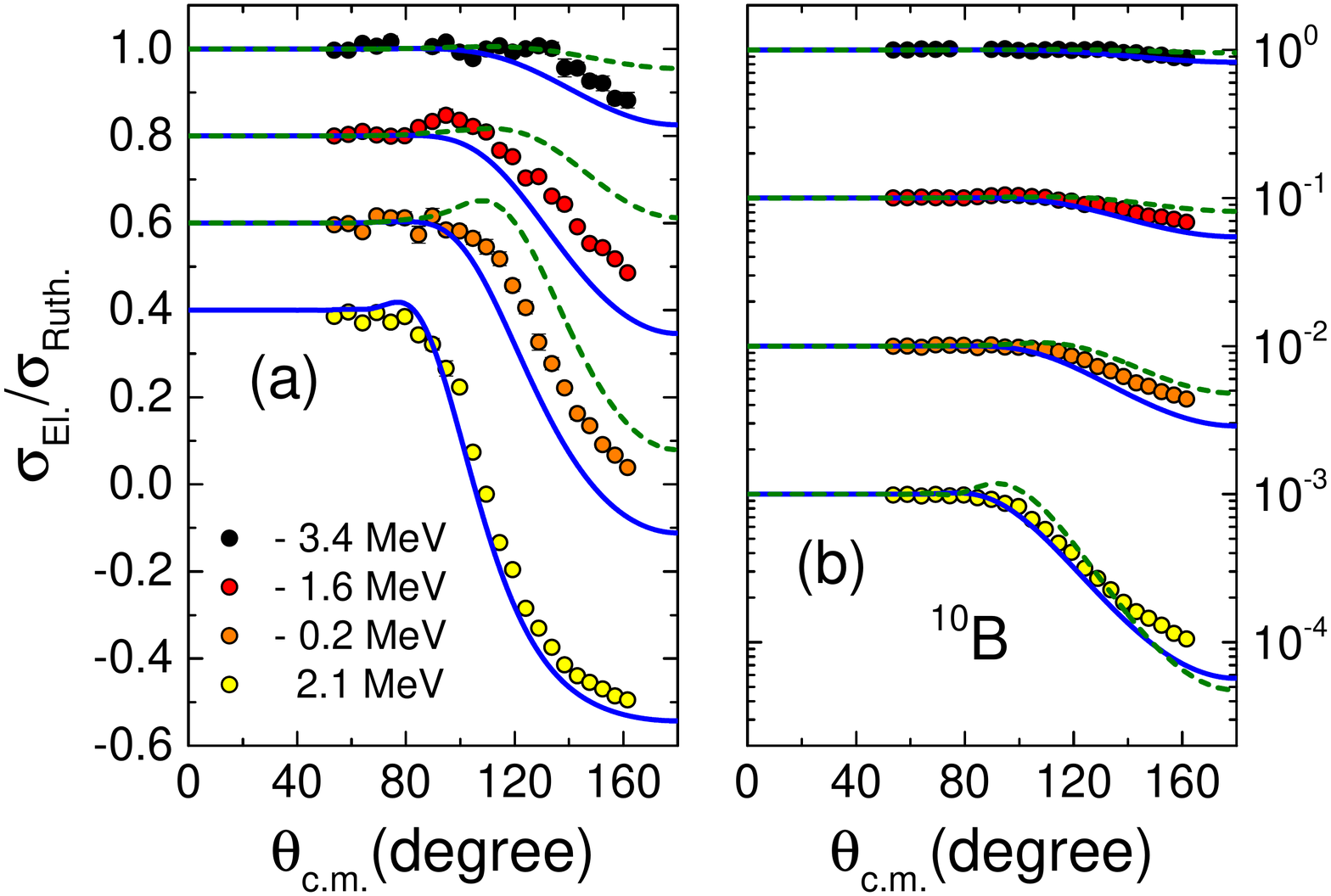}
\caption{\label{dens} (Color online) The same as Fig.5, for 
$^{10}$B + $^{120}$Sn. Data were extracted from \cite{Gas18,Alv18}.}
\end{minipage}\hspace{2pc}%
\end{figure}

\section{Comparison of the behavior of the optical potential for different 
projectiles}      
\label{compara}      

As commented in the previous section, the SSA provides an overall reasonable 
description of the complete data set studied here. Even so, small deviations
between data and theoretical predictions are observed. In this section, we 
assume Equation (4) with two adjustable parameters, $N_R$ and $N_I$, in order 
to fit the data more accurately, and compare the behavior of the corresponding 
OP parameter values obtained for different projectiles.

\subsection{The uncertainties of the $N_R$ and $N_I$ values}

In this subsection, we discuss some ambiguity inherent to the extraction of the 
$N_R$ and $N_I$ best fit values and their respective uncertainties. For this 
purpose,  we have performed several calculations in order to verify the sensitivity 
of the data fit on variations of the $N_R$ and $N_I$ parameter values. Just as 
an example, we illustrate here the results obtained with the data set for 
$^{18}$O at $E_{Red}= 2.1$ MeV. The corresponding best data fit is obtained 
with $N_R= 0.739$ and $N_I= 0.877$, with reduced chi-square of $\chi^2 = 5.80$. 

In Fig. 12 (a), we present the values of $\chi^2$ as a function of $N_I$ for 
several (fixed) values of $N_R$. For each $N_R$, there is an optimum 
$N_I$ value that provides the smallest $\chi^2$. Here, we can observe the strong correlation 
between the $N_R$ and $N_I$ parameters. This correlation can be even better observed in Fig. 12 
(b), which presents the optimum $N_I$ value as a function of $N_R$. Clearly, for
larger $N_R$ values we have smaller values of optimum $N_I$. In Fig. 12 (d), we 
show 
the $\chi^2$ (obtained with the optimum $N_I$) as a function of $N_R$. In Fig. 
12 (c), we show three curves (in the $N_R$ - $N_I$ plane) that correspond to 
different levels of $\chi^2$ (and also the point that provides the best 
$\chi^2 = 5.80$ for this data set).

Within the context of the theory of errors, the uncertainty of an adjustable
parameter can be approximately estimated considering variations of the reduced 
chi-square by about $1/N$ around the minimum $\chi^2$ value (which should be 
close to 1), where $N$ is the number of data. The experimental angular 
distribution, adopted as an example, contains 18 data points, and therefore 
$1/N \approx 0.06$. Since the best $\chi^2=5.80$, one should consider the range 
$\chi^2 \le 5.86$ for the determination of the error bars of the parameters. 
Nevertheless, the OM is only a simple (in fact simplified) 
theoretical model to describe the experimental phenomenon, and one can not 
expect the theory of errors to work perfectly in this case. For instance, the 
best $\chi^2=5.80$ is very far from the expected value $\chi^2 \approx 1$ of 
the theory. 

Taking into account this point, in many works, the estimate of uncertainties of
the OM adjustable parameters is performed considering a different level of 
reduced chi-square, for instance, an increase of 10\% or 20\% relative to its 
minimum value. Nevertheless, oftentimes the correlation between the parameters 
(as that for $N_R$ with $N_I$) is not considered when determining uncertainties. 
In this case, the uncertainties can be largely underestimated. 

Just to illustrate this point, let us suppose that we choose the level 
$\chi^2=7$ (about 20\% above the best $\chi^2=5.80$) to determine the 
uncertainties. This level is represented by the dashed line in Fig. 12 (a). 
The solid red curve in this figure corresponds to the variation of $\chi^2$ as 
a function of $N_I$ for the fixed (and also the best fit value) $N_R=0.739$. If 
one neglects the 
$N_R$ - $N_I$ correlation, the uncertainty of the $N_I$ parameter is found 
according to the intersections of the solid red curve with the $\chi^2=7$ level 
(dashed line). The blue arrows in figure 12 (a) show the corresponding region 
of uncertainty: $0.86 \le N_I \le 0.90$ (relative uncertainty of about 4.5\%). 
However, an inspection of the curve corresponding to the $\chi^2=7$ level in 
Fig. 12 (c) shows that, when considering the $N_R$ - $N_I$ correlation, a better 
estimate for the uncertainty of $N_I$ is $0.76 \le N_I \le 1.01$, therefore a 
much larger range of about 28\% for the relative uncertainty. The same could be 
said about the $N_R$ uncertainty. The dashed line in Fig. 12 (d) also represents 
the level $\chi^2=7$. The corresponding $N_R$ region is $0.61 \le N_R \le 0.84$ 
(about 32\% of relative uncertainty in $N_R$). This region already contains the 
effect of the correlation (since the $\chi^2$ versus $N_R$ curve of Fig. 12 (b) 
was obtained considering the variation of the optimum $N_I$ value with $N_R$). 
The blue arrows that connect Figs. 12 (b) and (d) illustrate the effect of the 
correlation on the uncertainties of the $N_R$ and $N_I$ parameter values. In our
example, the consideration or not of the correlation affects the parameter
uncertainty values by a factor about 6.

\begin{figure}[h]
\begin{minipage}{20pc}
\includegraphics[width=22pc]{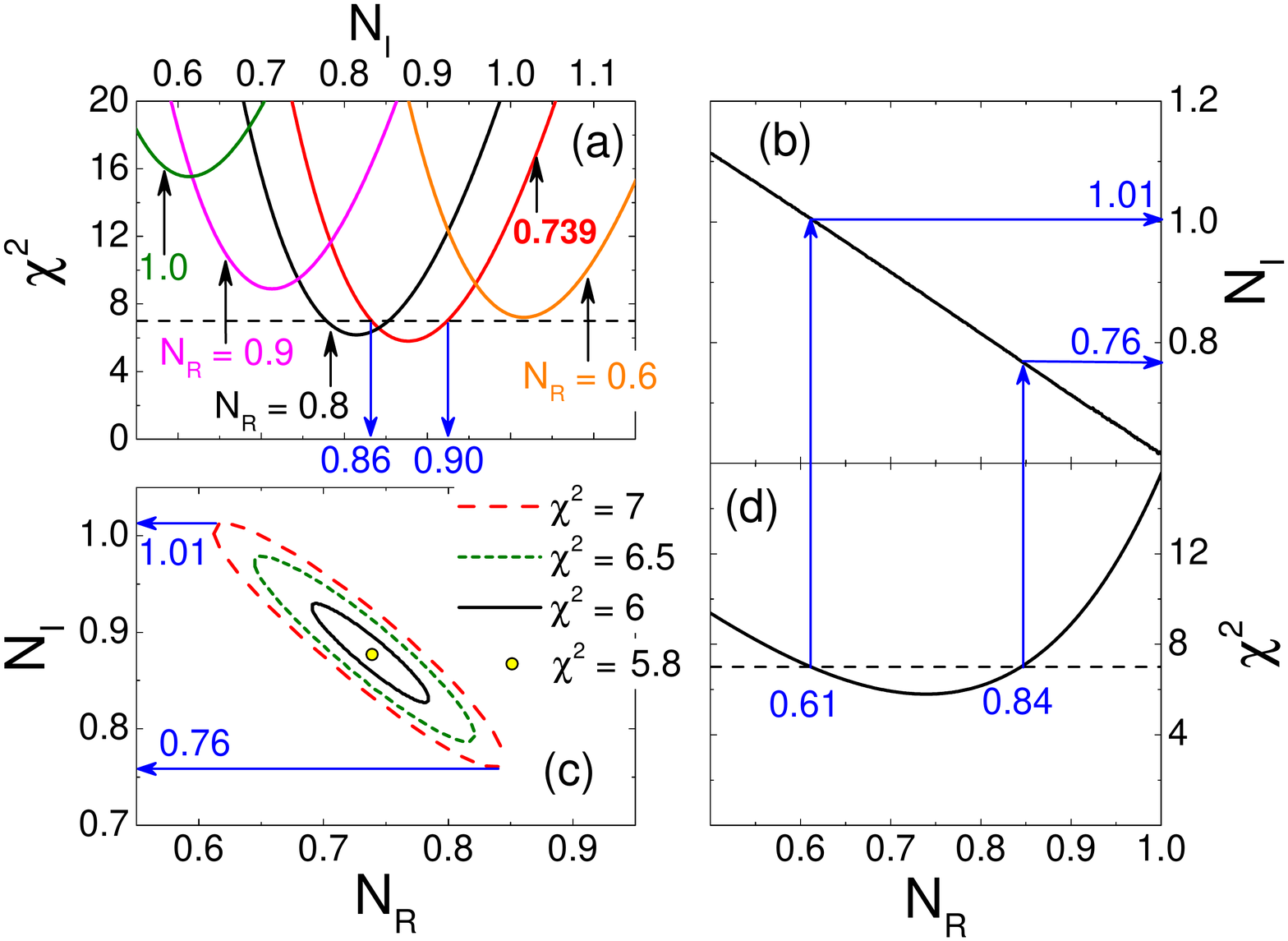}
\caption{\label{dens} (Color online) The figure shows results obtained by 
fitting an experimental angular distribution for $^{18}$O at $E_{Red}= 2.1$ MeV. 
(a): $\chi^2$ versus $N_I$ for some $N_R$ values. (b): Optimum $N_I$ value as 
a function of $N_R$. (c): Curves in the plane $N_R$ - $N_I$ that correspond to 
some levels of $\chi^2$. (d): $\chi^2$ obtained with the optimum $N_I$ as a 
function of $N_R$. See text for details.}
\end{minipage}\hspace{2pc}%
\end{figure}

Other important question can be raised here. What would be a good $\chi^2$ 
level to estimate uncertainties? In our
example, we chose 20\% above the best (minimum) $\chi^2$. The best $\chi^2$ is
obtained with $N_R= 0.739$ and $N_I= 0.877$, while the borders ($\chi^2=7$)
correspond to two possible pairs: $N_R= 0.61$ and $N_I= 1.01$ or 
$N_R= 0.84$ and $N_I= 0.76$. In Fig. 13 we present, in linear (a) and logarithmic (b) scales, the experimental angular 
distribution for $^{18}$O at $E_{Red}= 2.1$ MeV, and three theoretical curves.
Two of them, the solid black and dashed red lines, correspond to the best $\chi^2=5.80$ and to 
one case where $\chi^2=7$. These two lines are almost indistinguishable, 
indicating that this increase of 20\% in $\chi^2$ is probably too small to 
represent actual significance. The other curve (dotted blue lines in the figure) 
represents the result of a fit, in which $N_R= 1$ was fixed and only $N_I$ was 
considered as adjustable parameter. The corresponding optimum $N_I= 0.616$ was 
found, with $\chi^2=15.54$. Despite the difference of a factor of about three 
between the respective $\chi^2$ values, both OPs (of the best $\chi^2=5.80$ and 
that with $\chi^2=15.54$) provide a quite reasonable data fits (see the black and 
blue lines in Fig. 13). The large difference between the respective $\chi^2$ is 
mostly related to the fit in the backward angular region (in particular for 
the datum at the last angle $\theta_{c.m.} \approx 150^{\rm o}$). On the other 
hand, the fit with $\chi^2=15.54$ (dotted blue lines in the figure) clearly provides a slightly better data description in the rainbow region ($\theta_{c.m.} \approx 
90^{\rm o}$). Thus, one might ask: taking into account the physical behavior, 
does the fit with $\chi^2=5.80$ actually describe the experimental data in a better way than that with $\chi^2=15.54$?

\begin{figure}[h]
\begin{minipage}{20pc}
\includegraphics[width=20pc]{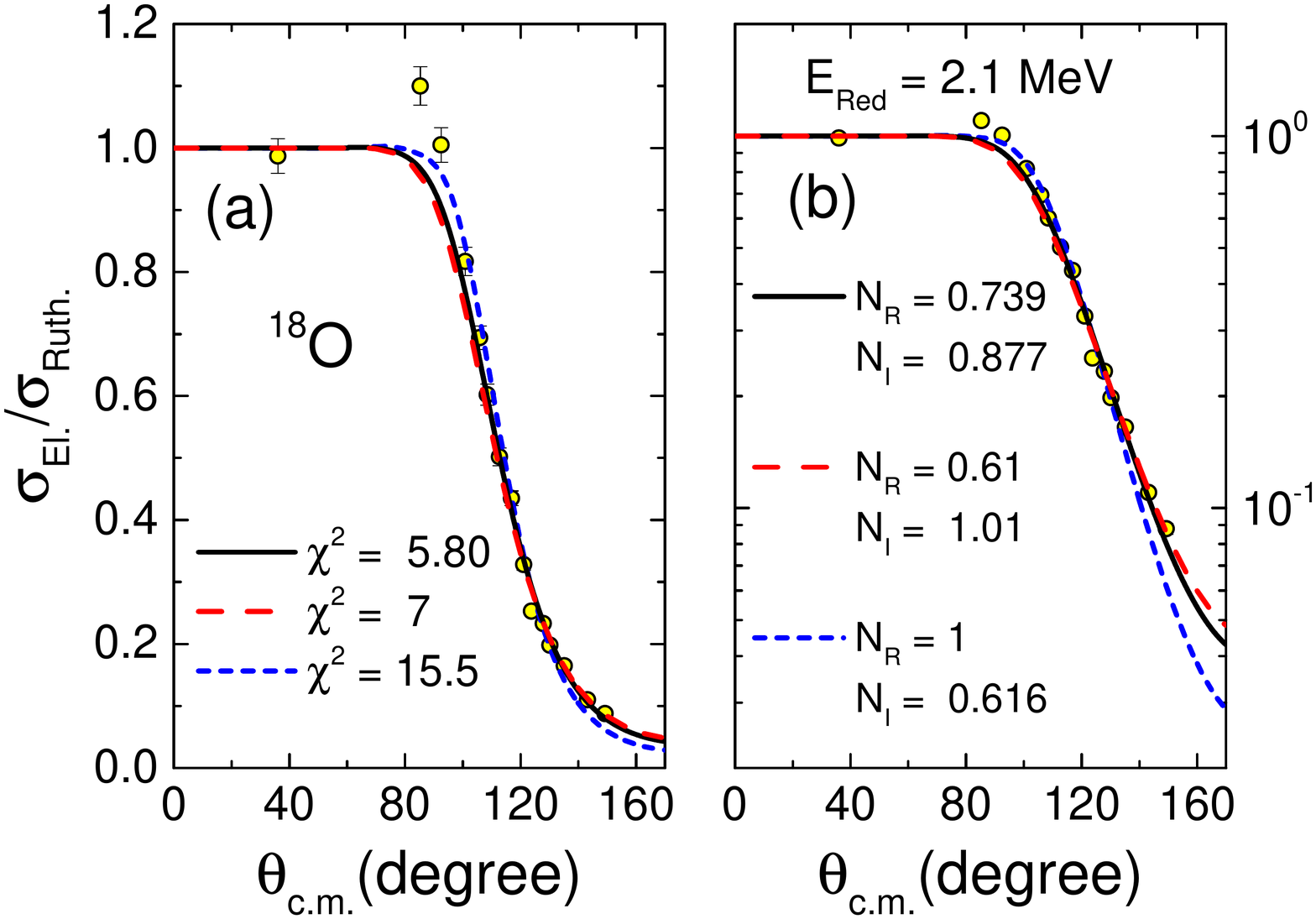}
\caption{\label{dens} (Color online) Experimental angular distribution for 
$^{18}$O at $E_{Red}= 2.1$ MeV, in linear (a) and
logarithmic (b) scales. The solid black lines represent the best data
fit ($\chi^2=5.80$) obtained within the OM. The dashed red lines were obtained with 
$N_R= 0.61$ and $N_I= 1.01$ that provide $\chi^2=7$. The dotted blue lines correspond 
to the result of the fit in which $N_R=1$ was fixed and only the $N_I$ value 
was adjusted ($\chi^2=15.54$).}
\end{minipage}\hspace{2pc}%
\end{figure}

Thus, uncertainties of adjustable parameter values obtained from OM data fits 
should be considered just as rough estimates. If the strong correlation between 
$N_R$ and $N_I$ is taken into account (and it should be), the uncertainties of 
these parameters become quite large. In addition, as commented in the previous 
paragraph, it is possible to obtain a quite reasonable description of the 
experimental angular distribution ($^{18}$O at $E_{Red}= 2.1$ MeV) assuming 
very different $N_R$ values. The reason for this behavior is also related to 
the correlation between the $N_R$ and $N_I$ parameters. As illustrated in Fig. 
13, the (best fit) pair $N_R= 0.739$ and $N_I= 0.877$ produces OM cross sections 
similar to those obtained with (fixed) $N_R= 1$ and (adjusted) $N_I= 0.616$ 
(despite the large difference of a factor of 3 in the corresponding $\chi^2$ 
values).

This behavior observed for the angular distribution of $^{18}$O at $E_{Red}= 
2.1$ MeV is also present in many other cases (projectiles and energies). The
correlation between $N_R$ and $N_I$ implies a wide ambiguity in
the determination of these parameter values, when simultaneously adjusted within 
the context of the OM data fits. In principle, the effect of the polarization 
due to inelastic channels would affect both: the real and imaginary parts of the 
OP. Even so, in order to avoid this question of correlation and consequent 
ambiguity, from now on we assume $N_R=1$ in the OM calculations, and adjust only 
the $N_I$ parameter value in the data fits. 

\subsection{The sub-barrier region}

When comparing data for different systems, it is important to take into account 
the region of energy considered. Thus, in this section we compare $N_I$ values 
obtained for different projectiles in approximately the same region of reduced 
energy. 

As illustrated in Figs. 3 (a) and 11, at energies below the barrier, both
systems, with $^{16}$O and $^{10}$B, present data with behavior in between the 
theoretical results of OIA (internal absorption = weak surface absorption) and 
SSA (strong surface absorption). In Fig. 14, we present the results obtained
through OM data fits, for $^{16}$O (a) and $^{10}$B (b). The figure also shows the $N_I$ values obtained for each angular 
distribution. The $^{16}$O case presents the expected 
behavior of strongly bound nuclei: vanishing surface absorption at 4 MeV below 
the barrier ($N_I = 0.09$), and increasing $N_I = 0.25$ and 0.50 values when
approaching the barrier. On the other hand, the $N_I$ values for $^{10}$B are
quite similar (about 0.35) for the three energies below the barrier, indicating
that the surface absorption does not decrease significantly even at sub-barrier 
energies.

\begin{figure}[h]
\begin{minipage}{20pc}
\includegraphics[width=20pc]{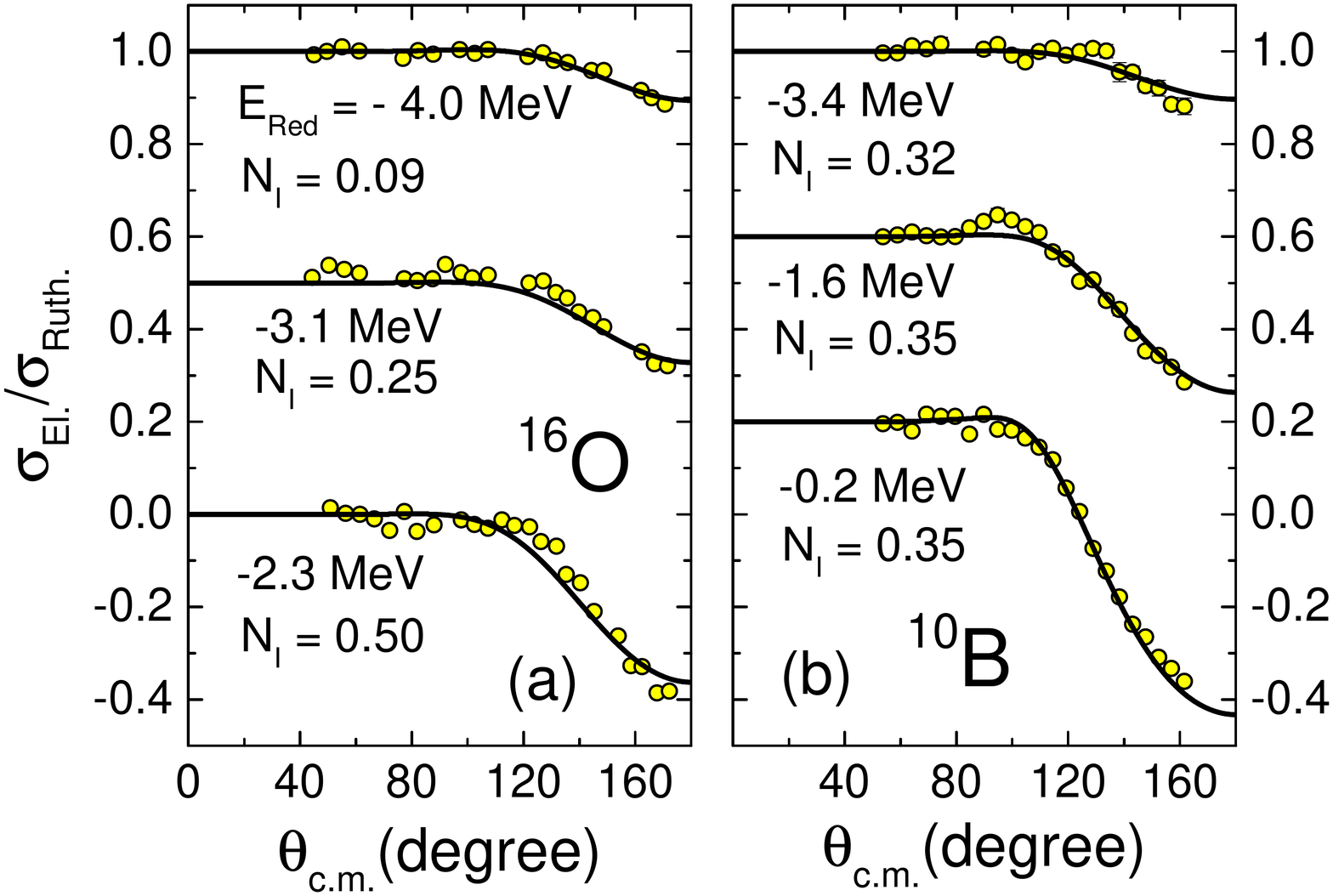}
\caption{\label{dens} (Color online) Experimental angular distributions for 
$^{16}$O and $^{10}$B and corresponding theoretical OM cross sections obtained
from data fits. The reduced energies and $N_I$ values are indicated in the 
figure.}
\end{minipage}\hspace{2pc}%
\end{figure}

In Fig. 15 (a), we present results for the weakly bound $^{6}$Li, $^{9}$Be and $^{10}$B projectiles, at energies about 1.6 MeV below the barrier.
In pannel (b), we have  $^{7}$Li (instead $^{6}$Li) and again 
$^{9}$Be and $^{10}$B, at energies about 0.5 MeV below the barrier. Even in this low energy region, the three projectiles present non vanishing $N_I$
values, the largest being those for $^{9}$Be ($N_I \approx 1.3$), followed by those for the lithium isotopes (about 0.7), and the small one (0.35) being that for $^{10}$B. This behavior indicates more absorption at sub-barrier energies,
probably due to the break-up process, for $^{6,7}$Li and $^{9}$Be \cite{Ara18, Luo13}.

\begin{figure}[h]
\begin{minipage}{20pc}
\includegraphics[width=20pc]{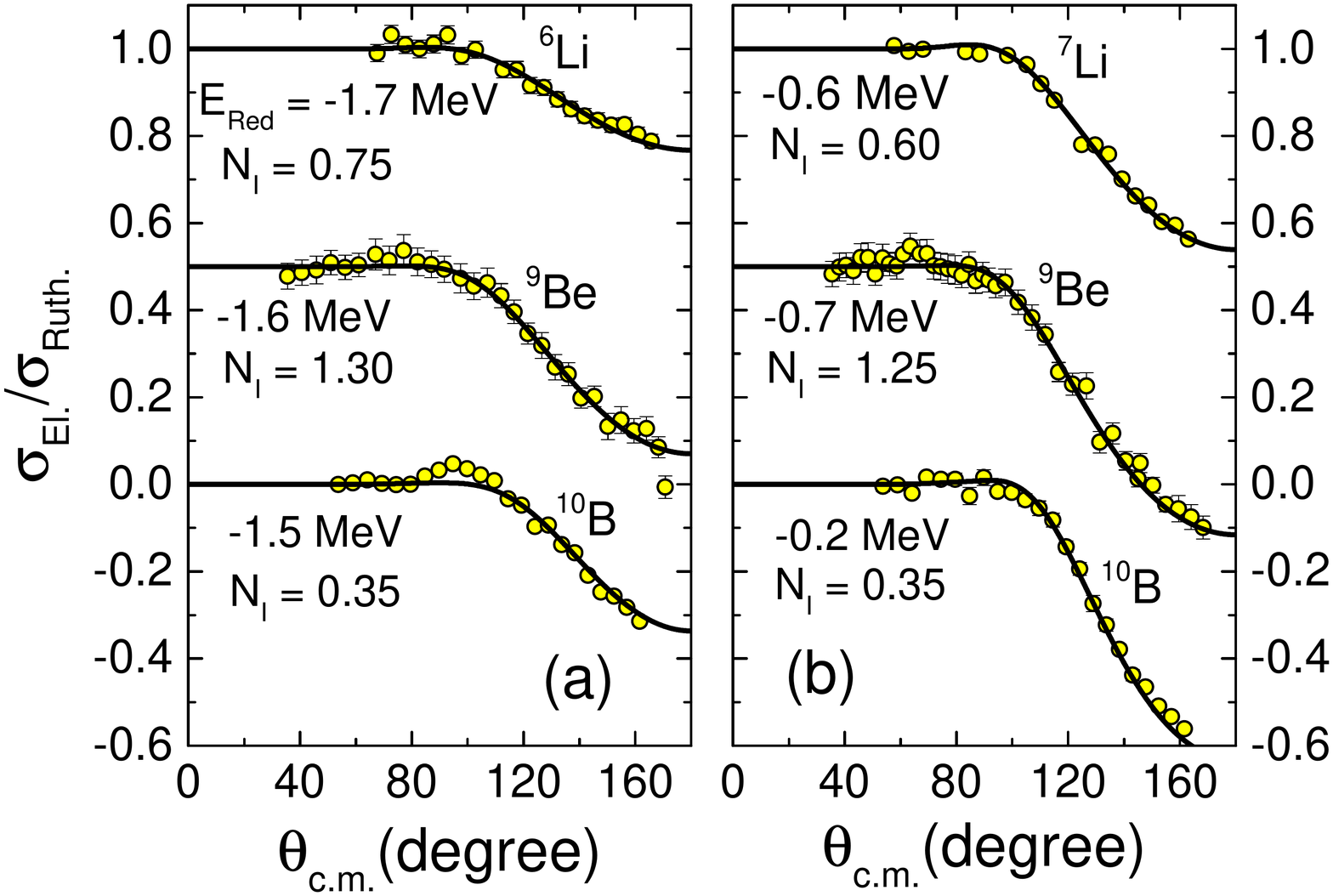}
\caption{\label{dens} (Color online) The same as Fig. 14, for other systems and
energies.}
\end{minipage}\hspace{2pc}%
\end{figure}

\subsection{The above-barrier region}

Now we analyze angular distributions at energies above the barrier. Again we 
present comparison of data only in similar reduced energy regions. For a good 
appreciation of the results, the figures contain both linear and logarithmic 
scales.

Fig. 16 presents angular distributions for $^7$Li and $^9$Be at $E_{Red} 
\approx 1.5$ MeV. The $N_I$ values of about 0.6 for $^7$Li and 1.2 for $^9$Be 
are quite similar to those obtained at sub-barrier energies (see Fig. 15).

\begin{figure}[h]
\begin{minipage}{20pc}
\includegraphics[width=20pc]{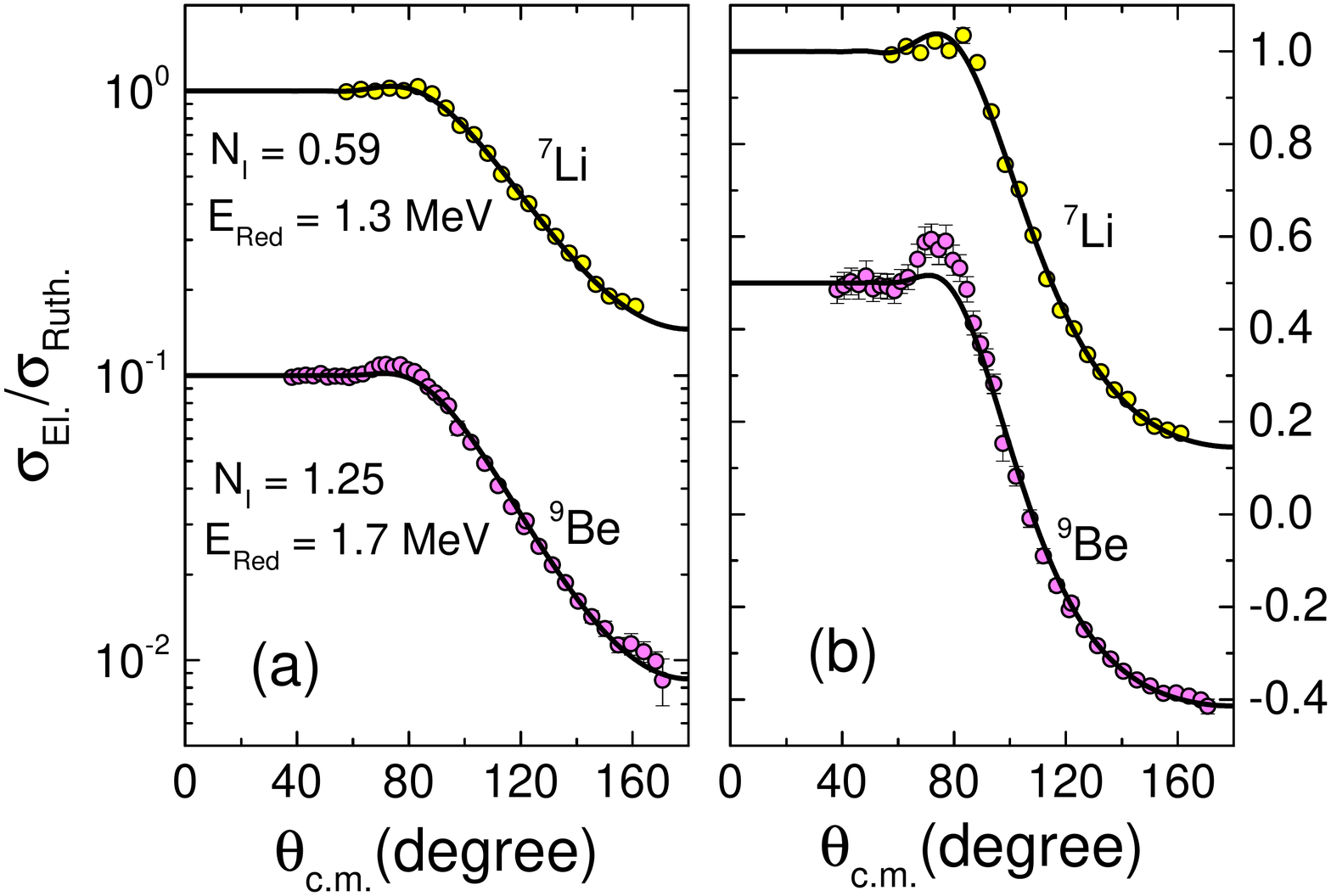}
\caption{\label{dens} (Color online) Experimental angular distributions for 
$^7$Li and $^9$Be, in logarithmic (a) and linear (b) scales. The black lines represent the OM fits (performed by adjusting only 
the $N_I$ parameter). The corresponding $N_I$ and reduced energy values 
are indicated in the figure.}
\end{minipage}\hspace{2pc}%
\end{figure}

Figure 17 presents results for $^6$He, $^6$Li, $^7$Li and $^9$Be at about 3.5 MeV
above the barrier. The best fit $N_I=4$ obtained for $^6$He is a very large
value. However, we point out that, due to the large error bars of the cross
section data, the sensitivity 
of the $\chi^2$ to the $N_I$ parameter value is very weak for this angular 
distribution, and much smaller $N_I$ values also provide a good data fit. The 
$N_I$ values obtained for the weakly bound $^6$Li, $^7$Li and $^9$Be nuclei are 
large, again indicating strong surface absorption in these cases.

\begin{figure}[h]
\begin{minipage}{20pc}
\includegraphics[width=20pc]{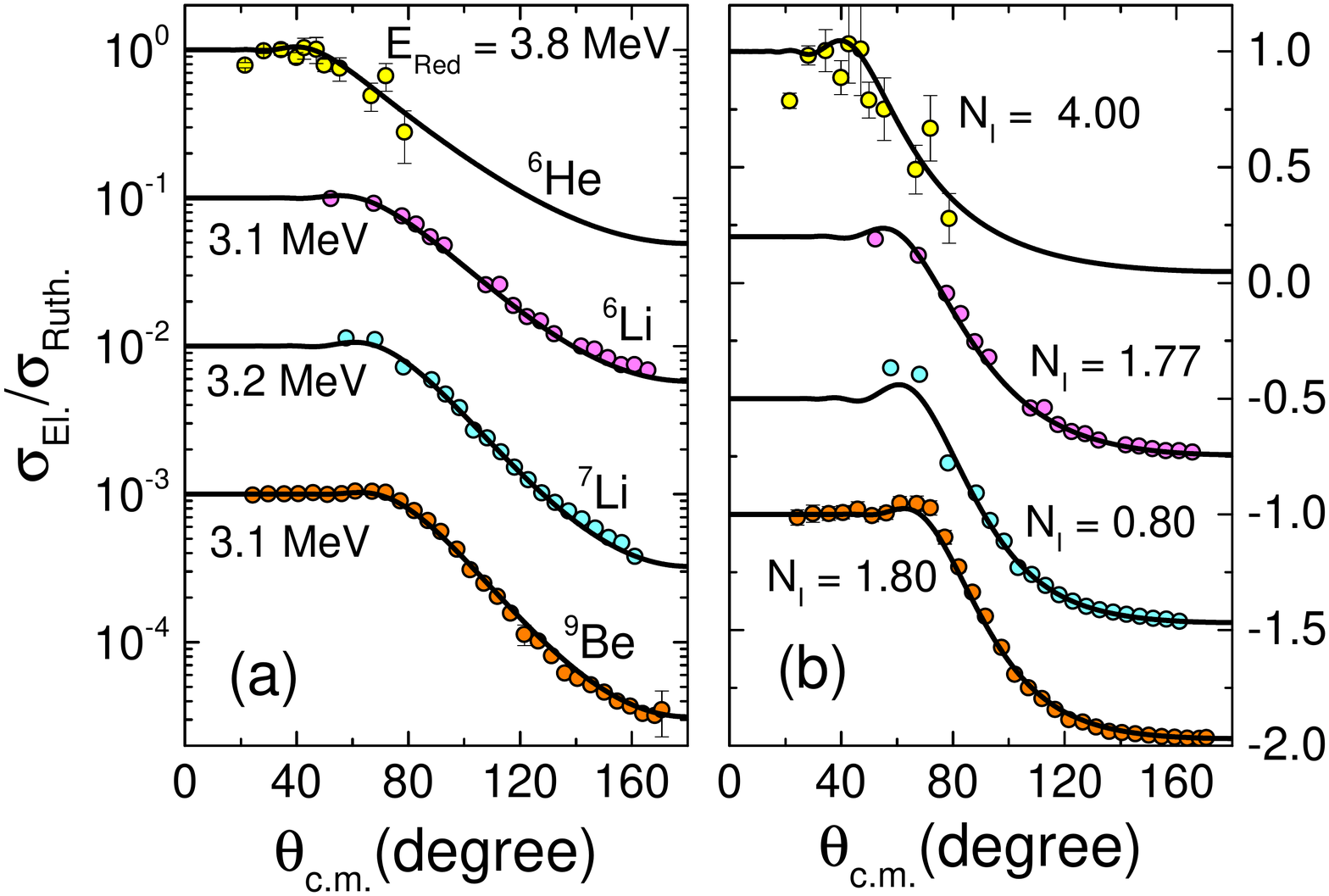}
\caption{\label{dens} (Color online) The same of Fig. 16, for other projectiles
and energies.}
\end{minipage}\hspace{2pc}%
\end{figure}

Figure 18 presents results for $^6$He, $^6$Li and $^7$Li (two energies) at 
$E_{Red} \approx 6$ MeV. The $^6$He and $^6$Li OM fits result in $N_I$ values
larger than 1. The two energies for $^7$Li provide, consistently, similar values
around $N_I \approx 0.89$.

\begin{figure}[h]
\begin{minipage}{20pc}
\includegraphics[width=20pc]{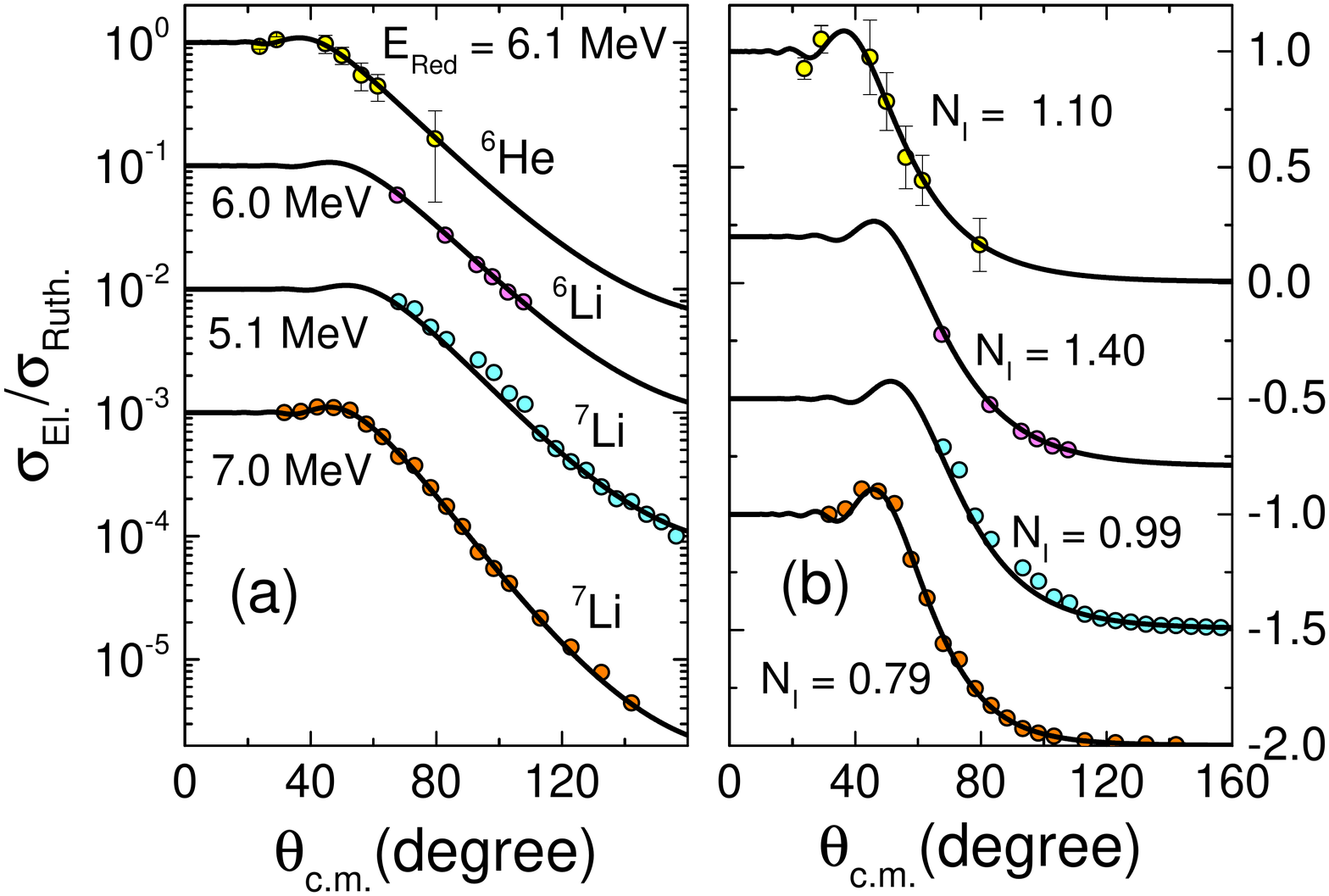}
\caption{\label{dens} (Color online) The same of Fig. 16, for other projectiles
and energies.}
\end{minipage}\hspace{2pc}%
\end{figure}

Finally, Fig. 19 presents results for the strongly bound $^4$He and the weakly
bound $^9$Be nuclei, at very high energies $E_{Red} \approx 20$ MeV. A striking
difference of  about one order of magnitude is observed for the corresponding 
$N_I$ values: 0.20 and 1.90. 

\begin{figure}[h]
\begin{minipage}{20pc}
\includegraphics[width=20pc]{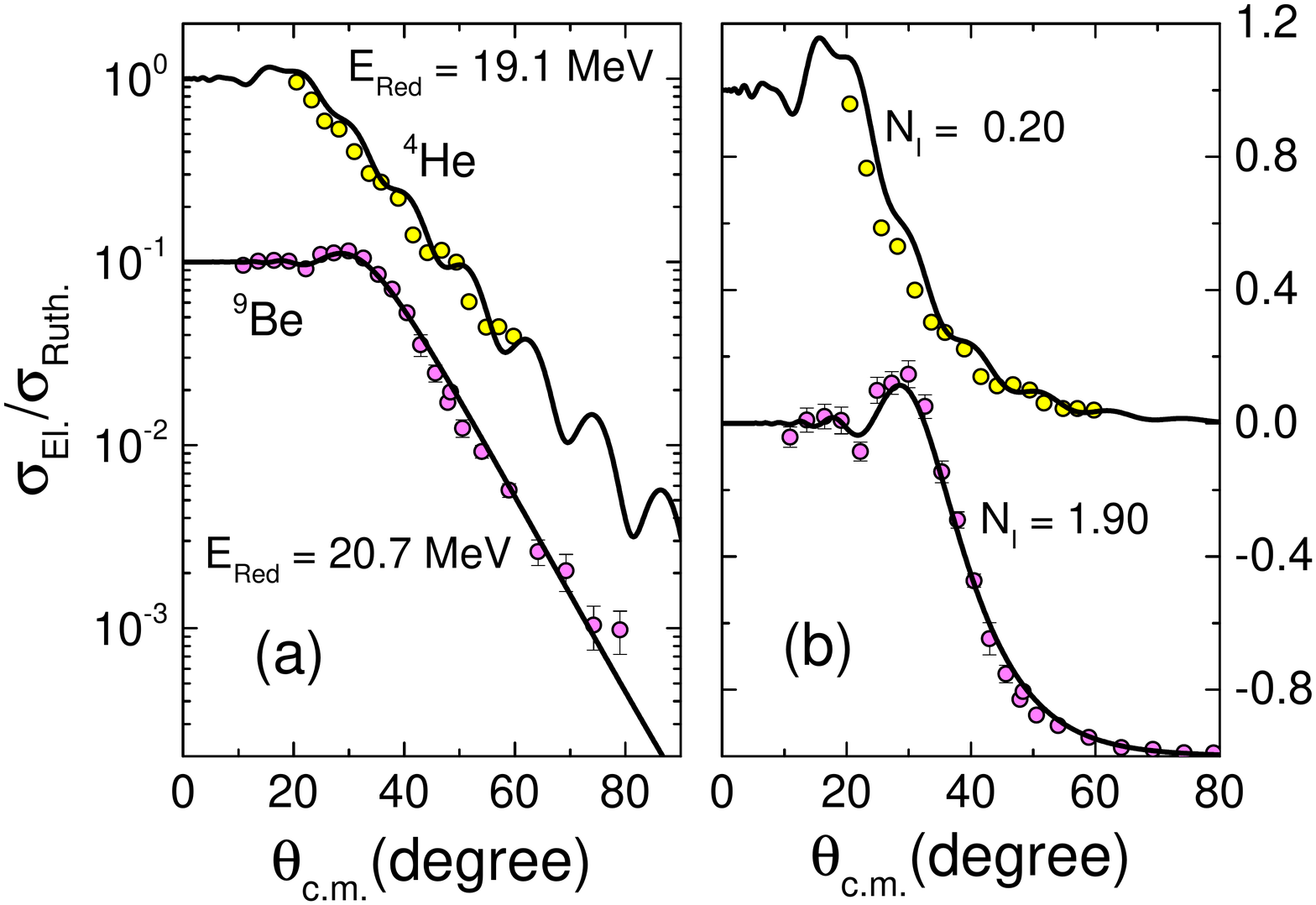}
\caption{\label{dens} (Color online) The same of Fig. 16, for other projectiles
and energies.}
\end{minipage}\hspace{2pc}%
\end{figure}

In Fig. 20, we show the $N_I$ values as a function of the reduced energy for several projectiles. We have not included the results for $^6$He
and $^4$He, at low energies, because the $\chi^2$ for these distributions are not very sensitive to the $N_I$ values, due to the large error bars of the experimental cross sections. The solid lines in this figure are only guides for the eyes. The dashed line corresponds to the standard $N_I=0.78$ value.
Considering only the behavior of the weakly bound nuclei, the figure indicates increasing $N_I$ parameter values in the following order: $^{10}$B, $^7$Li, $^6$Li and $^9$Be.

\begin{figure}[h]
\begin{minipage}{20pc}
\includegraphics[width=20pc]{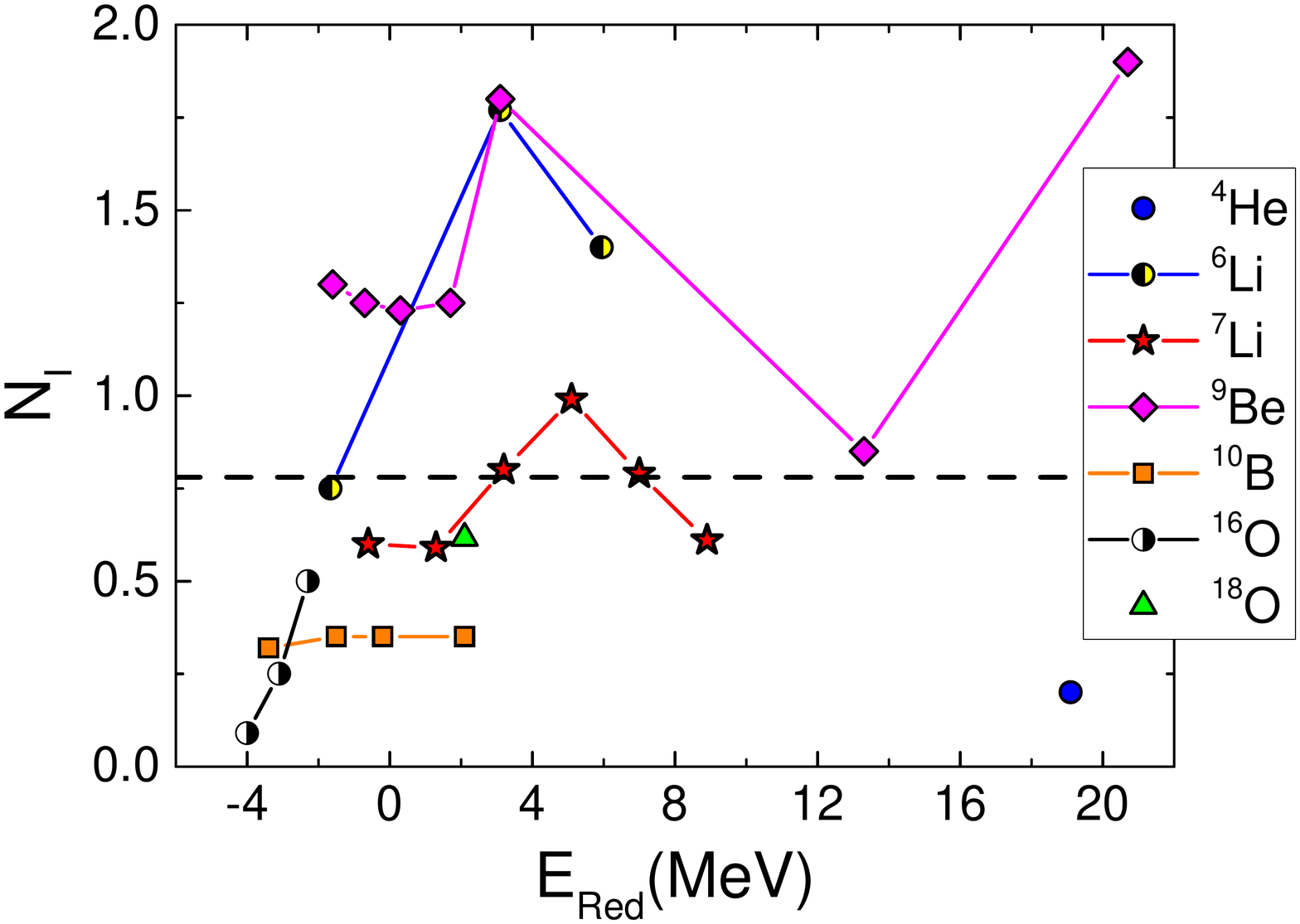}
\caption{\label{dens} (Color online) $N_I$ values obtained from OM data analyses
as a function of the reduced energy for several projectiles. The solid lines in 
the figure are only guides for the eyes, while the dashed line corresponds to 
the standard $N_I=0.78$.}
\end{minipage}\hspace{2pc}%
\end{figure}

\section{Conclusions}

In this paper, we have presented new data for the elastic scattering of $^6$Li +
$^{120}$Sn at $E_{\rm LAB}=$ 19, 24 and 27 MeV. The corresponding angular
distributions were considered together with other elastic scattering data of
several projectiles on the same target nucleus. The complete data set  was 
systematically analyzed within the context of the OM. We have demonstrated that 
the SPP in the context of the standard SSA provides a quite reasonable description 
of the data for all systems, without the necessity of any adjustable parameter. We 
have obtained more accurate agreement between data and theoretical cross 
sections by considering adjustable OP strengths in order to improve the data 
fits.

We have illustrated the strong correlation between the real and imaginary
adjustable strength factors ($N_R$ and $N_I$) in an example with one angular
distribution. If this correlation is taken into account, the uncertainties of the
$N_R$ and $N_I$ best fit values become very large. In addition, different pairs
of these parameters, with corresponding $\chi^2$ values that differ by a factor
as large as 3, provide rather similar theoretical angular distributions that
agree well with the data. This behavior is also found for other projectiles and
energies. In order to avoid this ambiguity, we have assumed the SPP for the real
part of the OP, with fixed standard $N_R = 1$, and adjusted only the $N_I$ 
parameter value in the OM data fits.

As observed in Figs. 14 to 19, the theoretical cross sections obtained through
OM fits with only one free parameter ($N_I$) are in quite good agreement with
the data for all systems and energies. We have studied the behavior of the best
fit $N_I$ value in different energy regions, and compared results obtained for
the various projectiles. The weakly bound $^{6,7}$Li, $^9$Be and $^{10}$B
projectiles present significant $N_I$ values at sub-barrier energies, indicating 
strong surface absorption even in this low energy region, a characteristic
probably related to the break-up process. Still considering these nuclei, 
increasing $N_I$ parameter values are observed in the following order: 
$^{10}$B, $^7$Li, $^6$Li and $^9$Be. This order is related to the binding 
energy of these nuclei (presented as $Q$ values in Table I). This suggests
a clear correlation between the break-up probability and the absorption of 
flux from the elastic channel.

\newpage
\begin{acknowledgments} 
This work was supported by the Ministry of Science, Innovation and Universities of Spain, through the project PGC2018-096994-B-C21. This work was also partially supported by the Spanish Ministry of Economy and Competitiveness, the European Regional Development Fund (FEDER), under Project $\rm N^o$  FIS2017-88410-P and by the European Union's Horizon 2020 research and innovation program, under Grant Agreement $\rm N^o$ 654002. This work has also been partially supported by Funda\c{c}\~{a}o de Amparo \`{a} Pesquisa do Estado de S\~{a}o Paulo (FAPESP) 
Proc. $\rm N^o$ 2018/09998-8 and 2017/05660-0, Conselho Nacional de Desenvolvimento Cient\'{i}fico e Tecnol\'{o}gico (CNPq) Proc. $\rm N^o$ 407096/2017-5 and 306433/2017-6, and, finally, it is a part of the project INCT-FNA Proc. $\rm N^o$ 464898/2014-5. A. Arazi acknowledges Grant $\rm N^o$ PIP00786CO from CONICET. D. A. Torres. and F. Ramirez acknowledge support from Colciencias under contract 110165842984.
\end{acknowledgments}      

\vspace{1.0cm}
\bibliographystyle{apsrev4-1}
\bibliography{biblio}

\begin{thebibliography}{56}%
\makeatletter
\providecommand \@ifxundefined [1]{%
 \@ifx{#1\undefined}
}%
\providecommand \@ifnum [1]{%
 \ifnum #1\expandafter \@firstoftwo
 \else \expandafter \@secondoftwo
 \fi
}%
\providecommand \@ifx [1]{%
 \ifx #1\expandafter \@firstoftwo
 \else \expandafter \@secondoftwo
 \fi
}%
\providecommand \natexlab [1]{#1}%
\providecommand \enquote  [1]{``#1''}%
\providecommand \bibnamefont  [1]{#1}%
\providecommand \bibfnamefont [1]{#1}%
\providecommand \citenamefont [1]{#1}%
\providecommand \href@noop [0]{\@secondoftwo}%
\providecommand \href [0]{\begingroup \@sanitize@url \@href}%
\providecommand \@href[1]{\@@startlink{#1}\@@href}%
\providecommand \@@href[1]{\endgroup#1\@@endlink}%
\providecommand \@sanitize@url [0]{\catcode `\\12\catcode `\$12\catcode
  `\&12\catcode `\#12\catcode `\^12\catcode `\_12\catcode `\%12\relax}%
\providecommand \@@startlink[1]{}%
\providecommand \@@endlink[0]{}%
\providecommand \url  [0]{\begingroup\@sanitize@url \@url }%
\providecommand \@url [1]{\endgroup\@href {#1}{\urlprefix }}%
\providecommand \urlprefix  [0]{URL }%
\providecommand \Eprint [0]{\href }%
\providecommand \doibase [0]{http://dx.doi.org/}%
\providecommand \selectlanguage [0]{\@gobble}%
\providecommand \bibinfo  [0]{\@secondoftwo}%
\providecommand \bibfield  [0]{\@secondoftwo}%
\providecommand \translation [1]{[#1]}%
\providecommand \BibitemOpen [0]{}%
\providecommand \bibitemStop [0]{}%
\providecommand \bibitemNoStop [0]{.\EOS\space}%
\providecommand \EOS [0]{\spacefactor3000\relax}%
\providecommand \BibitemShut  [1]{\csname bibitem#1\endcsname}%
\let\auto@bib@innerbib\@empty
\bibitem [{\citenamefont {Horiuchi}(2013)}]{Hori13}%
  \BibitemOpen
  \bibfield  {author} {\bibinfo {author} {\bibfnamefont {H.}~\bibnamefont
  {Horiuchi}},\ }\href@noop {} {\bibfield  {journal} {\bibinfo  {journal} {J.
  Phys. Conf. Ser.}\ }\textbf {\bibinfo {volume} {436}},\ \bibinfo {pages} {1}
  (\bibinfo {year} {2013})}\BibitemShut {NoStop}%
\bibitem [{\citenamefont {Luong}\ \emph {et~al.}(2013)\citenamefont {Luong},
  \citenamefont {Dasgupta}, \citenamefont {Hinde}, \citenamefont {du~Rietz},
  \citenamefont {Rafiei}, \citenamefont {Lin}, \citenamefont {Evers},\ and\
  \citenamefont {D\'{i}az-Torres}}]{Luo13}%
  \BibitemOpen
  \bibfield  {author} {\bibinfo {author} {\bibfnamefont {D.~H.}\ \bibnamefont
  {Luong}}, \bibinfo {author} {\bibfnamefont {M.}~\bibnamefont {Dasgupta}},
  \bibinfo {author} {\bibfnamefont {D.~J.}\ \bibnamefont {Hinde}}, \bibinfo
  {author} {\bibfnamefont {R.}~\bibnamefont {du~Rietz}}, \bibinfo {author}
  {\bibfnamefont {R.}~\bibnamefont {Rafiei}}, \bibinfo {author} {\bibfnamefont
  {C.~J.}\ \bibnamefont {Lin}}, \bibinfo {author} {\bibfnamefont
  {M.}~\bibnamefont {Evers}}, \ and\ \bibinfo {author} {\bibfnamefont
  {A.}~\bibnamefont {D\'{i}az-Torres}},\ }\href@noop {} {\bibfield  {journal}
  {\bibinfo  {journal} {Phys. Rev. C}\ }\textbf {\bibinfo {volume} {88}},\
  \bibinfo {pages} {034609} (\bibinfo {year} {2013})}\BibitemShut {NoStop}%
\bibitem [{\citenamefont {Escrig}\ \emph {et~al.}(2007)\citenamefont {Escrig},
  \citenamefont {S\'{a}nchez-Ben\'{i}tez}, \citenamefont {Moro}, \citenamefont
  {Alvarez}, \citenamefont {Andr\'{e}s}, \citenamefont {Angulo}, \citenamefont
  {Borge}, \citenamefont {Cabrera}, \citenamefont {Cherubini}, \citenamefont
  {Demaret}, \citenamefont {Espino}, \citenamefont {Figuera}, \citenamefont
  {Freer}, \citenamefont {Garc\'{i}a-Ramos}, \citenamefont {G\'{o}mez-Camacho},
  \citenamefont {Gulino}, \citenamefont {Kakuee}, \citenamefont {Martel},
  \citenamefont {Metelko}, \citenamefont {P\'{e}rez-Bernal}, \citenamefont
  {Rahighi}, \citenamefont {Rusek}, \citenamefont {Smirnov}, \citenamefont
  {Tengblad},\ and\ \citenamefont {Ziman}}]{Esc07}%
  \BibitemOpen
  \bibfield  {author} {\bibinfo {author} {\bibfnamefont {D.}~\bibnamefont
  {Escrig}}, \bibinfo {author} {\bibfnamefont {A.}~\bibnamefont
  {S\'{a}nchez-Ben\'{i}tez}}, \bibinfo {author} {\bibfnamefont
  {A.}~\bibnamefont {Moro}}, \bibinfo {author} {\bibfnamefont {M.~A.~G.}\
  \bibnamefont {Alvarez}}, \bibinfo {author} {\bibfnamefont {M.}~\bibnamefont
  {Andr\'{e}s}}, \bibinfo {author} {\bibfnamefont {C.}~\bibnamefont {Angulo}},
  \bibinfo {author} {\bibfnamefont {M.}~\bibnamefont {Borge}}, \bibinfo
  {author} {\bibfnamefont {J.}~\bibnamefont {Cabrera}}, \bibinfo {author}
  {\bibfnamefont {S.}~\bibnamefont {Cherubini}}, \bibinfo {author}
  {\bibfnamefont {P.}~\bibnamefont {Demaret}}, \bibinfo {author} {\bibfnamefont
  {J.}~\bibnamefont {Espino}}, \bibinfo {author} {\bibfnamefont
  {P.}~\bibnamefont {Figuera}}, \bibinfo {author} {\bibfnamefont
  {M.}~\bibnamefont {Freer}}, \bibinfo {author} {\bibfnamefont
  {J.}~\bibnamefont {Garc\'{i}a-Ramos}}, \bibinfo {author} {\bibfnamefont
  {J.}~\bibnamefont {G\'{o}mez-Camacho}}, \bibinfo {author} {\bibfnamefont
  {M.}~\bibnamefont {Gulino}}, \bibinfo {author} {\bibfnamefont
  {O.}~\bibnamefont {Kakuee}}, \bibinfo {author} {\bibfnamefont
  {I.}~\bibnamefont {Martel}}, \bibinfo {author} {\bibfnamefont
  {C.}~\bibnamefont {Metelko}}, \bibinfo {author} {\bibfnamefont
  {F.}~\bibnamefont {P\'{e}rez-Bernal}}, \bibinfo {author} {\bibfnamefont
  {J.}~\bibnamefont {Rahighi}}, \bibinfo {author} {\bibfnamefont
  {K.}~\bibnamefont {Rusek}}, \bibinfo {author} {\bibfnamefont
  {D.}~\bibnamefont {Smirnov}}, \bibinfo {author} {\bibfnamefont
  {O.}~\bibnamefont {Tengblad}}, \ and\ \bibinfo {author} {\bibfnamefont
  {V.}~\bibnamefont {Ziman}},\ }\href@noop {} {\bibfield  {journal} {\bibinfo
  {journal} {Nucl. Phys. A}\ }\textbf {\bibinfo {volume} {792}},\ \bibinfo
  {pages} {2 } (\bibinfo {year} {2007})}\BibitemShut {NoStop}%
\bibitem [{\citenamefont {Di~Pietro}\ \emph {et~al.}(2004)\citenamefont
  {Di~Pietro}, \citenamefont {Figuera}, \citenamefont {Amorini}, \citenamefont
  {Angulo}, \citenamefont {Cardella}, \citenamefont {Cherubini}, \citenamefont
  {Davinson}, \citenamefont {Leanza}, \citenamefont {Lu}, \citenamefont
  {Mahmud}, \citenamefont {Milin}, \citenamefont {Musumarra}, \citenamefont
  {Ninane}, \citenamefont {Papa}, \citenamefont {Pellegriti}, \citenamefont
  {Raabe}, \citenamefont {Rizzo}, \citenamefont {Ruiz}, \citenamefont
  {Shotter}, \citenamefont {Soi\ifmmode~\acute{c}\else \'{c}\fi{}},
  \citenamefont {Tudisco},\ and\ \citenamefont {Weissman}}]{DiPietro2004}%
  \BibitemOpen
  \bibfield  {author} {\bibinfo {author} {\bibfnamefont {A.}~\bibnamefont
  {Di~Pietro}}, \bibinfo {author} {\bibfnamefont {P.}~\bibnamefont {Figuera}},
  \bibinfo {author} {\bibfnamefont {F.}~\bibnamefont {Amorini}}, \bibinfo
  {author} {\bibfnamefont {C.}~\bibnamefont {Angulo}}, \bibinfo {author}
  {\bibfnamefont {G.}~\bibnamefont {Cardella}}, \bibinfo {author}
  {\bibfnamefont {S.}~\bibnamefont {Cherubini}}, \bibinfo {author}
  {\bibfnamefont {T.}~\bibnamefont {Davinson}}, \bibinfo {author}
  {\bibfnamefont {D.}~\bibnamefont {Leanza}}, \bibinfo {author} {\bibfnamefont
  {J.}~\bibnamefont {Lu}}, \bibinfo {author} {\bibfnamefont {H.}~\bibnamefont
  {Mahmud}}, \bibinfo {author} {\bibfnamefont {M.}~\bibnamefont {Milin}},
  \bibinfo {author} {\bibfnamefont {A.}~\bibnamefont {Musumarra}}, \bibinfo
  {author} {\bibfnamefont {A.}~\bibnamefont {Ninane}}, \bibinfo {author}
  {\bibfnamefont {M.}~\bibnamefont {Papa}}, \bibinfo {author} {\bibfnamefont
  {M.~G.}\ \bibnamefont {Pellegriti}}, \bibinfo {author} {\bibfnamefont
  {R.}~\bibnamefont {Raabe}}, \bibinfo {author} {\bibfnamefont
  {F.}~\bibnamefont {Rizzo}}, \bibinfo {author} {\bibfnamefont
  {C.}~\bibnamefont {Ruiz}}, \bibinfo {author} {\bibfnamefont {A.~C.}\
  \bibnamefont {Shotter}}, \bibinfo {author} {\bibfnamefont {N.}~\bibnamefont
  {Soi\ifmmode~\acute{c}\else \'{c}\fi{}}}, \bibinfo {author} {\bibfnamefont
  {S.}~\bibnamefont {Tudisco}}, \ and\ \bibinfo {author} {\bibfnamefont
  {L.}~\bibnamefont {Weissman}},\ }\href@noop {} {\bibfield  {journal}
  {\bibinfo  {journal} {Phys. Rev. C}\ }\textbf {\bibinfo {volume} {69}},\
  \bibinfo {pages} {044613} (\bibinfo {year} {2004})}\BibitemShut {NoStop}%
\bibitem [{\citenamefont {Fern{\'a}ndez-Garc{\'i}a}\ \emph
  {et~al.}(2013)\citenamefont {Fern{\'a}ndez-Garc{\'i}a}, \citenamefont
  {Cubero}, \citenamefont {Rodr\'{\i}guez-Gallardo}, \citenamefont {Acosta},
  \citenamefont {Alcorta}, \citenamefont {Alvarez}, \citenamefont {Borge},
  \citenamefont {Buchmann}, \citenamefont {Diget}, \citenamefont {Falou},
  \citenamefont {Fulton}, \citenamefont {Fynbo}, \citenamefont {Galaviz},
  \citenamefont {G\'omez-Camacho}, \citenamefont {Kanungo}, \citenamefont
  {Lay}, \citenamefont {Madurga}, \citenamefont {Martel}, \citenamefont {Moro},
  \citenamefont {Mukha}, \citenamefont {Nilsson}, \citenamefont
  {S\'anchez-Ben\'{\i}tez}, \citenamefont {Shotter}, \citenamefont {Tengblad},\
  and\ \citenamefont {Walden}}]{jp11li}%
  \BibitemOpen
  \bibfield  {author} {\bibinfo {author} {\bibfnamefont {J.~P.}\ \bibnamefont
  {Fern{\'a}ndez-Garc{\'i}a}}, \bibinfo {author} {\bibfnamefont
  {M.}~\bibnamefont {Cubero}}, \bibinfo {author} {\bibfnamefont
  {M.}~\bibnamefont {Rodr\'{\i}guez-Gallardo}}, \bibinfo {author}
  {\bibfnamefont {L.}~\bibnamefont {Acosta}}, \bibinfo {author} {\bibfnamefont
  {M.}~\bibnamefont {Alcorta}}, \bibinfo {author} {\bibfnamefont {M.~A.~G.}\
  \bibnamefont {Alvarez}}, \bibinfo {author} {\bibfnamefont {M.~J.~G.}\
  \bibnamefont {Borge}}, \bibinfo {author} {\bibfnamefont {L.}~\bibnamefont
  {Buchmann}}, \bibinfo {author} {\bibfnamefont {C.~A.}\ \bibnamefont {Diget}},
  \bibinfo {author} {\bibfnamefont {H.~A.}\ \bibnamefont {Falou}}, \bibinfo
  {author} {\bibfnamefont {B.~R.}\ \bibnamefont {Fulton}}, \bibinfo {author}
  {\bibfnamefont {H.~O.~U.}\ \bibnamefont {Fynbo}}, \bibinfo {author}
  {\bibfnamefont {D.}~\bibnamefont {Galaviz}}, \bibinfo {author} {\bibfnamefont
  {J.}~\bibnamefont {G\'omez-Camacho}}, \bibinfo {author} {\bibfnamefont
  {R.}~\bibnamefont {Kanungo}}, \bibinfo {author} {\bibfnamefont {J.~A.}\
  \bibnamefont {Lay}}, \bibinfo {author} {\bibfnamefont {M.}~\bibnamefont
  {Madurga}}, \bibinfo {author} {\bibfnamefont {I.}~\bibnamefont {Martel}},
  \bibinfo {author} {\bibfnamefont {A.~M.}\ \bibnamefont {Moro}}, \bibinfo
  {author} {\bibfnamefont {I.}~\bibnamefont {Mukha}}, \bibinfo {author}
  {\bibfnamefont {T.}~\bibnamefont {Nilsson}}, \bibinfo {author} {\bibfnamefont
  {A.~M.}\ \bibnamefont {S\'anchez-Ben\'{\i}tez}}, \bibinfo {author}
  {\bibfnamefont {A.}~\bibnamefont {Shotter}}, \bibinfo {author} {\bibfnamefont
  {O.}~\bibnamefont {Tengblad}}, \ and\ \bibinfo {author} {\bibfnamefont
  {P.}~\bibnamefont {Walden}},\ }\href@noop {} {\bibfield  {journal} {\bibinfo
  {journal} {Phys. Rev. Lett.}\ }\textbf {\bibinfo {volume} {110}},\ \bibinfo
  {pages} {142701} (\bibinfo {year} {2013})}\BibitemShut {NoStop}%
\bibitem [{\citenamefont {Tilley}\ \emph {et~al.}(2002)\citenamefont {Tilley},
  \citenamefont {Cheves}, \citenamefont {Godwin}, \citenamefont {Hale},
  \citenamefont {Hofmann}, \citenamefont {Kelley}, \citenamefont {Sheu},\ and\
  \citenamefont {Weller}}]{Till02}%
  \BibitemOpen
  \bibfield  {author} {\bibinfo {author} {\bibfnamefont {D.~R.}\ \bibnamefont
  {Tilley}}, \bibinfo {author} {\bibfnamefont {C.~M.}\ \bibnamefont {Cheves}},
  \bibinfo {author} {\bibfnamefont {J.~L.}\ \bibnamefont {Godwin}}, \bibinfo
  {author} {\bibfnamefont {G.~M.}\ \bibnamefont {Hale}}, \bibinfo {author}
  {\bibfnamefont {H.~M.}\ \bibnamefont {Hofmann}}, \bibinfo {author}
  {\bibfnamefont {J.~H.}\ \bibnamefont {Kelley}}, \bibinfo {author}
  {\bibfnamefont {C.~G.}\ \bibnamefont {Sheu}}, \ and\ \bibinfo {author}
  {\bibfnamefont {H.~R.}\ \bibnamefont {Weller}},\ }\href@noop {} {\bibfield
  {journal} {\bibinfo  {journal} {Nucl. Phys. A}\ }\textbf {\bibinfo {volume}
  {708}},\ \bibinfo {pages} {3} (\bibinfo {year} {2002})}\BibitemShut {NoStop}%
\bibitem [{\citenamefont {Chaussidon}\ \emph {et~al.}(2006)\citenamefont
  {Chaussidon}, \citenamefont {Robert},\ and\ \citenamefont
  {McKeegan}}]{Chau06}%
  \BibitemOpen
  \bibfield  {author} {\bibinfo {author} {\bibfnamefont {M.}~\bibnamefont
  {Chaussidon}}, \bibinfo {author} {\bibfnamefont {F.}~\bibnamefont {Robert}},
  \ and\ \bibinfo {author} {\bibfnamefont {K.~D.}\ \bibnamefont {McKeegan}},\
  }\href@noop {} {\bibfield  {journal} {\bibinfo  {journal} {Geochimica et
  Cosmochimica Acta}\ }\textbf {\bibinfo {volume} {70}},\ \bibinfo {pages} {224
  } (\bibinfo {year} {2006})}\BibitemShut {NoStop}%
\bibitem [{\citenamefont {Lodders}(2003)}]{Lod03}%
  \BibitemOpen
  \bibfield  {author} {\bibinfo {author} {\bibfnamefont {K.}~\bibnamefont
  {Lodders}},\ }\href@noop {} {\bibfield  {journal} {\bibinfo  {journal}
  {Astrophysical Journal}\ }\textbf {\bibinfo {volume} {591}},\ \bibinfo
  {pages} {1220} (\bibinfo {year} {2003})}\BibitemShut {NoStop}%
\bibitem [{\citenamefont {Ost}\ \emph {et~al.}(1972)\citenamefont {Ost},
  \citenamefont {Speth}, \citenamefont {Pfeiffer},\ and\ \citenamefont
  {Bethge}}]{Ost72}%
  \BibitemOpen
  \bibfield  {author} {\bibinfo {author} {\bibfnamefont {R.}~\bibnamefont
  {Ost}}, \bibinfo {author} {\bibfnamefont {E.}~\bibnamefont {Speth}}, \bibinfo
  {author} {\bibfnamefont {K.~O.}\ \bibnamefont {Pfeiffer}}, \ and\ \bibinfo
  {author} {\bibfnamefont {K.}~\bibnamefont {Bethge}},\ }\href@noop {}
  {\bibfield  {journal} {\bibinfo  {journal} {Phys. Rev. C}\ }\textbf {\bibinfo
  {volume} {5}},\ \bibinfo {pages} {1835} (\bibinfo {year} {1972})}\BibitemShut
  {NoStop}%
\bibitem [{\citenamefont {Casal}\ \emph {et~al.}(2014)\citenamefont {Casal},
  \citenamefont {Rodr\'{\i}guez-Gallardo}, \citenamefont {Arias},\ and\
  \citenamefont {Thompson}}]{Cas14}%
  \BibitemOpen
  \bibfield  {author} {\bibinfo {author} {\bibfnamefont {J.}~\bibnamefont
  {Casal}}, \bibinfo {author} {\bibfnamefont {M.}~\bibnamefont
  {Rodr\'{\i}guez-Gallardo}}, \bibinfo {author} {\bibfnamefont {J.~M.}\
  \bibnamefont {Arias}}, \ and\ \bibinfo {author} {\bibfnamefont {I.~J.}\
  \bibnamefont {Thompson}},\ }\href@noop {} {\bibfield  {journal} {\bibinfo
  {journal} {Phys. Rev. C}\ }\textbf {\bibinfo {volume} {90}},\ \bibinfo
  {pages} {044304} (\bibinfo {year} {2014})}\BibitemShut {NoStop}%
\bibitem [{\citenamefont {Arazi}\ \emph {et~al.}(2018)\citenamefont {Arazi},
  \citenamefont {Casal}, \citenamefont {Rodr\'{\i}guez-Gallardo}, \citenamefont
  {Arias}, \citenamefont {Lichtenth\"aler~Filho}, \citenamefont {Abriola},
  \citenamefont {Capurro}, \citenamefont {Cardona}, \citenamefont {Carnelli},
  \citenamefont {de~Barbar\'a}, \citenamefont {Fern\'andez~Niello},
  \citenamefont {Figueira}, \citenamefont {Fimiani}, \citenamefont {Hojman},
  \citenamefont {Mart\'{\i}}, \citenamefont {Mart\'{\i}nez~Heimman},\ and\
  \citenamefont {Pacheco}}]{Ara18}%
  \BibitemOpen
  \bibfield  {author} {\bibinfo {author} {\bibfnamefont {A.}~\bibnamefont
  {Arazi}}, \bibinfo {author} {\bibfnamefont {J.}~\bibnamefont {Casal}},
  \bibinfo {author} {\bibfnamefont {M.}~\bibnamefont
  {Rodr\'{\i}guez-Gallardo}}, \bibinfo {author} {\bibfnamefont {J.~M.}\
  \bibnamefont {Arias}}, \bibinfo {author} {\bibfnamefont {R.}~\bibnamefont
  {Lichtenth\"aler~Filho}}, \bibinfo {author} {\bibfnamefont {D.}~\bibnamefont
  {Abriola}}, \bibinfo {author} {\bibfnamefont {O.~A.}\ \bibnamefont
  {Capurro}}, \bibinfo {author} {\bibfnamefont {M.~A.}\ \bibnamefont
  {Cardona}}, \bibinfo {author} {\bibfnamefont {P.~F.~F.}\ \bibnamefont
  {Carnelli}}, \bibinfo {author} {\bibfnamefont {E.}~\bibnamefont
  {de~Barbar\'a}}, \bibinfo {author} {\bibfnamefont {J.}~\bibnamefont
  {Fern\'andez~Niello}}, \bibinfo {author} {\bibfnamefont {J.~M.}\ \bibnamefont
  {Figueira}}, \bibinfo {author} {\bibfnamefont {L.}~\bibnamefont {Fimiani}},
  \bibinfo {author} {\bibfnamefont {D.}~\bibnamefont {Hojman}}, \bibinfo
  {author} {\bibfnamefont {G.~V.}\ \bibnamefont {Mart\'{\i}}}, \bibinfo
  {author} {\bibfnamefont {D.}~\bibnamefont {Mart\'{\i}nez~Heimman}}, \ and\
  \bibinfo {author} {\bibfnamefont {A.~J.}\ \bibnamefont {Pacheco}},\
  }\href@noop {} {\bibfield  {journal} {\bibinfo  {journal} {Phys. Rev. C}\
  }\textbf {\bibinfo {volume} {97}},\ \bibinfo {pages} {044609} (\bibinfo
  {year} {2018})}\BibitemShut {NoStop}%
\bibitem [{\citenamefont {Alvarez}\ \emph {et~al.}(2018)\citenamefont
  {Alvarez}, \citenamefont {Rodr\'{\i}guez-Gallardo}, \citenamefont {Gasques},
  \citenamefont {Chamon}, \citenamefont {Oliveira}, \citenamefont {Scarduelli},
  \citenamefont {Freitas}, \citenamefont {Rossi}, \citenamefont {Zagatto},
  \citenamefont {Rangel}, \citenamefont {Lubian},\ and\ \citenamefont
  {Padron}}]{Alv18}%
  \BibitemOpen
  \bibfield  {author} {\bibinfo {author} {\bibfnamefont {M.~A.~G.}\
  \bibnamefont {Alvarez}}, \bibinfo {author} {\bibfnamefont {M.}~\bibnamefont
  {Rodr\'{\i}guez-Gallardo}}, \bibinfo {author} {\bibfnamefont {L.~R.}\
  \bibnamefont {Gasques}}, \bibinfo {author} {\bibfnamefont {L.~C.}\
  \bibnamefont {Chamon}}, \bibinfo {author} {\bibfnamefont {J.~R.~B.}\
  \bibnamefont {Oliveira}}, \bibinfo {author} {\bibfnamefont {V.}~\bibnamefont
  {Scarduelli}}, \bibinfo {author} {\bibfnamefont {A.~S.}\ \bibnamefont
  {Freitas}}, \bibinfo {author} {\bibfnamefont {E.~S.}\ \bibnamefont {Rossi}},
  \bibinfo {author} {\bibfnamefont {V.~A.~B.}\ \bibnamefont {Zagatto}},
  \bibinfo {author} {\bibfnamefont {J.}~\bibnamefont {Rangel}}, \bibinfo
  {author} {\bibfnamefont {J.}~\bibnamefont {Lubian}}, \ and\ \bibinfo {author}
  {\bibfnamefont {I.}~\bibnamefont {Padron}},\ }\href@noop {} {\bibfield
  {journal} {\bibinfo  {journal} {Phys. Rev. C}\ }\textbf {\bibinfo {volume}
  {98}},\ \bibinfo {pages} {024621} (\bibinfo {year} {2018})}\BibitemShut
  {NoStop}%
\bibitem [{\citenamefont {S\'{a}nchez-Ben\'{i}tez}\ \emph
  {et~al.}(2008)\citenamefont {S\'{a}nchez-Ben\'{i}tez}, \citenamefont
  {Escrig}, \citenamefont {Alvarez}, \citenamefont {Andr\'{e}s}, \citenamefont
  {Angulo}, \citenamefont {Borge}, \citenamefont {Cabrera}, \citenamefont
  {Cherubini}, \citenamefont {Demaret}, \citenamefont {Espino}, \citenamefont
  {Figuera}, \citenamefont {Freer}, \citenamefont {Garc\'{i}a-Ramos},
  \citenamefont {G\'{o}mez-Camacho}, \citenamefont {Gulino}, \citenamefont
  {Kakuee}, \citenamefont {Martel}, \citenamefont {Metelko}, \citenamefont
  {Moro}, \citenamefont {P\'{e}rez-Bernal}, \citenamefont {Rahighi},
  \citenamefont {Rusek}, \citenamefont {Smirnov}, \citenamefont {Tengblad},
  \citenamefont {Duppen},\ and\ \citenamefont {Ziman}}]{San08}%
  \BibitemOpen
  \bibfield  {author} {\bibinfo {author} {\bibfnamefont {A.~M.}\ \bibnamefont
  {S\'{a}nchez-Ben\'{i}tez}}, \bibinfo {author} {\bibfnamefont
  {D.}~\bibnamefont {Escrig}}, \bibinfo {author} {\bibfnamefont {M.~A.~G.}\
  \bibnamefont {Alvarez}}, \bibinfo {author} {\bibfnamefont {M.~V.}\
  \bibnamefont {Andr\'{e}s}}, \bibinfo {author} {\bibfnamefont
  {C.}~\bibnamefont {Angulo}}, \bibinfo {author} {\bibfnamefont {M.~J.~G.}\
  \bibnamefont {Borge}}, \bibinfo {author} {\bibfnamefont {J.}~\bibnamefont
  {Cabrera}}, \bibinfo {author} {\bibfnamefont {S.}~\bibnamefont {Cherubini}},
  \bibinfo {author} {\bibfnamefont {P.}~\bibnamefont {Demaret}}, \bibinfo
  {author} {\bibfnamefont {J.~M.}\ \bibnamefont {Espino}}, \bibinfo {author}
  {\bibfnamefont {P.}~\bibnamefont {Figuera}}, \bibinfo {author} {\bibfnamefont
  {M.}~\bibnamefont {Freer}}, \bibinfo {author} {\bibfnamefont
  {J.}~\bibnamefont {Garc\'{i}a-Ramos}}, \bibinfo {author} {\bibfnamefont
  {J.}~\bibnamefont {G\'{o}mez-Camacho}}, \bibinfo {author} {\bibfnamefont
  {M.}~\bibnamefont {Gulino}}, \bibinfo {author} {\bibfnamefont
  {O.}~\bibnamefont {Kakuee}}, \bibinfo {author} {\bibfnamefont
  {I.}~\bibnamefont {Martel}}, \bibinfo {author} {\bibfnamefont
  {C.}~\bibnamefont {Metelko}}, \bibinfo {author} {\bibfnamefont
  {A.}~\bibnamefont {Moro}}, \bibinfo {author} {\bibfnamefont {F.}~\bibnamefont
  {P\'{e}rez-Bernal}}, \bibinfo {author} {\bibfnamefont {J.}~\bibnamefont
  {Rahighi}}, \bibinfo {author} {\bibfnamefont {K.}~\bibnamefont {Rusek}},
  \bibinfo {author} {\bibfnamefont {D.}~\bibnamefont {Smirnov}}, \bibinfo
  {author} {\bibfnamefont {O.}~\bibnamefont {Tengblad}}, \bibinfo {author}
  {\bibfnamefont {P.~V.}\ \bibnamefont {Duppen}}, \ and\ \bibinfo {author}
  {\bibfnamefont {V.}~\bibnamefont {Ziman}},\ }\href@noop {} {\bibfield
  {journal} {\bibinfo  {journal} {Nucl. Phys. A}\ }\textbf {\bibinfo {volume}
  {803}},\ \bibinfo {pages} {30 } (\bibinfo {year} {2008})}\BibitemShut
  {NoStop}%
\bibitem [{\citenamefont {Acosta}\ \emph {et~al.}(2009)\citenamefont {Acosta}
  \emph {et~al.}}]{Aco09}%
  \BibitemOpen
  \bibfield  {author} {\bibinfo {author} {\bibfnamefont {L.}~\bibnamefont
  {Acosta}} \emph {et~al.},\ }\href@noop {} {\bibfield  {journal} {\bibinfo
  {journal} {Eur. Phys. J. A}\ }\textbf {\bibinfo {volume} {42}},\ \bibinfo
  {pages} {461} (\bibinfo {year} {2009})}\BibitemShut {NoStop}%
\bibitem [{\citenamefont {Acosta}\ \emph {et~al.}(2011)\citenamefont {Acosta},
  \citenamefont {S\'anchez-Ben\'{i}tez}, \citenamefont {G\'omez}, \citenamefont
  {Martel}, \citenamefont {P\'erez-Bernal}, \citenamefont {Pizarro},
  \citenamefont {Rodr\'{i}guez-Quintero}, \citenamefont {Rusek}, \citenamefont
  {Alvarez}, \citenamefont {Andr\'es}, \citenamefont {Espino}, \citenamefont
  {Fern\'andez-Garc\'{i}a}, \citenamefont {G\'omez-Camacho}, \citenamefont
  {Moro}, \citenamefont {Angulo}, \citenamefont {Cabrera}, \citenamefont
  {Casarejos}, \citenamefont {Demaret}, \citenamefont {Borge}, \citenamefont
  {Escrig}, \citenamefont {Tengblad}, \citenamefont {Cherubini}, \citenamefont
  {Figuera}, \citenamefont {Gulino}, \citenamefont {Freer}, \citenamefont
  {Metelko}, \citenamefont {Ziman}, \citenamefont {Raabe}, \citenamefont
  {Mukha}, \citenamefont {Smirnov}, \citenamefont {Kakuee},\ and\ \citenamefont
  {Rahighi}}]{Aco11}%
  \BibitemOpen
  \bibfield  {author} {\bibinfo {author} {\bibfnamefont {L.}~\bibnamefont
  {Acosta}}, \bibinfo {author} {\bibfnamefont {A.~M.}\ \bibnamefont
  {S\'anchez-Ben\'{i}tez}}, \bibinfo {author} {\bibfnamefont {M.~E.}\
  \bibnamefont {G\'omez}}, \bibinfo {author} {\bibfnamefont {I.}~\bibnamefont
  {Martel}}, \bibinfo {author} {\bibfnamefont {F.}~\bibnamefont
  {P\'erez-Bernal}}, \bibinfo {author} {\bibfnamefont {F.}~\bibnamefont
  {Pizarro}}, \bibinfo {author} {\bibfnamefont {J.}~\bibnamefont
  {Rodr\'{i}guez-Quintero}}, \bibinfo {author} {\bibfnamefont {K.}~\bibnamefont
  {Rusek}}, \bibinfo {author} {\bibfnamefont {M.~A.~G.}\ \bibnamefont
  {Alvarez}}, \bibinfo {author} {\bibfnamefont {M.~V.}\ \bibnamefont
  {Andr\'es}}, \bibinfo {author} {\bibfnamefont {J.~M.}\ \bibnamefont
  {Espino}}, \bibinfo {author} {\bibfnamefont {J.~P.}\ \bibnamefont
  {Fern\'andez-Garc\'{i}a}}, \bibinfo {author} {\bibfnamefont {J.}~\bibnamefont
  {G\'omez-Camacho}}, \bibinfo {author} {\bibfnamefont {A.~M.}\ \bibnamefont
  {Moro}}, \bibinfo {author} {\bibfnamefont {C.}~\bibnamefont {Angulo}},
  \bibinfo {author} {\bibfnamefont {J.}~\bibnamefont {Cabrera}}, \bibinfo
  {author} {\bibfnamefont {E.}~\bibnamefont {Casarejos}}, \bibinfo {author}
  {\bibfnamefont {P.}~\bibnamefont {Demaret}}, \bibinfo {author} {\bibfnamefont
  {M.~J.~G.}\ \bibnamefont {Borge}}, \bibinfo {author} {\bibfnamefont
  {D.}~\bibnamefont {Escrig}}, \bibinfo {author} {\bibfnamefont
  {O.}~\bibnamefont {Tengblad}}, \bibinfo {author} {\bibfnamefont
  {S.}~\bibnamefont {Cherubini}}, \bibinfo {author} {\bibfnamefont
  {P.}~\bibnamefont {Figuera}}, \bibinfo {author} {\bibfnamefont
  {M.}~\bibnamefont {Gulino}}, \bibinfo {author} {\bibfnamefont
  {M.}~\bibnamefont {Freer}}, \bibinfo {author} {\bibfnamefont
  {C.}~\bibnamefont {Metelko}}, \bibinfo {author} {\bibfnamefont
  {V.}~\bibnamefont {Ziman}}, \bibinfo {author} {\bibfnamefont
  {R.}~\bibnamefont {Raabe}}, \bibinfo {author} {\bibfnamefont
  {I.}~\bibnamefont {Mukha}}, \bibinfo {author} {\bibfnamefont
  {D.}~\bibnamefont {Smirnov}}, \bibinfo {author} {\bibfnamefont {O.~R.}\
  \bibnamefont {Kakuee}}, \ and\ \bibinfo {author} {\bibfnamefont
  {J.}~\bibnamefont {Rahighi}},\ }\href@noop {} {\bibfield  {journal} {\bibinfo
   {journal} {Phys. Rev. C}\ }\textbf {\bibinfo {volume} {84}},\ \bibinfo
  {pages} {044604} (\bibinfo {year} {2011})}\BibitemShut {NoStop}%
\bibitem [{\citenamefont {Rafiei}\ \emph {et~al.}(2010)\citenamefont {Rafiei},
  \citenamefont {du~Rietz}, \citenamefont {Luong}, \citenamefont {Hinde},
  \citenamefont {Dasgupta}, \citenamefont {Evers},\ and\ \citenamefont
  {D\'{i}az-Torres}}]{Raf10}%
  \BibitemOpen
  \bibfield  {author} {\bibinfo {author} {\bibfnamefont {R.}~\bibnamefont
  {Rafiei}}, \bibinfo {author} {\bibfnamefont {R.}~\bibnamefont {du~Rietz}},
  \bibinfo {author} {\bibfnamefont {D.~H.}\ \bibnamefont {Luong}}, \bibinfo
  {author} {\bibfnamefont {D.~J.}\ \bibnamefont {Hinde}}, \bibinfo {author}
  {\bibfnamefont {M.}~\bibnamefont {Dasgupta}}, \bibinfo {author}
  {\bibfnamefont {M.}~\bibnamefont {Evers}}, \ and\ \bibinfo {author}
  {\bibfnamefont {A.}~\bibnamefont {D\'{i}az-Torres}},\ }\href@noop {}
  {\bibfield  {journal} {\bibinfo  {journal} {Phys. Rev. C}\ }\textbf {\bibinfo
  {volume} {81}},\ \bibinfo {pages} {024601} (\bibinfo {year}
  {2010})}\BibitemShut {NoStop}%
\bibitem [{\citenamefont {Luong}\ \emph {et~al.}(2011)\citenamefont {Luong},
  \citenamefont {Dasgupta}, \citenamefont {Hinde}, \citenamefont {du~Rietz},
  \citenamefont {Rafiei}, \citenamefont {Lin}, \citenamefont {Evers},\ and\
  \citenamefont {D\'{i}az-Torres}}]{Luo11}%
  \BibitemOpen
  \bibfield  {author} {\bibinfo {author} {\bibfnamefont {D.~H.}\ \bibnamefont
  {Luong}}, \bibinfo {author} {\bibfnamefont {M.}~\bibnamefont {Dasgupta}},
  \bibinfo {author} {\bibfnamefont {D.}~\bibnamefont {Hinde}}, \bibinfo
  {author} {\bibfnamefont {R.}~\bibnamefont {du~Rietz}}, \bibinfo {author}
  {\bibfnamefont {R.}~\bibnamefont {Rafiei}}, \bibinfo {author} {\bibfnamefont
  {C.}~\bibnamefont {Lin}}, \bibinfo {author} {\bibfnamefont {M.}~\bibnamefont
  {Evers}}, \ and\ \bibinfo {author} {\bibfnamefont {A.}~\bibnamefont
  {D\'{i}az-Torres}},\ }\href@noop {} {\bibfield  {journal} {\bibinfo
  {journal} {Phys. Lett. B}\ }\textbf {\bibinfo {volume} {695}},\ \bibinfo
  {pages} {105 } (\bibinfo {year} {2011})}\BibitemShut {NoStop}%
\bibitem [{\citenamefont {Kalkal}\ \emph {et~al.}(2016)\citenamefont {Kalkal},
  \citenamefont {Simpson}, \citenamefont {Luong}, \citenamefont {Cook},
  \citenamefont {Dasgupta}, \citenamefont {Hinde}, \citenamefont {Carter},
  \citenamefont {Jeung}, \citenamefont {Mohanto}, \citenamefont {Palshetkar},
  \citenamefont {Prasad}, \citenamefont {Rafferty}, \citenamefont {Simenel},
  \citenamefont {Vo-Phuoc}, \citenamefont {Williams}, \citenamefont {Gasques},
  \citenamefont {Gomes},\ and\ \citenamefont {Linares}}]{Kal16}%
  \BibitemOpen
  \bibfield  {author} {\bibinfo {author} {\bibfnamefont {S.}~\bibnamefont
  {Kalkal}}, \bibinfo {author} {\bibfnamefont {E.~C.}\ \bibnamefont {Simpson}},
  \bibinfo {author} {\bibfnamefont {D.~H.}\ \bibnamefont {Luong}}, \bibinfo
  {author} {\bibfnamefont {K.~J.}\ \bibnamefont {Cook}}, \bibinfo {author}
  {\bibfnamefont {M.}~\bibnamefont {Dasgupta}}, \bibinfo {author}
  {\bibfnamefont {D.~J.}\ \bibnamefont {Hinde}}, \bibinfo {author}
  {\bibfnamefont {I.~P.}\ \bibnamefont {Carter}}, \bibinfo {author}
  {\bibfnamefont {D.~Y.}\ \bibnamefont {Jeung}}, \bibinfo {author}
  {\bibfnamefont {G.}~\bibnamefont {Mohanto}}, \bibinfo {author} {\bibfnamefont
  {C.~S.}\ \bibnamefont {Palshetkar}}, \bibinfo {author} {\bibfnamefont
  {E.}~\bibnamefont {Prasad}}, \bibinfo {author} {\bibfnamefont {D.~C.}\
  \bibnamefont {Rafferty}}, \bibinfo {author} {\bibfnamefont {C.}~\bibnamefont
  {Simenel}}, \bibinfo {author} {\bibfnamefont {K.}~\bibnamefont {Vo-Phuoc}},
  \bibinfo {author} {\bibfnamefont {E.}~\bibnamefont {Williams}}, \bibinfo
  {author} {\bibfnamefont {L.~R.}\ \bibnamefont {Gasques}}, \bibinfo {author}
  {\bibfnamefont {P.~R.~S.}\ \bibnamefont {Gomes}}, \ and\ \bibinfo {author}
  {\bibfnamefont {R.}~\bibnamefont {Linares}},\ }\href@noop {} {\bibfield
  {journal} {\bibinfo  {journal} {Phys. Rev. C}\ }\textbf {\bibinfo {volume}
  {93}},\ \bibinfo {pages} {044605} (\bibinfo {year} {2016})}\BibitemShut
  {NoStop}%
\bibitem [{\citenamefont {Zagatto}\ \emph {et~al.}(2017)\citenamefont
  {Zagatto}, \citenamefont {Lubian}, \citenamefont {Gasques}, \citenamefont
  {Alvarez}, \citenamefont {Chamon}, \citenamefont {Oliveira}, \citenamefont
  {Alc\'antara-N\'u\~nez}, \citenamefont {Medina}, \citenamefont {Scarduelli},
  \citenamefont {Freitas}, \citenamefont {Padron}, \citenamefont {Rossi},\ and\
  \citenamefont {Shorto}}]{Zag17}%
  \BibitemOpen
  \bibfield  {author} {\bibinfo {author} {\bibfnamefont {V.~A.~B.}\
  \bibnamefont {Zagatto}}, \bibinfo {author} {\bibfnamefont {J.}~\bibnamefont
  {Lubian}}, \bibinfo {author} {\bibfnamefont {L.~R.}\ \bibnamefont {Gasques}},
  \bibinfo {author} {\bibfnamefont {M.~A.~G.}\ \bibnamefont {Alvarez}},
  \bibinfo {author} {\bibfnamefont {L.~C.}\ \bibnamefont {Chamon}}, \bibinfo
  {author} {\bibfnamefont {J.~R.~B.}\ \bibnamefont {Oliveira}}, \bibinfo
  {author} {\bibfnamefont {J.~A.}\ \bibnamefont {Alc\'antara-N\'u\~nez}},
  \bibinfo {author} {\bibfnamefont {N.~H.}\ \bibnamefont {Medina}}, \bibinfo
  {author} {\bibfnamefont {V.}~\bibnamefont {Scarduelli}}, \bibinfo {author}
  {\bibfnamefont {A.}~\bibnamefont {Freitas}}, \bibinfo {author} {\bibfnamefont
  {I.}~\bibnamefont {Padron}}, \bibinfo {author} {\bibfnamefont {E.~S.}\
  \bibnamefont {Rossi}}, \ and\ \bibinfo {author} {\bibfnamefont {J.~M.~B.}\
  \bibnamefont {Shorto}},\ }\href@noop {} {\bibfield  {journal} {\bibinfo
  {journal} {Phys. Rev. C}\ }\textbf {\bibinfo {volume} {95}},\ \bibinfo
  {pages} {064614} (\bibinfo {year} {2017})}\BibitemShut {NoStop}%
\bibitem [{\citenamefont {Gasques}\ \emph {et~al.}(2018)\citenamefont
  {Gasques}, \citenamefont {Freitas}, \citenamefont {Chamon}, \citenamefont
  {Oliveira}, \citenamefont {Medina}, \citenamefont {Scarduelli}, \citenamefont
  {Rossi}, \citenamefont {Alvarez}, \citenamefont {Zagatto}, \citenamefont
  {Lubian}, \citenamefont {Nobre}, \citenamefont {Padron},\ and\ \citenamefont
  {Carlson}}]{Gas18}%
  \BibitemOpen
  \bibfield  {author} {\bibinfo {author} {\bibfnamefont {L.~R.}\ \bibnamefont
  {Gasques}}, \bibinfo {author} {\bibfnamefont {A.~S.}\ \bibnamefont
  {Freitas}}, \bibinfo {author} {\bibfnamefont {L.~C.}\ \bibnamefont {Chamon}},
  \bibinfo {author} {\bibfnamefont {J.~R.~B.}\ \bibnamefont {Oliveira}},
  \bibinfo {author} {\bibfnamefont {N.~H.}\ \bibnamefont {Medina}}, \bibinfo
  {author} {\bibfnamefont {V.}~\bibnamefont {Scarduelli}}, \bibinfo {author}
  {\bibfnamefont {E.~S.}\ \bibnamefont {Rossi}}, \bibinfo {author}
  {\bibfnamefont {M.~A.~G.}\ \bibnamefont {Alvarez}}, \bibinfo {author}
  {\bibfnamefont {V.~A.~B.}\ \bibnamefont {Zagatto}}, \bibinfo {author}
  {\bibfnamefont {J.}~\bibnamefont {Lubian}}, \bibinfo {author} {\bibfnamefont
  {G.~P.~A.}\ \bibnamefont {Nobre}}, \bibinfo {author} {\bibfnamefont
  {I.}~\bibnamefont {Padron}}, \ and\ \bibinfo {author} {\bibfnamefont {B.~V.}\
  \bibnamefont {Carlson}},\ }\href@noop {} {\bibfield  {journal} {\bibinfo
  {journal} {Phys. Rev. C}\ }\textbf {\bibinfo {volume} {97}},\ \bibinfo
  {pages} {034629} (\bibinfo {year} {2018})}\BibitemShut {NoStop}%
\bibitem [{\citenamefont {Chamon}\ \emph {et~al.}(2002)\citenamefont {Chamon},
  \citenamefont {Carlson}, \citenamefont {Gasques}, \citenamefont {Pereira},
  \citenamefont {De~Conti}, \citenamefont {Alvarez}, \citenamefont {Hussein},
  \citenamefont {C\^andido~Ribeiro}, \citenamefont {Rossi},\ and\ \citenamefont
  {Silva}}]{Cha02}%
  \BibitemOpen
  \bibfield  {author} {\bibinfo {author} {\bibfnamefont {L.~C.}\ \bibnamefont
  {Chamon}}, \bibinfo {author} {\bibfnamefont {B.~V.}\ \bibnamefont {Carlson}},
  \bibinfo {author} {\bibfnamefont {L.~R.}\ \bibnamefont {Gasques}}, \bibinfo
  {author} {\bibfnamefont {D.}~\bibnamefont {Pereira}}, \bibinfo {author}
  {\bibfnamefont {C.}~\bibnamefont {De~Conti}}, \bibinfo {author}
  {\bibfnamefont {M.~A.~G.}\ \bibnamefont {Alvarez}}, \bibinfo {author}
  {\bibfnamefont {M.~S.}\ \bibnamefont {Hussein}}, \bibinfo {author}
  {\bibfnamefont {M.~A.}\ \bibnamefont {C\^andido~Ribeiro}}, \bibinfo {author}
  {\bibfnamefont {E.~S.}\ \bibnamefont {Rossi}}, \ and\ \bibinfo {author}
  {\bibfnamefont {C.~P.}\ \bibnamefont {Silva}},\ }\href@noop {} {\bibfield
  {journal} {\bibinfo  {journal} {Phys. Rev. C}\ }\textbf {\bibinfo {volume}
  {66}},\ \bibinfo {pages} {014610} (\bibinfo {year} {2002})}\BibitemShut
  {NoStop}%
\bibitem [{\citenamefont {Mohr}\ \emph {et~al.}(2010)\citenamefont {Mohr},
  \citenamefont {de~Faria}, \citenamefont {Lichtenthaler}, \citenamefont
  {Pires}, \citenamefont {Guimar{\~a}es}, \citenamefont {L{\'e}pine-Szily},
  \citenamefont {Mendes}, \citenamefont {Arazi}, \citenamefont {Barioni},
  \citenamefont {Morcelle},\ and\ \citenamefont {Morais}}]{Mohr10}%
  \BibitemOpen
  \bibfield  {author} {\bibinfo {author} {\bibfnamefont {P.}~\bibnamefont
  {Mohr}}, \bibinfo {author} {\bibfnamefont {P.~N.}\ \bibnamefont {de~Faria}},
  \bibinfo {author} {\bibfnamefont {R.}~\bibnamefont {Lichtenthaler}}, \bibinfo
  {author} {\bibfnamefont {K.~C.~C.}\ \bibnamefont {Pires}}, \bibinfo {author}
  {\bibfnamefont {V.}~\bibnamefont {Guimar{\~a}es}}, \bibinfo {author}
  {\bibfnamefont {A.}~\bibnamefont {L{\'e}pine-Szily}}, \bibinfo {author}
  {\bibfnamefont {D.~R.}\ \bibnamefont {Mendes}}, \bibinfo {author}
  {\bibfnamefont {A.}~\bibnamefont {Arazi}}, \bibinfo {author} {\bibfnamefont
  {A.}~\bibnamefont {Barioni}}, \bibinfo {author} {\bibfnamefont
  {V.}~\bibnamefont {Morcelle}}, \ and\ \bibinfo {author} {\bibfnamefont
  {M.~C.}\ \bibnamefont {Morais}},\ }\href@noop {} {\bibfield  {journal}
  {\bibinfo  {journal} {Phys. Rev. C}\ }\textbf {\bibinfo {volume} {82}},\
  \bibinfo {pages} {044606} (\bibinfo {year} {2010})}\BibitemShut {NoStop}%
\bibitem [{\citenamefont {Kumabe}\ \emph {et~al.}(1968)\citenamefont {Kumabe},
  \citenamefont {Ogata}, \citenamefont {Kim}, \citenamefont {Inoue},
  \citenamefont {Okuma},\ and\ \citenamefont {Matoba}}]{Kum68}%
  \BibitemOpen
  \bibfield  {author} {\bibinfo {author} {\bibfnamefont {I.}~\bibnamefont
  {Kumabe}}, \bibinfo {author} {\bibfnamefont {H.}~\bibnamefont {Ogata}},
  \bibinfo {author} {\bibfnamefont {T.-H.}\ \bibnamefont {Kim}}, \bibinfo
  {author} {\bibfnamefont {M.}~\bibnamefont {Inoue}}, \bibinfo {author}
  {\bibfnamefont {Y.}~\bibnamefont {Okuma}}, \ and\ \bibinfo {author}
  {\bibfnamefont {M.}~\bibnamefont {Matoba}},\ }\href@noop {} {\bibfield
  {journal} {\bibinfo  {journal} {J. Phys. Soc. Jpn.}\ }\textbf {\bibinfo
  {volume} {25}},\ \bibinfo {pages} {14} (\bibinfo {year} {1968})}\BibitemShut
  {NoStop}%
\bibitem [{\citenamefont {Silva}\ \emph {et~al.}(2001)\citenamefont {Silva},
  \citenamefont {Alvarez}, \citenamefont {Chamon}, \citenamefont {Pereira},
  \citenamefont {Rao}, \citenamefont {Jr.}, \citenamefont {Gasques},
  \citenamefont {Santo}, \citenamefont {Anjos}, \citenamefont {Lubian},
  \citenamefont {Gomes}, \citenamefont {Muri}, \citenamefont {Carlson},
  \citenamefont {Kailas}, \citenamefont {Chatterjee}, \citenamefont {Singh},
  \citenamefont {Shrivastava}, \citenamefont {Mahata},\ and\ \citenamefont
  {Santra}}]{Cel01}%
  \BibitemOpen
  \bibfield  {author} {\bibinfo {author} {\bibfnamefont {C.}~\bibnamefont
  {Silva}}, \bibinfo {author} {\bibfnamefont {M.}~\bibnamefont {Alvarez}},
  \bibinfo {author} {\bibfnamefont {L.}~\bibnamefont {Chamon}}, \bibinfo
  {author} {\bibfnamefont {D.}~\bibnamefont {Pereira}}, \bibinfo {author}
  {\bibfnamefont {M.}~\bibnamefont {Rao}}, \bibinfo {author} {\bibfnamefont
  {E.~R.}\ \bibnamefont {Jr.}}, \bibinfo {author} {\bibfnamefont
  {L.}~\bibnamefont {Gasques}}, \bibinfo {author} {\bibfnamefont
  {M.}~\bibnamefont {Santo}}, \bibinfo {author} {\bibfnamefont
  {R.}~\bibnamefont {Anjos}}, \bibinfo {author} {\bibfnamefont
  {J.}~\bibnamefont {Lubian}}, \bibinfo {author} {\bibfnamefont
  {P.}~\bibnamefont {Gomes}}, \bibinfo {author} {\bibfnamefont
  {C.}~\bibnamefont {Muri}}, \bibinfo {author} {\bibfnamefont {B.}~\bibnamefont
  {Carlson}}, \bibinfo {author} {\bibfnamefont {S.}~\bibnamefont {Kailas}},
  \bibinfo {author} {\bibfnamefont {A.}~\bibnamefont {Chatterjee}}, \bibinfo
  {author} {\bibfnamefont {P.}~\bibnamefont {Singh}}, \bibinfo {author}
  {\bibfnamefont {A.}~\bibnamefont {Shrivastava}}, \bibinfo {author}
  {\bibfnamefont {K.}~\bibnamefont {Mahata}}, \ and\ \bibinfo {author}
  {\bibfnamefont {S.}~\bibnamefont {Santra}},\ }\href@noop {} {\bibfield
  {journal} {\bibinfo  {journal} {Nucl. Phys. A}\ }\textbf {\bibinfo {volume}
  {679}},\ \bibinfo {pages} {287 } (\bibinfo {year} {2001})}\BibitemShut
  {NoStop}%
\bibitem [{\citenamefont {Bohlen}\ \emph {et~al.}(1975)\citenamefont {Bohlen},
  \citenamefont {Hildenbrand}, \citenamefont {Gobbi},\ and\ \citenamefont
  {Kubo}}]{Boh75}%
  \BibitemOpen
  \bibfield  {author} {\bibinfo {author} {\bibfnamefont {H.~G.}\ \bibnamefont
  {Bohlen}}, \bibinfo {author} {\bibfnamefont {K.~D.}\ \bibnamefont
  {Hildenbrand}}, \bibinfo {author} {\bibfnamefont {A.}~\bibnamefont {Gobbi}},
  \ and\ \bibinfo {author} {\bibfnamefont {K.~I.}\ \bibnamefont {Kubo}},\
  }\href@noop {} {\bibfield  {journal} {\bibinfo  {journal} {Z. Phys. A Atoms
  and Nuclei}\ }\textbf {\bibinfo {volume} {273}},\ \bibinfo {pages} {211 }
  (\bibinfo {year} {1975})}\BibitemShut {NoStop}%
\bibitem [{\citenamefont {de~Faria}\ \emph {et~al.}(2010)\citenamefont
  {de~Faria}, \citenamefont {Lichtenth{\"a}ler}, \citenamefont {Pires},
  \citenamefont {Moro}, \citenamefont {L{\'e}pine-Szily}, \citenamefont
  {Guimar{\~a}es}, \citenamefont {Mendes}, \citenamefont {Arazi}, \citenamefont
  {Rodr{\'i}guez-Gallardo}, \citenamefont {Barioni}, \citenamefont {Morcelle},
  \citenamefont {Morais}, \citenamefont {Camargo}, \citenamefont
  {Alc{\'a}ntara~Nu{\~n}ez},\ and\ \citenamefont {Assun{\c c}{\~a}o}}]{Far10}%
  \BibitemOpen
  \bibfield  {author} {\bibinfo {author} {\bibfnamefont {P.~N.}\ \bibnamefont
  {de~Faria}}, \bibinfo {author} {\bibfnamefont {R.}~\bibnamefont
  {Lichtenth{\"a}ler}}, \bibinfo {author} {\bibfnamefont {K.~C.~C.}\
  \bibnamefont {Pires}}, \bibinfo {author} {\bibfnamefont {A.~M.}\ \bibnamefont
  {Moro}}, \bibinfo {author} {\bibfnamefont {A.}~\bibnamefont
  {L{\'e}pine-Szily}}, \bibinfo {author} {\bibfnamefont {V.}~\bibnamefont
  {Guimar{\~a}es}}, \bibinfo {author} {\bibfnamefont {D.~R.~J.}\ \bibnamefont
  {Mendes}}, \bibinfo {author} {\bibfnamefont {A.}~\bibnamefont {Arazi}},
  \bibinfo {author} {\bibfnamefont {M.}~\bibnamefont {Rodr{\'i}guez-Gallardo}},
  \bibinfo {author} {\bibfnamefont {A.}~\bibnamefont {Barioni}}, \bibinfo
  {author} {\bibfnamefont {V.}~\bibnamefont {Morcelle}}, \bibinfo {author}
  {\bibfnamefont {M.~C.}\ \bibnamefont {Morais}}, \bibinfo {author}
  {\bibfnamefont {O.}~\bibnamefont {Camargo}}, \bibinfo {author} {\bibfnamefont
  {J.}~\bibnamefont {Alc{\'a}ntara~Nu{\~n}ez}}, \ and\ \bibinfo {author}
  {\bibfnamefont {M.}~\bibnamefont {Assun{\c c}{\~a}o}},\ }\href@noop {}
  {\bibfield  {journal} {\bibinfo  {journal} {Phys. Rev. C}\ }\textbf {\bibinfo
  {volume} {81}},\ \bibinfo {pages} {044605} (\bibinfo {year}
  {2010})}\BibitemShut {NoStop}%
\bibitem [{\citenamefont {Appannababu}\ \emph {et~al.}(2019)\citenamefont
  {Appannababu}, \citenamefont {Lichtenth{\"a}ler}, \citenamefont {Alvarez},
  \citenamefont {Rodr{\'i}guez-Gallardo}, \citenamefont {L{\'e}pine-Szily},
  \citenamefont {Pires}, \citenamefont {Santos}, \citenamefont {Silva},
  \citenamefont {de~Faria}, \citenamefont {Guimar{\~a}es}, \citenamefont
  {Zevallos}, \citenamefont {Scarduelli}, \citenamefont {Assun{\c c}{\~a}o},
  \citenamefont {Shorto}, \citenamefont {Barioni}, \citenamefont
  {Alc{\'a}ntara-Nu{\~n}ez},\ and\ \citenamefont {Morcelle.}}]{Appa19}%
  \BibitemOpen
  \bibfield  {author} {\bibinfo {author} {\bibfnamefont {S.}~\bibnamefont
  {Appannababu}}, \bibinfo {author} {\bibfnamefont {R.}~\bibnamefont
  {Lichtenth{\"a}ler}}, \bibinfo {author} {\bibfnamefont {M.~A.~G.}\
  \bibnamefont {Alvarez}}, \bibinfo {author} {\bibfnamefont {M.}~\bibnamefont
  {Rodr{\'i}guez-Gallardo}}, \bibinfo {author} {\bibfnamefont {A.}~\bibnamefont
  {L{\'e}pine-Szily}}, \bibinfo {author} {\bibfnamefont {K.~C.~C.}\
  \bibnamefont {Pires}}, \bibinfo {author} {\bibfnamefont {O.~C.~B.}\
  \bibnamefont {Santos}}, \bibinfo {author} {\bibfnamefont {U.~U.}\
  \bibnamefont {Silva}}, \bibinfo {author} {\bibfnamefont {P.~N.}\ \bibnamefont
  {de~Faria}}, \bibinfo {author} {\bibfnamefont {V.}~\bibnamefont
  {Guimar{\~a}es}}, \bibinfo {author} {\bibfnamefont {E.~O.~N.}\ \bibnamefont
  {Zevallos}}, \bibinfo {author} {\bibfnamefont {V.}~\bibnamefont
  {Scarduelli}}, \bibinfo {author} {\bibfnamefont {M.}~\bibnamefont {Assun{\c
  c}{\~a}o}}, \bibinfo {author} {\bibfnamefont {J.~M.~B.}\ \bibnamefont
  {Shorto}}, \bibinfo {author} {\bibfnamefont {A.}~\bibnamefont {Barioni}},
  \bibinfo {author} {\bibfnamefont {J.}~\bibnamefont
  {Alc{\'a}ntara-Nu{\~n}ez}}, \ and\ \bibinfo {author} {\bibfnamefont
  {V.}~\bibnamefont {Morcelle.}},\ }\href@noop {} {\bibfield  {journal}
  {\bibinfo  {journal} {Phys. Rev. C}\ }\textbf {\bibinfo {volume} {99}},\
  \bibinfo {pages} {014601} (\bibinfo {year} {2019})}\BibitemShut {NoStop}%
\bibitem [{\citenamefont {Zerva}\ \emph {et~al.}(2012)\citenamefont {Zerva}
  \emph {et~al.}}]{Zer12}%
  \BibitemOpen
  \bibfield  {author} {\bibinfo {author} {\bibfnamefont {K.}~\bibnamefont
  {Zerva}} \emph {et~al.},\ }\href@noop {} {\bibfield  {journal} {\bibinfo
  {journal} {Eur. Phys. J. A}\ }\textbf {\bibinfo {volume} {48}},\ \bibinfo
  {pages} {102} (\bibinfo {year} {2012})}\BibitemShut {NoStop}%
\bibitem [{\citenamefont {Kundu}\ \emph {et~al.}(2017)\citenamefont {Kundu},
  \citenamefont {Santra}, \citenamefont {Pal}, \citenamefont {Chattopadhyay},
  \citenamefont {Tripathi}, \citenamefont {Roy}, \citenamefont {Nag},
  \citenamefont {Nayak}, \citenamefont {Saxena},\ and\ \citenamefont
  {Kailas}}]{Kun17}%
  \BibitemOpen
  \bibfield  {author} {\bibinfo {author} {\bibfnamefont {A.}~\bibnamefont
  {Kundu}}, \bibinfo {author} {\bibfnamefont {S.}~\bibnamefont {Santra}},
  \bibinfo {author} {\bibfnamefont {A.}~\bibnamefont {Pal}}, \bibinfo {author}
  {\bibfnamefont {D.}~\bibnamefont {Chattopadhyay}}, \bibinfo {author}
  {\bibfnamefont {R.}~\bibnamefont {Tripathi}}, \bibinfo {author}
  {\bibfnamefont {B.~J.}\ \bibnamefont {Roy}}, \bibinfo {author} {\bibfnamefont
  {T.~N.}\ \bibnamefont {Nag}}, \bibinfo {author} {\bibfnamefont {B.~K.}\
  \bibnamefont {Nayak}}, \bibinfo {author} {\bibfnamefont {A.}~\bibnamefont
  {Saxena}}, \ and\ \bibinfo {author} {\bibfnamefont {S.}~\bibnamefont
  {Kailas}},\ }\href@noop {} {\bibfield  {journal} {\bibinfo  {journal} {Phys.
  Rev. C}\ }\textbf {\bibinfo {volume} {95}},\ \bibinfo {pages} {034615}
  (\bibinfo {year} {2017})}\BibitemShut {NoStop}%
\bibitem [{\citenamefont {Satchler}\ and\ \citenamefont {Love}(1979)}]{Sat79}%
  \BibitemOpen
  \bibfield  {author} {\bibinfo {author} {\bibfnamefont {G.~R.}\ \bibnamefont
  {Satchler}}\ and\ \bibinfo {author} {\bibfnamefont {W.~G.}\ \bibnamefont
  {Love}},\ }\href@noop {} {\bibfield  {journal} {\bibinfo  {journal} {Phys.
  Rep.}\ }\textbf {\bibinfo {volume} {55}},\ \bibinfo {pages} {183 } (\bibinfo
  {year} {1979})}\BibitemShut {NoStop}%
\bibitem [{\citenamefont {Satchler}(1983)}]{Sat83}%
  \BibitemOpen
  \bibfield  {author} {\bibinfo {author} {\bibfnamefont {G.~R.}\ \bibnamefont
  {Satchler}},\ }\href@noop {} {\emph {\bibinfo {title} {{Direct Nuclear
  Reactions}}}}\ (\bibinfo  {publisher} {Clarendon Press, Oxford},\ \bibinfo
  {year} {1983})\BibitemShut {NoStop}%
\bibitem [{\citenamefont {Satchler}(1991)}]{Sat91}%
  \BibitemOpen
  \bibfield  {author} {\bibinfo {author} {\bibfnamefont {G.~R.}\ \bibnamefont
  {Satchler}},\ }\href@noop {} {\bibfield  {journal} {\bibinfo  {journal}
  {Phys. Rep.}\ }\textbf {\bibinfo {volume} {199}},\ \bibinfo {pages} {147}
  (\bibinfo {year} {1991})}\BibitemShut {NoStop}%
\bibitem [{\citenamefont {Satchler}\ \emph {et~al.}(1987)\citenamefont
  {Satchler}, \citenamefont {Nagarajan}, \citenamefont {Lilley},\ and\
  \citenamefont {Thompson}}]{Sat87}%
  \BibitemOpen
  \bibfield  {author} {\bibinfo {author} {\bibfnamefont {G.~R.}\ \bibnamefont
  {Satchler}}, \bibinfo {author} {\bibfnamefont {M.~A.}\ \bibnamefont
  {Nagarajan}}, \bibinfo {author} {\bibfnamefont {J.~S.}\ \bibnamefont
  {Lilley}}, \ and\ \bibinfo {author} {\bibfnamefont {I.~J.}\ \bibnamefont
  {Thompson}},\ }\href@noop {} {\bibfield  {journal} {\bibinfo  {journal} {Ann.
  Phys. (New York)}\ }\textbf {\bibinfo {volume} {178}},\ \bibinfo {pages}
  {110} (\bibinfo {year} {1987})}\BibitemShut {NoStop}%
\bibitem [{\citenamefont {Khoa}(1988)}]{Khoa88}%
  \BibitemOpen
  \bibfield  {author} {\bibinfo {author} {\bibfnamefont {D.~T.}\ \bibnamefont
  {Khoa}},\ }\href@noop {} {\bibfield  {journal} {\bibinfo  {journal} {Nucl.
  Phys. A}\ }\textbf {\bibinfo {volume} {484}},\ \bibinfo {pages} {376 }
  (\bibinfo {year} {1988})}\BibitemShut {NoStop}%
\bibitem [{\citenamefont {Brandan}\ and\ \citenamefont
  {Satchler}(1988)}]{Bran88}%
  \BibitemOpen
  \bibfield  {author} {\bibinfo {author} {\bibfnamefont {M.~E.}\ \bibnamefont
  {Brandan}}\ and\ \bibinfo {author} {\bibfnamefont {G.~R.}\ \bibnamefont
  {Satchler}},\ }\href@noop {} {\bibfield  {journal} {\bibinfo  {journal}
  {Nucl. Phys. A}\ }\textbf {\bibinfo {volume} {487}},\ \bibinfo {pages} {477 }
  (\bibinfo {year} {1988})}\BibitemShut {NoStop}%
\bibitem [{\citenamefont {Lenzi}\ \emph {et~al.}(1989)\citenamefont {Lenzi},
  \citenamefont {Vitturi},\ and\ \citenamefont {Zardi}}]{Len89}%
  \BibitemOpen
  \bibfield  {author} {\bibinfo {author} {\bibfnamefont {S.~M.}\ \bibnamefont
  {Lenzi}}, \bibinfo {author} {\bibfnamefont {A.}~\bibnamefont {Vitturi}}, \
  and\ \bibinfo {author} {\bibfnamefont {F.}~\bibnamefont {Zardi}},\
  }\href@noop {} {\bibfield  {journal} {\bibinfo  {journal} {Phys. Rev. C}\
  }\textbf {\bibinfo {volume} {40}},\ \bibinfo {pages} {2114} (\bibinfo {year}
  {1989})}\BibitemShut {NoStop}%
\bibitem [{\citenamefont {Satchler}(1994)}]{Sat94}%
  \BibitemOpen
  \bibfield  {author} {\bibinfo {author} {\bibfnamefont {G.~R.}\ \bibnamefont
  {Satchler}},\ }\href@noop {} {\bibfield  {journal} {\bibinfo  {journal}
  {Nucl. Phys. A}\ }\textbf {\bibinfo {volume} {579}},\ \bibinfo {pages} {241 }
  (\bibinfo {year} {1994})}\BibitemShut {NoStop}%
\bibitem [{\citenamefont {Brandan}\ and\ \citenamefont
  {Satchler}(1997{\natexlab{a}})}]{Bran97}%
  \BibitemOpen
  \bibfield  {author} {\bibinfo {author} {\bibfnamefont {M.~E.}\ \bibnamefont
  {Brandan}}\ and\ \bibinfo {author} {\bibfnamefont {G.~R.}\ \bibnamefont
  {Satchler}},\ }\href@noop {} {\bibfield  {journal} {\bibinfo  {journal}
  {Phys. Rep.}\ }\textbf {\bibinfo {volume} {285}},\ \bibinfo {pages} {143 }
  (\bibinfo {year} {1997}{\natexlab{a}})}\BibitemShut {NoStop}%
\bibitem [{\citenamefont {Brandan}\ and\ \citenamefont
  {Satchler}(1997{\natexlab{b}})}]{Bran97rep}%
  \BibitemOpen
  \bibfield  {author} {\bibinfo {author} {\bibfnamefont {M.~E.}\ \bibnamefont
  {Brandan}}\ and\ \bibinfo {author} {\bibfnamefont {G.~R.}\ \bibnamefont
  {Satchler}},\ }\href@noop {} {\bibfield  {journal} {\bibinfo  {journal}
  {Phys. Rep.}\ }\textbf {\bibinfo {volume} {285}},\ \bibinfo {pages} {143 }
  (\bibinfo {year} {1997}{\natexlab{b}})}\BibitemShut {NoStop}%
\bibitem [{\citenamefont {Pollarolo}\ \emph {et~al.}(1981)\citenamefont
  {Pollarolo}, \citenamefont {Broglia},\ and\ \citenamefont {Winther}}]{Pol81}%
  \BibitemOpen
  \bibfield  {author} {\bibinfo {author} {\bibfnamefont {G.}~\bibnamefont
  {Pollarolo}}, \bibinfo {author} {\bibfnamefont {R.~A.}\ \bibnamefont
  {Broglia}}, \ and\ \bibinfo {author} {\bibfnamefont {A.}~\bibnamefont
  {Winther}},\ }\href@noop {} {\bibfield  {journal} {\bibinfo  {journal}
  {Nuclear Physics A}\ }\textbf {\bibinfo {volume} {361}},\ \bibinfo {pages}
  {307 } (\bibinfo {year} {1981})}\BibitemShut {NoStop}%
\bibitem [{\citenamefont {Pollarolo}\ \emph {et~al.}(1983)\citenamefont
  {Pollarolo}, \citenamefont {Broglia},\ and\ \citenamefont {Winther}}]{Pol83}%
  \BibitemOpen
  \bibfield  {author} {\bibinfo {author} {\bibfnamefont {G.}~\bibnamefont
  {Pollarolo}}, \bibinfo {author} {\bibfnamefont {R.~A.}\ \bibnamefont
  {Broglia}}, \ and\ \bibinfo {author} {\bibfnamefont {A.}~\bibnamefont
  {Winther}},\ }\href@noop {} {\bibfield  {journal} {\bibinfo  {journal}
  {Nuclear Physics A}\ }\textbf {\bibinfo {volume} {406}},\ \bibinfo {pages}
  {369 } (\bibinfo {year} {1983})}\BibitemShut {NoStop}%
\bibitem [{\citenamefont {Sakuragi}(1987)}]{Sak87}%
  \BibitemOpen
  \bibfield  {author} {\bibinfo {author} {\bibfnamefont {Y.}~\bibnamefont
  {Sakuragi}},\ }\href@noop {} {\bibfield  {journal} {\bibinfo  {journal}
  {Phys. Rev. C}\ }\textbf {\bibinfo {volume} {35}},\ \bibinfo {pages} {2161}
  (\bibinfo {year} {1987})}\BibitemShut {NoStop}%
\bibitem [{\citenamefont {Alvarez}\ \emph {et~al.}(2003)\citenamefont
  {Alvarez}, \citenamefont {Chamon}, \citenamefont {Hussein}, \citenamefont
  {Pereira}, \citenamefont {Gasques}, \citenamefont {Rossi},\ and\
  \citenamefont {Silva}}]{Alv03}%
  \BibitemOpen
  \bibfield  {author} {\bibinfo {author} {\bibfnamefont {M.~A.~G.}\
  \bibnamefont {Alvarez}}, \bibinfo {author} {\bibfnamefont {L.~C.}\
  \bibnamefont {Chamon}}, \bibinfo {author} {\bibfnamefont {M.~S.}\
  \bibnamefont {Hussein}}, \bibinfo {author} {\bibfnamefont {D.}~\bibnamefont
  {Pereira}}, \bibinfo {author} {\bibfnamefont {L.~R.}\ \bibnamefont
  {Gasques}}, \bibinfo {author} {\bibfnamefont {E.~S.}\ \bibnamefont {Rossi}},
  \ and\ \bibinfo {author} {\bibfnamefont {C.~P.}\ \bibnamefont {Silva}},\
  }\href@noop {} {\bibfield  {journal} {\bibinfo  {journal} {Nucl. Phys. A}\
  }\textbf {\bibinfo {volume} {723}},\ \bibinfo {pages} {93} (\bibinfo {year}
  {2003})}\BibitemShut {NoStop}%
\bibitem [{\citenamefont {Pereira}\ \emph {et~al.}(2006)\citenamefont
  {Pereira}, \citenamefont {Rossi}, \citenamefont {Nobre}, \citenamefont
  {Chamon}, \citenamefont {Silva}, \citenamefont {Gasques}, \citenamefont
  {Alvarez}, \citenamefont {Ribas}, \citenamefont {Oliveira}, \citenamefont
  {Medina}, \citenamefont {Rao}, \citenamefont {Cybulska}, \citenamefont
  {Seale}, \citenamefont {Carlin}, \citenamefont {Gomes}, \citenamefont
  {Lubian},\ and\ \citenamefont {Anjos}}]{Per06}%
  \BibitemOpen
  \bibfield  {author} {\bibinfo {author} {\bibfnamefont {D.}~\bibnamefont
  {Pereira}}, \bibinfo {author} {\bibfnamefont {E.~S.}\ \bibnamefont {Rossi},
  \bibfnamefont {Jr.}}, \bibinfo {author} {\bibfnamefont {G.~P.~A.}\
  \bibnamefont {Nobre}}, \bibinfo {author} {\bibfnamefont {L.~C.}\ \bibnamefont
  {Chamon}}, \bibinfo {author} {\bibfnamefont {C.~P.}\ \bibnamefont {Silva}},
  \bibinfo {author} {\bibfnamefont {L.~R.}\ \bibnamefont {Gasques}}, \bibinfo
  {author} {\bibfnamefont {M.~A.~G.}\ \bibnamefont {Alvarez}}, \bibinfo
  {author} {\bibfnamefont {R.~V.}\ \bibnamefont {Ribas}}, \bibinfo {author}
  {\bibfnamefont {J.~R.~B.}\ \bibnamefont {Oliveira}}, \bibinfo {author}
  {\bibfnamefont {N.~H.}\ \bibnamefont {Medina}}, \bibinfo {author}
  {\bibfnamefont {M.~N.}\ \bibnamefont {Rao}}, \bibinfo {author} {\bibfnamefont
  {E.~W.}\ \bibnamefont {Cybulska}}, \bibinfo {author} {\bibfnamefont {W.~A.}\
  \bibnamefont {Seale}}, \bibinfo {author} {\bibfnamefont {N.}~\bibnamefont
  {Carlin}}, \bibinfo {author} {\bibfnamefont {P.~R.~S.}\ \bibnamefont
  {Gomes}}, \bibinfo {author} {\bibfnamefont {J.}~\bibnamefont {Lubian}}, \
  and\ \bibinfo {author} {\bibfnamefont {R.~M.}\ \bibnamefont {Anjos}},\
  }\href@noop {} {\bibfield  {journal} {\bibinfo  {journal} {Phys. Rev. C}\
  }\textbf {\bibinfo {volume} {74}},\ \bibinfo {pages} {034608} (\bibinfo
  {year} {2006})}\BibitemShut {NoStop}%
\bibitem [{\citenamefont {Fern{\'a}ndez-Garc{\'i}a}\ \emph
  {et~al.}(2010)\citenamefont {Fern{\'a}ndez-Garc{\'i}a}, \citenamefont
  {Rodr{\'i}guez-Gallardo}, \citenamefont {Alvarez},\ and\ \citenamefont
  {Moro}}]{Fer10}%
  \BibitemOpen
  \bibfield  {author} {\bibinfo {author} {\bibfnamefont {J.~P.}\ \bibnamefont
  {Fern{\'a}ndez-Garc{\'i}a}}, \bibinfo {author} {\bibfnamefont
  {M.}~\bibnamefont {Rodr{\'i}guez-Gallardo}}, \bibinfo {author} {\bibfnamefont
  {M.~A.~G.}\ \bibnamefont {Alvarez}}, \ and\ \bibinfo {author} {\bibfnamefont
  {A.~M.}\ \bibnamefont {Moro}},\ }\href@noop {} {\bibfield  {journal}
  {\bibinfo  {journal} {Nucl. Phys. A}\ }\textbf {\bibinfo {volume} {840}},\
  \bibinfo {pages} {19} (\bibinfo {year} {2010})}\BibitemShut {NoStop}%
\bibitem [{\citenamefont {Fern\'andez-Garc\'{\i}a}\ \emph
  {et~al.}(2015)\citenamefont {Fern\'andez-Garc\'{\i}a}, \citenamefont
  {Alvarez},\ and\ \citenamefont {Chamon}}]{Fer15}%
  \BibitemOpen
  \bibfield  {author} {\bibinfo {author} {\bibfnamefont {J.~P.}\ \bibnamefont
  {Fern\'andez-Garc\'{\i}a}}, \bibinfo {author} {\bibfnamefont {M.~A.~G.}\
  \bibnamefont {Alvarez}}, \ and\ \bibinfo {author} {\bibfnamefont {L.~C.}\
  \bibnamefont {Chamon}},\ }\href@noop {} {\bibfield  {journal} {\bibinfo
  {journal} {Phys. Rev. C}\ }\textbf {\bibinfo {volume} {92}},\ \bibinfo
  {pages} {014604} (\bibinfo {year} {2015})}\BibitemShut {NoStop}%
\bibitem [{\citenamefont {Gasques}\ \emph {et~al.}(2004)\citenamefont
  {Gasques}, \citenamefont {Chamon}, \citenamefont {Pereira}, \citenamefont
  {Alvarez}, \citenamefont {Rossi}, \citenamefont {Silva},\ and\ \citenamefont
  {Carlson}}]{Gas04}%
  \BibitemOpen
  \bibfield  {author} {\bibinfo {author} {\bibfnamefont {L.~R.}\ \bibnamefont
  {Gasques}}, \bibinfo {author} {\bibfnamefont {L.~C.}\ \bibnamefont {Chamon}},
  \bibinfo {author} {\bibfnamefont {D.}~\bibnamefont {Pereira}}, \bibinfo
  {author} {\bibfnamefont {M.~A.~G.}\ \bibnamefont {Alvarez}}, \bibinfo
  {author} {\bibfnamefont {E.~S.}\ \bibnamefont {Rossi}}, \bibinfo {author}
  {\bibfnamefont {C.~P.}\ \bibnamefont {Silva}}, \ and\ \bibinfo {author}
  {\bibfnamefont {B.~V.}\ \bibnamefont {Carlson}},\ }\href@noop {} {\bibfield
  {journal} {\bibinfo  {journal} {Phys. Rev. C}\ }\textbf {\bibinfo {volume}
  {69}},\ \bibinfo {pages} {034603} (\bibinfo {year} {2004})}\BibitemShut
  {NoStop}%
\bibitem [{\citenamefont {Canto}\ \emph {et~al.}(2009)\citenamefont {Canto},
  \citenamefont {Gomes}, \citenamefont {Lubian}, \citenamefont {Chamon},\ and\
  \citenamefont {Crema}}]{Can09}%
  \BibitemOpen
  \bibfield  {author} {\bibinfo {author} {\bibfnamefont {L.}~\bibnamefont
  {Canto}}, \bibinfo {author} {\bibfnamefont {P.}~\bibnamefont {Gomes}},
  \bibinfo {author} {\bibfnamefont {J.}~\bibnamefont {Lubian}}, \bibinfo
  {author} {\bibfnamefont {L.}~\bibnamefont {Chamon}}, \ and\ \bibinfo {author}
  {\bibfnamefont {E.}~\bibnamefont {Crema}},\ }\href@noop {} {\bibfield
  {journal} {\bibinfo  {journal} {Nucl. Phys. A}\ }\textbf {\bibinfo {volume}
  {821}},\ \bibinfo {pages} {51} (\bibinfo {year} {2009})}\BibitemShut
  {NoStop}%
\bibitem [{\citenamefont {Nobre}\ \emph
  {et~al.}(2007{\natexlab{a}})\citenamefont {Nobre}, \citenamefont {Silva},
  \citenamefont {Chamon},\ and\ \citenamefont {Carlson}}]{Nob76}%
  \BibitemOpen
  \bibfield  {author} {\bibinfo {author} {\bibfnamefont {G.~P.~A.}\
  \bibnamefont {Nobre}}, \bibinfo {author} {\bibfnamefont {C.~P.}\ \bibnamefont
  {Silva}}, \bibinfo {author} {\bibfnamefont {L.~C.}\ \bibnamefont {Chamon}}, \
  and\ \bibinfo {author} {\bibfnamefont {B.~V.}\ \bibnamefont {Carlson}},\
  }\href@noop {} {\bibfield  {journal} {\bibinfo  {journal} {Phys. Rev. C}\
  }\textbf {\bibinfo {volume} {76}},\ \bibinfo {pages} {024605} (\bibinfo
  {year} {2007}{\natexlab{a}})}\BibitemShut {NoStop}%
\bibitem [{\citenamefont {Nobre}\ \emph
  {et~al.}(2007{\natexlab{b}})\citenamefont {Nobre}, \citenamefont {Chamon},
  \citenamefont {Carlson}, \citenamefont {Thompson},\ and\ \citenamefont
  {Gasques}}]{Nob78}%
  \BibitemOpen
  \bibfield  {author} {\bibinfo {author} {\bibfnamefont {G.}~\bibnamefont
  {Nobre}}, \bibinfo {author} {\bibfnamefont {L.}~\bibnamefont {Chamon}},
  \bibinfo {author} {\bibfnamefont {B.}~\bibnamefont {Carlson}}, \bibinfo
  {author} {\bibfnamefont {I.}~\bibnamefont {Thompson}}, \ and\ \bibinfo
  {author} {\bibfnamefont {L.}~\bibnamefont {Gasques}},\ }\href@noop {}
  {\bibfield  {journal} {\bibinfo  {journal} {Nucl. Phys. A}\ }\textbf
  {\bibinfo {volume} {786}},\ \bibinfo {pages} {90} (\bibinfo {year}
  {2007}{\natexlab{b}})}\BibitemShut {NoStop}%
\bibitem [{\citenamefont {Nobre}\ \emph
  {et~al.}(2007{\natexlab{c}})\citenamefont {Nobre}, \citenamefont {Chamon},
  \citenamefont {Gasques}, \citenamefont {Carlson},\ and\ \citenamefont
  {Thompson}}]{Nob07}%
  \BibitemOpen
  \bibfield  {author} {\bibinfo {author} {\bibfnamefont {G.~P.~A.}\
  \bibnamefont {Nobre}}, \bibinfo {author} {\bibfnamefont {L.~C.}\ \bibnamefont
  {Chamon}}, \bibinfo {author} {\bibfnamefont {L.~R.}\ \bibnamefont {Gasques}},
  \bibinfo {author} {\bibfnamefont {B.~V.}\ \bibnamefont {Carlson}}, \ and\
  \bibinfo {author} {\bibfnamefont {I.~J.}\ \bibnamefont {Thompson}},\
  }\href@noop {} {\bibfield  {journal} {\bibinfo  {journal} {Phys. Rev. C}\
  }\textbf {\bibinfo {volume} {75}},\ \bibinfo {pages} {044606} (\bibinfo
  {year} {2007}{\natexlab{c}})}\BibitemShut {NoStop}%
\bibitem [{\citenamefont {Feshbach}(1992)}]{Fesh92}%
  \BibitemOpen
  \bibfield  {author} {\bibinfo {author} {\bibfnamefont {H.}~\bibnamefont
  {Feshbach}},\ }\href@noop {} {\bibfield  {journal} {\bibinfo  {journal}
  {Theoretical Nuclear Physics, Wiley, New York, 1992}\ }\textbf {\bibinfo
  {volume} {1}},\ \bibinfo {pages} {1} (\bibinfo {year} {1992})}\BibitemShut
  {NoStop}%
\bibitem [{\citenamefont {Carlson}\ and\ \citenamefont
  {Hirata}(2000)}]{Carl00}%
  \BibitemOpen
  \bibfield  {author} {\bibinfo {author} {\bibfnamefont {B.~V.}\ \bibnamefont
  {Carlson}}\ and\ \bibinfo {author} {\bibfnamefont {D.}~\bibnamefont
  {Hirata}},\ }\href@noop {} {\bibfield  {journal} {\bibinfo  {journal} {Phys.
  Rev. C}\ }\textbf {\bibinfo {volume} {62}},\ \bibinfo {pages} {054310}
  (\bibinfo {year} {2000})}\BibitemShut {NoStop}%
\bibitem [{\citenamefont {Lalazissis}\ \emph {et~al.}(1997)\citenamefont
  {Lalazissis}, \citenamefont {Konig},\ and\ \citenamefont {Ring}}]{Lala02}%
  \BibitemOpen
  \bibfield  {author} {\bibinfo {author} {\bibfnamefont {G.~A.}\ \bibnamefont
  {Lalazissis}}, \bibinfo {author} {\bibfnamefont {J.}~\bibnamefont {Konig}}, \
  and\ \bibinfo {author} {\bibfnamefont {P.}~\bibnamefont {Ring}},\ }\href@noop
  {} {\bibfield  {journal} {\bibinfo  {journal} {Phys. Rev. C}\ }\textbf
  {\bibinfo {volume} {55}},\ \bibinfo {pages} {540} (\bibinfo {year}
  {1997})}\BibitemShut {NoStop}%
\bibitem [{\citenamefont {Nik\ifmmode \check{s}\else
  \v{s}\fi{}i\ifmmode~\acute{c}\else \'{c}\fi{}}\ \emph
  {et~al.}(2002)\citenamefont {Nik\ifmmode \check{s}\else
  \v{s}\fi{}i\ifmmode~\acute{c}\else \'{c}\fi{}}, \citenamefont {Vretenar},
  \citenamefont {Finelli},\ and\ \citenamefont {Ring}}]{Nik02}%
  \BibitemOpen
  \bibfield  {author} {\bibinfo {author} {\bibfnamefont {T.}~\bibnamefont
  {Nik\ifmmode \check{s}\else \v{s}\fi{}i\ifmmode~\acute{c}\else \'{c}\fi{}}},
  \bibinfo {author} {\bibfnamefont {D.}~\bibnamefont {Vretenar}}, \bibinfo
  {author} {\bibfnamefont {P.}~\bibnamefont {Finelli}}, \ and\ \bibinfo
  {author} {\bibfnamefont {P.}~\bibnamefont {Ring}},\ }\href@noop {} {\bibfield
   {journal} {\bibinfo  {journal} {Phys. Rev. C}\ }\textbf {\bibinfo {volume}
  {66}},\ \bibinfo {pages} {024306} (\bibinfo {year} {2002})}\BibitemShut
  {NoStop}%
\bibitem [{\citenamefont {Alkhazov}\ \emph {et~al.}(1997)\citenamefont
  {Alkhazov}, \citenamefont {Andronenko}, \citenamefont {Dobrovolsky},
  \citenamefont {Egelhof}, \citenamefont {Gavrilov}, \citenamefont {Geissel},
  \citenamefont {Irnich}, \citenamefont {Khanzadeev}, \citenamefont {Korolev},
  \citenamefont {Lobodenko}, \citenamefont {M{\"u}nzenberg}, \citenamefont
  {Mutterer}, \citenamefont {Neumaier}, \citenamefont {Nickel}, \citenamefont
  {Schwab}, \citenamefont {Seliverstov}, \citenamefont {Suzuki}, \citenamefont
  {Theobald}, \citenamefont {Timofeev}, \citenamefont {Vorobyov},\ and\
  \citenamefont {Yatsoura}}]{Alk97}%
  \BibitemOpen
  \bibfield  {author} {\bibinfo {author} {\bibfnamefont {G.~D.}\ \bibnamefont
  {Alkhazov}}, \bibinfo {author} {\bibfnamefont {M.~N.}\ \bibnamefont
  {Andronenko}}, \bibinfo {author} {\bibfnamefont {A.~V.}\ \bibnamefont
  {Dobrovolsky}}, \bibinfo {author} {\bibfnamefont {P.}~\bibnamefont
  {Egelhof}}, \bibinfo {author} {\bibfnamefont {G.~E.}\ \bibnamefont
  {Gavrilov}}, \bibinfo {author} {\bibfnamefont {H.}~\bibnamefont {Geissel}},
  \bibinfo {author} {\bibfnamefont {H.}~\bibnamefont {Irnich}}, \bibinfo
  {author} {\bibfnamefont {A.~V.}\ \bibnamefont {Khanzadeev}}, \bibinfo
  {author} {\bibfnamefont {G.~A.}\ \bibnamefont {Korolev}}, \bibinfo {author}
  {\bibfnamefont {A.~A.}\ \bibnamefont {Lobodenko}}, \bibinfo {author}
  {\bibfnamefont {G.}~\bibnamefont {M{\"u}nzenberg}}, \bibinfo {author}
  {\bibfnamefont {M.}~\bibnamefont {Mutterer}}, \bibinfo {author}
  {\bibfnamefont {S.~R.}\ \bibnamefont {Neumaier}}, \bibinfo {author}
  {\bibfnamefont {F.}~\bibnamefont {Nickel}}, \bibinfo {author} {\bibfnamefont
  {W.}~\bibnamefont {Schwab}}, \bibinfo {author} {\bibfnamefont {D.~M.}\
  \bibnamefont {Seliverstov}}, \bibinfo {author} {\bibfnamefont
  {T.}~\bibnamefont {Suzuki}}, \bibinfo {author} {\bibfnamefont {J.~P.}\
  \bibnamefont {Theobald}}, \bibinfo {author} {\bibfnamefont {N.~A.}\
  \bibnamefont {Timofeev}}, \bibinfo {author} {\bibfnamefont {A.~A.}\
  \bibnamefont {Vorobyov}}, \ and\ \bibinfo {author} {\bibfnamefont {V.~I.}\
  \bibnamefont {Yatsoura}},\ }\href@noop {} {\bibfield  {journal} {\bibinfo
  {journal} {Phys. Rev. Lett.}\ }\textbf {\bibinfo {volume} {78}},\ \bibinfo
  {pages} {2313} (\bibinfo {year} {1997})}\BibitemShut {NoStop}%
\end{thebibliography}%

\end{document}